\documentclass[prb,superscriptaddress, preprintnumbers, twocolumn, letterpaper]{revtex4}

\usepackage{times}
\usepackage{amsmath}
\usepackage{amsthm}
\usepackage{amssymb}
\usepackage{amsbsy}
\usepackage{amsfonts}
\usepackage{bm}
\usepackage{graphicx}
\usepackage{multirow}
\usepackage[all,pdftex,cmtip]{xy}

\begin{document}


\title{Interactions along an Entanglement Cut in 2+1D Abelian Topological Phases}

\author{Jennifer Cano}
\affiliation{Department of Physics, University of California, Santa Barbara,
California 93106, USA}
\author{Taylor L. Hughes}
\affiliation{Department of Physics, Institute for Condensed Matter Theory,
University of Illinois, Urbana IL 61801, USA}
\author{Michael Mulligan}
\affiliation{Department of Physics, Stanford University, Stanford, CA 94305, USA}
\affiliation{Kavli Institute for Theoretical Physics, University of California, Santa Barbara, California 93106, USA}

\begin{abstract}
A given fractional quantum Hall state may admit multiple, distinct edge phases on its boundary.
We explore the implications that multiple edge phases have for the entanglement spectrum and entropy of a given bulk state.
We describe the precise manner in which the entanglement spectrum depends upon local (tunneling) interactions along an entanglement cut and throughout the bulk.
The sensitivity to local conditions near the entanglement cut appears not only in gross features of the spectrum, but can also manifest itself in an additive, positive constant correction to the topological entanglement entropy, i.e., it increases its magnitude.
A natural interpretation for this result is that the tunneling interactions across an entanglement cut can function as a barrier to certain types of quasiparticle transport across the cut, thereby, lowering the total entanglement between the two regions.
\end{abstract}

\maketitle

\section{Introduction}

\subsection{Motivation}

The bulk-boundary correspondence in topologically ordered systems relates a given bulk state to its possible boundary excitations.\cite{wittenjones, Elitzur89, Wengapless}
This correspondence is remarkable because it effectively implies that the bulk and boundary data are one and the same and that one might learn something about the properties of the bulk of a system just by studying the boundary.
However, in the context of 2+1D Abelian topological phases or, equivalently,
\footnote{Here, we make use of our ability to describe the long-wavelength (topological) properties of any 2+1D Abelian topological phase or Abelian fractional quantum Hall state using Abelian Chern-Simons theory.} 
Abelian fractional quantum Hall (FQH) states -- the focus of study of this paper -- it has recently been stressed,\cite{PlamadealaE8, generalstableequivalence} following previous work,\cite{frohlichzee91, frohlichthiran94, frohlichmany95, frohlichstuderthiran, Kao99} that the map connecting the possible edge excitations to a given bulk state is many-to-one.
In other words, a bulk state may admit multiple, distinct edge phases. The distinction between the edge phases can be quite dramatic, for example, in some cases a bulk phase may admit a purely fermionic edge and a distinct, purely bosonic, edge description.
\footnote{A `bosonic' edge is one for which there is no quasiparticle creation operator that commutes with all others operators, but anti-commutes with itself, i.e., all operators have integer spin.
A `fermionic' edge has at least one operator with half-integer spin.} 

The underlying reason that multiple edge terminations can occur lies in the possibility that different edge interactions can, if sufficiently strong, drive the low-lying edge excitations into distinct low-energy phases.
The $\nu=2/3$ FQH state is a well known example of a state that can host either a clean, or disorder-dominated, gapless edge phase.\cite{KFPrandom, KFrandom}
Similarly, the toric code\cite{kitaevtoriccode} admits distinct gapped edge phases as well as intervening gapless critical points.\cite{bravyi1998,Haldanestability, wangwenboundarydegen, levinprotected, classtops}
(see Refs. [\onlinecite{KitaevKonggappedboundaries, FuchsSchweigertValentino, KapustinSaulinatopboundary, Kapustingsdegen, LanWangWen}] for a more mathematical discussion of possible boundary interactions in various non-chiral 2+1D topological phases).
On the other hand, fully chiral examples can occur in QH systems at both integer, $\nu = 8$ and $12$, and fractional fillings lying at $\nu = 8/7$ and $12/11$, among others.\cite{PlamadealaE8, generalstableequivalence}
To be precise, an edge is said to be fully chiral if all edge excitations are, say, left-moving with no right-movers; chiral, if there are more, say, left-movers than right-movers; and non-chiral, if there are an equal number of left-movers and right-movers.
Remarkably, {\it every} Laughlin state state at inverse filling $\nu^{-1} = 2m+1$ admits more than one edge phase, however, only one phase is guaranteed to be fully chiral.\cite{generalstableequivalence}
Distinct fully chiral edge terminations can (nearly always
\footnote{One exception occurs for short-range entangled, bosonic systems with 16 fully-chiral edge modes: the exponents of operators in the two distinct edge phases are determined by the $E_8 \oplus E_8$ and ${\rm Spin}(32)/\mathbb{Z}_2$ lattices, respectively, which happen to have identical spectra and degeneracy.
Higher-point correlation functions are needed to differentiate these two edge phases.})
be differentiated by experimentally-measurable tunneling exponents.\cite{generalstableequivalence}
These exponents characterize the differential conductance, say, across a Hall sample or into the edge from a Fermi liquid lead, at a quantum point contact.  
Distinct non-chiral phases may also possess distinguishing signatures.\cite{bosonedgephases}

Additional insight into the nature of a particular QH state is furnished by the topological entanglement entropy and, more generally, the entanglement spectrum (both defined below).\cite{LevinWenentropy, KitaevPreskillentropy, hammakitaev, lihaldane}
Such entanglement measures are sensitive to long-ranged correlations within a state that may be otherwise invisible to local order parameters.
Interestingly, arguments suggest that the physical edge spectrum can be reconstructed from the (bulk) entanglement spectrum which is only a function of the ground-state wavefunction.\cite{lihaldane, Fidkowski2010,qikatsuraludwig,chandran2011, swinglesenthil}
Thus, since there is an expected bulk-boundary entanglement correspondence, it is natural to ask how the entanglement spectrum is compatible with the different possible edge phases of a particular state. For example, given a spatial cut of a particular bulk wavefunction, to which distinct edge theory does the entanglement spectrum correspond? The distinct edge phases of a given bulk phase will have distinguishing spectral features in their energies (at least in many cases) so the answer to this question is likely to be non-trivial since we would expect the entanglement spectra to also have distinguishing characteristics. This question provided the motivation for this work, but as will be shown below, our results have additional remarkable implications. 

Thus, in this article, we study the entanglement entropy and the entanglement spectrum for 2+1D Abelian topological phases that admit multiple edge phases or multiple gapped interfaces.
As anticipated from the physics that may occur at an actual boundary of the system, we describe how {\it both} the entanglement entropy and the entanglement spectrum are sensitive to the possible interactions occurring near an entangling cut, and in general, throughout the entire bulk, which is reminiscent of recent work.\cite{ChandranKhemaniSondhi, CasiniHuertaRosabal, OhmoriTachikawa}
In particular, we find that the entanglement entropy can receive a constant (negative-definite) sub-leading correction that depends upon the bulk state under consideration \emph{and} the interactions occurring in the neighborhood of the entanglement cut.
We primarily focus upon fully chiral states, however, our results and techniques apply equally well to non-chiral systems and we provide a few examples of these as well. Before we provide a summary of our results, let us review the concepts of quantum entanglement to provide the relevant context for our work.

\subsection{Review of Entanglement Entropy and Spectrum}

Given a state $|\psi\rangle$ and a bipartition of the Hilbert space, ${\cal H} = {\cal H}_A \otimes {\cal H}_B$, the entanglement entropy is equal to the von-Neumann entropy of the reduced density matrix, $\rho_A = {\rm Tr}_B(|\psi\rangle \langle \psi |)$:
\begin{align}
S(\rho_A) = - {\rm Tr}_A\Big(\rho_A \log(\rho_A)\Big), 
\end{align}
where the subscript $A$ on the ${\rm Tr}$ operation indicates the subspace over which the trace is performed.
If $|\psi\rangle$ is a pure state then $S(\rho_A) = S(\rho_B)$.
In this paper, we shall be solely concerned with a Hilbert space division associated to the degrees of freedom living in spatially distinct regions: $A$ and, its complement, $B$.
\footnote{There is no ambiguity in this Hilbert space division given the explicit lattice construction of our states.} 
Therefore, the \emph{entanglement cut} can be identified with the boundary between regions $A$ and $B$.

The entanglement entropy of a gapped system in 2+1D takes the scaling form:
\begin{align}
S(\rho_A) = \alpha \ell - \gamma,
\end{align}
where subdominant corrections in the $\ell \rightarrow \infty$ limit have been suppressed, and we assume that region $A$ has a smooth boundary.
The linear size $\ell$ of region $A$ is assumed to obey $\xi \ll \ell \ll {\cal L}$, where $\xi$ is the correlation length (which is finite due to the bulk gap) and ${\cal L}$ is the linear size of the system.
The non-universal constant $\alpha$ depends upon the regularization of the theory -- for instance, it can be a function of the lattice cutoff.
However, the {\it universal} constant $\gamma$  is called the topological entanglement entropy (TEE) and is known to be a robust feature of a gapped phase.
The value of the TEE partially characterizes the topological phase and, when the region $A$ has the shape of a disk with smooth boundary, takes the celebrated value\cite{LevinWenentropy, KitaevPreskillentropy, hammakitaev}:
\begin{align}
\label{standardTEE}
\gamma = {1 \over 2} \log\Big({\cal D}^2\Big) = {1 \over 2} \log\Big(\sum_{a=1}^M d_a^2\Big).
\end{align}
Here the total quantum dimension ${\cal D}$ of a particular phase can be expressed in terms of the sum of the squares of the individual quantum dimensions $d_a$ associated to the emergent quasiparticle excitations of the phase, which we have labeled by $a.$
The individual quantum dimensions $d_a$  determine the asymptotic growth of the Hilbert space dimension of a configuration of N-quasiparticles of type $a$ via ${\rm dim}(H^{(N)}_a) \sim d_a^N$.
For further details about these quantities, see Ref. [\onlinecite{nonabelianreview}].

In the literature so far, it has been shown that as long as region $A$ has topology of a smooth disk, the TEE is given by the expression in Eqn. (\ref{standardTEE}).
For more elaborate topologies -- for instance, when the state is defined on the torus, and region $A$ contains a non-contractible cycle -- the constant sub-leading term in the entanglement entropy may take a different form which captures additional identifying properties of the phase.
These constant sub-leading `corrections' depend upon the topology of the total space, the topology of $A$, the particular linear combination of degenerate ground states, and the modular ${\cal S}$-matrix of the underlying topological field theory.\cite{fradkintopologiesent, groverbraiding}
Additional constant corrections can arise from a departure from the strict $\xi/\ell \rightarrow 0$ limit and when the boundary of $A$ has corners, i.e. when the boundary is not smooth.\cite{PapanikolaouRamanFradkin} However, from the quite general arguments in, for example, Ref. [\onlinecite{fradkintopologiesent,groverbraiding}], one expects that when the boundary is smooth, and the system is gapped, then any constant, sub-leading contribution to the entanglement entropy in 2+1D has a topological origin. 

The entanglement spectrum is defined to be the eigenvalues of the entanglement Hamiltonian $H_A$
\footnote{The entanglement Hamiltonian also goes by the name of the modular Hamiltonian in the context of axiomatic quantum field theory.} 
that is defined by writing the density matrix in the form:
\begin{align}
\rho_A = {e^{- H_A} \over Z_A},
\end{align}
where 
\begin{align}
Z_A = {\rm Tr}_A \Big( e^{- H_A}\Big).
\end{align} 
With this definition, the entanglement entropy is identified with the thermal entropy of the ensemble defined by $\rho_A$ at a temperature equal to unity.

Over the past five years, there has been an enormous outpouring of research on identifying phases of matter, especially topological phases, using the entanglement spectrum. 
The initial work focused on (fractional) quantum Hall states.\cite{lihaldane,Regnault2009,Thomale2010,lauchli2010,regnault2011,sterdyniak2011,hermanns2011,sterdyniak2011a,papic2011,schliemann2011,sterdyniak2012,dubail2012,Dubail2012more, Simon2012, Simon2013}
This work was soon extended to the study of topological insulators and other symmetry-protected topological (SPT) phases,\cite{Fidkowski2010,Prodan2010,Pollman2010,Turner2010,Hughes2011,alex2011,turner2011a,gilbert2012,fang2013}  more general topologically ordered phases,\cite{fradkintopologiesent,Flammia2009,Yao2010,dubail2011,groverbraiding,mondragon2014} and even disordered systems.\cite{refael2004,jia2008,Prodan2010,gilbert2012,chen2012,mondragon2013,pouranvari2013,mondragon2014a,pouranvari2014}
As we will mention further below, one major outcome of this research was the strong evidence for an entanglement-edge correspondence, i.e., the low-energy entanglement spectrum is connected with the low-energy spectrum of the physical edge states.\cite{KitaevPreskillentropy, lihaldane, Fidkowski2010,qikatsuraludwig,chandran2011,swinglesenthil}

\subsection{Summary of Results}

In this article, we utilize the coupled wire construction\cite{KaneMukhopadhyayLubensky, TeoKanewires, thomalewires} to study the entanglement entropy and entanglement spectrum of Abelian topological phases on the cylinder, following the methods described in Refs. [\onlinecite{lundgrenentanglement, chenentanglement}] (see Ref. [\onlinecite{FurukawaKim}] for a related calculation). Using this construction, and a smooth entanglement cut between coupled wires, we illustrate quite generally that there exist additional sub-leading, constant corrections to the entanglement entropy. 
Our primary focus is upon fully-chiral phases built from local fermions, however we expect our methods to readily generalize to non-chiral, non-Abelian, and/or bosonic states. We give extra examples of some of these below as well, although we leave the consideration of non-Abelian states to future work. Let us now provide a summary of our results before moving on to the detailed calculations and examples.

\subsubsection{Entanglement Entropy}

Interestingly, we find that {\it both} the entropy and the spectrum are sensitive to the interactions near an entanglement cut that runs parallel to the wires.
In particular, we find that $\gamma$ can receive a positive constant correction that depends upon the physical interactions near the would-be entanglement cut.
This positive correction to $\gamma$ means that the entanglement entropy $S(\rho_A)$ decreases, and the magnitude of the TEE increases.
We believe these corrections to be universal and analogous to similar sub-leading corrections that can occur in 1+1D conformal field theory.\cite{CardyCalabreseentropy, schollwockbcentangle,affleckbcentangle, lauchlibcentropy, OhmoriTachikawa}

The dependence of the entanglement entropy on the interactions near an entangling cut arises as follows within our approach.
The states that we study are made by sewing together a collection of parallel 1+1D wires each hosting a non-chiral Luttinger liquid (see Fig. \ref{fig:wirearray}).\cite{KaneMukhopadhyayLubensky, TeoKanewires}
Two wires are said to be sewn together if the right-moving modes on the first wire and the left-moving modes on the second are perturbed by a gap-generating perturbation within this right-left sector such that the low-energy degrees of freedom are completely gapped.
A gapped, 2+1D state may then be obtained by populating the entire plane with an array of such coupled parallel wires.

\begin{figure}[h!]
  \centering
\includegraphics[width=.4\linewidth]{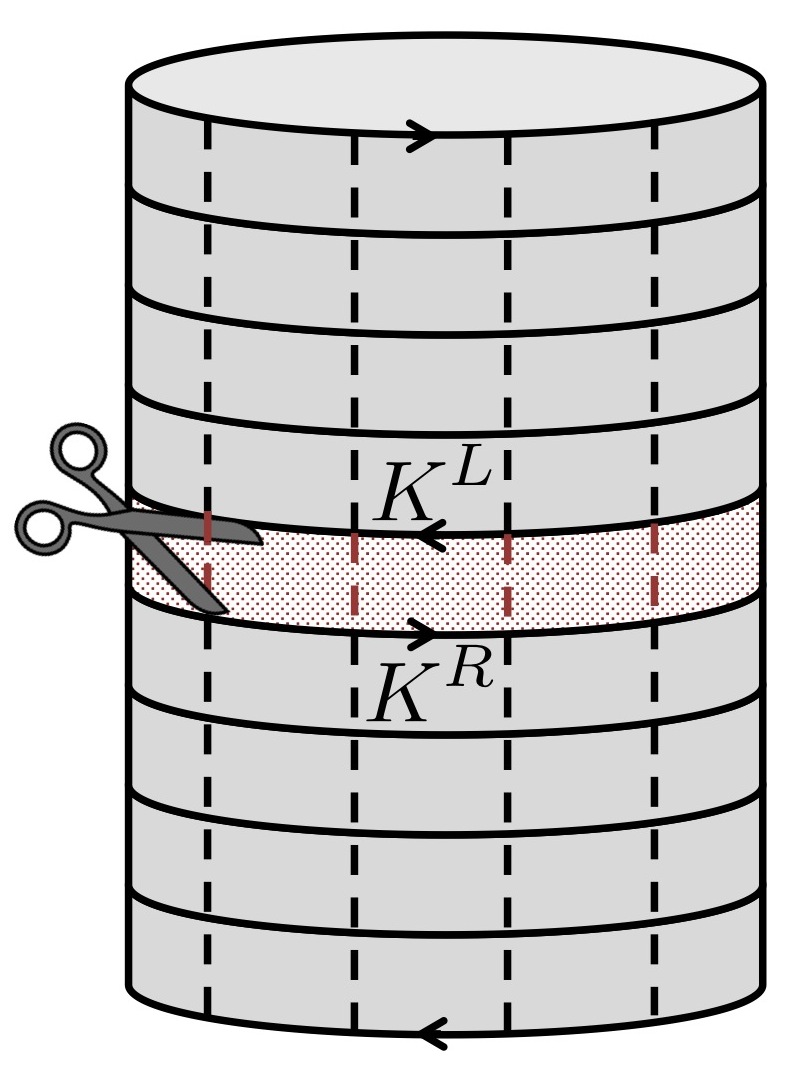}\\
\caption{An entanglement cut between any two wires (horizontal lines) is sensitive to the phases, denoted here by $K^{R/L}$, of the edge modes occurring at the cut and the interactions across the cut. 
Interactions between wires are denoted by dashed lines.
Interactions across the entanglement cut are represented with red dashed lines to distinguish them from interactions in the bulk.
In general, these interactions need not be the same as those in the bulk.}\label{fig:wirecut}
\end{figure}

An entanglement cut between two wires is sensitive to the particular interactions that couple those wires together, as illustrated in Fig \ref{fig:wirecut}.
From this point of view, a correction to the entanglement entropy is not totally unexpected, since
the tunneling perturbations determine how quasiparticles are constrained to move between wires.
We show that there exist choices of tunneling terms for which only a restricted set of local quasiparticles can pass at any one time, e.g. consider an electronic system in which electron-pair tunneling is favored over single-electron tunneling.
When such tunneling terms are the most dominant coupling between the wires along which we choose to partition the system, then there exists a barrier to the arbitrary movement of quasiparticles across the cut.
In fact, quasiparticles that are prohibited from moving between the subsystems across the cut may or may not be restricted within their respective subsystems, and both cases lead to interesting consequences.
If the tunneling interactions localize certain quasiparticles within their respective subregions, we have more information about the state in question, e.g., there is individual number conservation of the localized quasiparticles within each subregion, as opposed to the total system.
It is natural to expect that the entanglement between two such regions decreases when such tunneling perturbations become dominant across the entanglement cut and this is precisely what we find.

\begin{figure}[h!]
  \centering
\includegraphics[width=.4\linewidth]{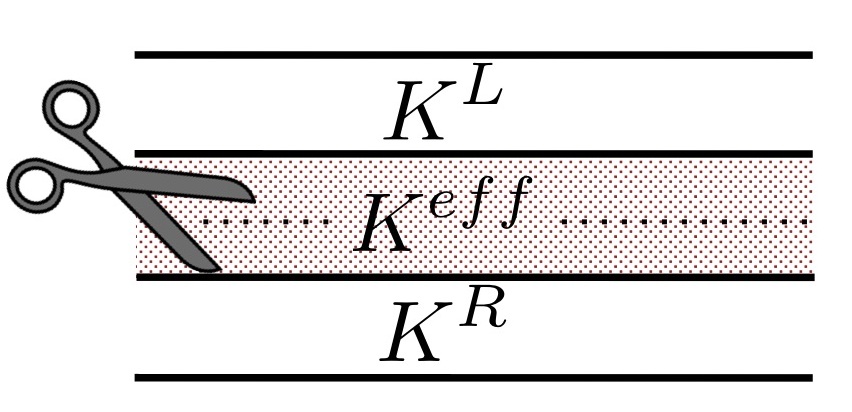}\\
\caption{The effect of the tunneling interactions across the cut on the entanglement entropy can be understood by introducing the $K^{\rm eff}$-matrix.
It is the fundamental quasiparticles of a theory defined by $K^{\rm eff}$ that may tunnel unimpeded across the entanglement cut.}\label{fig:effectivephase}
\end{figure}

To better understand the form of the constant correction to the entanglement entropy, it is helpful to provide an alternative interpretation of the topological entanglement entropy.
For an Abelian topological phase defined by a $K$-matrix, the topological entanglement entropy (for a disk geometry with smooth boundary or the topology that we study) takes the value $\gamma_K = \log \sqrt{|\det(K)|}$.
Within the wire construction, the $K$-matrix also determines the precise form of local tunneling operators that transfer (fundamental) electrons and, possibly their composites, across the entanglement cut.
Thus, we may interpret the topological entanglement entropy as a, rather coarse, measure of the fundamental quasiparticles that can tunnel across the cut through its dependence upon the $K$-matrix.
\footnote{As we will review momentarily, the $K$-matrix is equal to the Gram matrix of the underlying fundamental quasiparticle lattice that is determined by the particular topological phase under study. 
The determinant of the $K$-matrix is equal to the volume of the unit cell of the fundamental quasiparticle, i.e., electron, lattice.\cite{Read90, WenZee92}
In the $\nu=1/3$ Laughlin state, the quasiparticle lattice has volume $1/3$, while the fundamental quasiparticle or electron lattice has volume $3$.}

As we show in Eqn. (\ref{entropygeneral}), the {\it total} constant sub-leading term in the entanglement entropy -- including the corrections we find -- can be written in terms of an {\it effective} $K$-matrix, which is different from the original and determined by the neighboring topological phases and most dominant interactions between edge modes along the entanglement cut -- see Fig. \ref{fig:wirecut}.
The effective $K$-matrix incorporates any additional restrictions imposed by the structure of the theory that describes the edge modes along the entanglement cut, or in general how the bulk phase is sewn together -- see Fig. \ref{fig:effectivephase}.
It is the `fundamental' quasiparticles, defined with respect to the effective $K$-matrix, that may tunnel unimpeded across the entanglement cut. 
Consistent with the above reasoning, we find that the constant sub-leading term in the entropy takes the form:
\begin{align}
- \gamma_{K^{\rm eff}} = - \log\sqrt{|\det(K^{\rm eff})|}.
\end{align}


We study examples where this phenomenon occurs at both integer fillings, $\nu = 2, \nu = 4$, and $\nu=8$ and fractional fillings, $\nu = 4/3$ and $\nu=15/8$. We also study cases of interfaces between inequivalent topological phases that support \emph{gapped} interfaces, and show that in those cases a cut at the interface also yields a non-vanishing sub-leading constant contribution. 
We believe this phenomenon to be rather generic as our examples readily generalize to many other Abelian states as well.

\subsubsection{Caveats}

We note that such constant corrections to the entropy do not contradict the seminal results in Refs. [\onlinecite{LevinWenentropy, KitaevPreskillentropy}].
The interactions we study occur along the entire entanglement cut instead of only at select points along the cut.
In fact, we view our result as bolstering the general sentiment of such works in the sense that the corrections we find highlight additional characteristics of a given topological phase that entanglement can probe.

Two states may be identified if there exists a finite-depth local unitary operator that transforms one state into the other, consistent with any symmetries that are to be preserved by the two states.\cite{vidalmonotones, chenguwenunitaries, Hastingslocality}
Such transformations do not affect the universal entanglement properties of a given state.
We do not believe the constant correction to the entanglement entropy that we find is in conflict with this intuition.
Recall that the entanglement entropy is computed in the limit where the length of the entangling surface ${\cal O}(\ell) \rightarrow \infty$.
The corrections we find are due to interactions that occur along the entire entanglement cut.
Therefore, we do not expect the state where such interactions are absent to be connected to the state where such interactions are present by a local unitary of finite depth. 

It is tempting to interpret our result as being due to the nucleation of a `strip' of Hall fluid defined by the effective $K$-matrix across the wires where the unconventional interactions are dominant (see Fig. \ref{fig:effectivephase}).
The entanglement cut runs through this strip and the sub-leading constant term is in exact agreement with the expected topological entanglement entropy of such a fluid, $\gamma_{K^{\rm eff}} = \log\sqrt{|\det(K^{\rm eff})|}$.
Indeed, we may populate the entire plane by wires whose mutual interactions are incorporated into a $K^{\rm eff}$.
The resulting 2D phase has the same observable properties, e.g., charge and thermal Hall conductance, but with perhaps a different set of low-energy local quasiparticles.
However, as we show in an Appendix \ref{nu2appendix}, such a 2D state does {\it not} admit deconfined quasiparticles with the statistics implied by $K^{\rm eff}$ and so the the `strips' do not coalesce into a 2D fluid defined by $K^{\rm eff}$; rather, the quasiparticle statistics of such a state are identical with those of the theory defined by the un-perturbed $K$-matrix.

We obtain our results by concentrating on a finite set of tunneling interactions that we assume to be most dominant.
Certainly, there is no general reason to restrict our attention to a finite set of interactions, as any interaction not forbidden by symmetry can appear in the action describing the edge modes at the entanglement cut.
However, we do not expect small perturbations to significantly alter the gapped ground state that is determined by the dominant interactions, and so the form of the entanglement spectrum and entropy that we find should be preserved at leading order.

One might also be worried that our results are an artifact of the wire construction. This construction is essentially in the limit when the correlation length transverse to the wires vanishes and the boundary states are ultra-localized on the edge. From this we are able to show that including interactions exactly at the cut itself is enough to modify the TEE. In a more realistic model with a finite correlation length we expect that we would have to extend the range of interactions to at least a finite-strip with width larger than the correlation length. Other than this caveat we do not expect the results of our calculations to differ from what would be calculated in other models realizing the respective phases.

\subsubsection{Entanglement Spectrum}

Previous work\cite{KitaevPreskillentropy, lihaldane,Regnault2009, Thomale2010, lauchli2010,regnault2011,sterdyniak2011,hermanns2011,sterdyniak2011a, papic2011,schliemann2011,dubail2012,sterdyniak2012, qikatsuraludwig, Dubail2012more, Simon2012, Simon2013, Fidkowski2010,Prodan2010,Pollman2010,Turner2010,Hughes2011,alex2011,turner2011a,gilbert2012,fang2013,chandran2011, swinglesenthil, Yao2010, dubail2011, refael2004,jia2008,chen2012,mondragon2013,pouranvari2013,mondragon2014a,pouranvari2014} suggests a close connection between the entanglement spectrum and the edge spectrum.
These works directly imply that the various edge phases available to a given bulk should be visible in the entanglement spectrum.
In particular, the experimentally-measurable tunneling exponents, which may be used to distinguish different edge phases by measuring the scaling dimensions of vertex operators within the particular edge theory, are related to the eigenvalues of the entanglement Hamiltonian within particular (flux) sectors of the theory.
We make the connection between the edge spectrum and the entanglement spectrum more precise in the context of edge phases with more than one edge mode.

Given a list of spectral eigenvalues, in general it is unclear how to decode the list to deduce the phase of the edge using the analytic expression for the entanglement spectrum except at fine-tuned points.
However, it is possible to compare the behavior of such lists under global perturbations of the bulk state, if they are known to transform in distinct ways.
To this end, we imagine putting our states on a spatial cylinder.
We then compare spectra associated to distinct edge phases before and after insertion of flux through the cylinder.
The behavior under flux insertion allows some distinct edge phases to be distinguished in an unambiguous way, e.g. completely fermionic and completely bosonic edge phases can be distinguished by threading $\pi$-flux through the hole of the cylinder.\cite{groverbraiding, tuzhangqimomentumone, zaletelflux}

\subsection{Outline}

The remainder of the paper is organized as follows.
In Sec. II, we review the construction of (Abelian) Hall states using a collection of coupled wires.
This section is pedagogical, however, we highlight certain aspects of this construction which are especially important for our study.
In Sec. III, we calculate the entanglement spectrum and entropy that result from cutting the state at the interface between two wires.
This section generalizes the beautiful work in Refs. [\onlinecite{lundgrenentanglement, chenentanglement}], and makes explicit how interactions near the entanglement cut can affect the spectrum.
The next three sections elucidate the technology developed in the previous section through examples.
In Sec. IV, we study filling fractions $\nu=2,4,$ and $\nu=4/3$, which illustrate how the tunneling interactions across the cut can lead to constant, sub-leading corrections to the entanglement entropy even in the states with no topological order, i.e. $\nu=2, 4.$
In Sec. V, we consider filling fractions $\nu=8$ and $\nu=8/15$, for which more than one edge phase can occur at the entanglement cut, and describe how these distinct phases manifest themselves in the entanglement spectrum and entropy.
In Sec. VI, we consider the entanglement at an interface between two Laughlin states at filling $\nu=1/k^R$ and $\nu = 1/k^L$.
In Sec. VII, we briefly describe how $\pi$-flux can be used to distinguish certain entanglement cuts. 
In Sec. VIII, we conclude and discuss a number of questions we hope to address in the future.
Additionally, our paper contains nine glorious appendices containing technical details used in the main text.

\section{Coupled Wire Construction of Fractional Quantum Hall States}

In this section, we review the construction of FQH states (of fermions) using a collection of coupled 1D Luttinger liquids.\cite{KaneMukhopadhyayLubensky, TeoKanewires}
The coupled wire approach is powerful because it allows us to study non-perturbative interactions between distinct wires using bosonization techniques.
Additionally, the calculation of the entanglement spectrum of a particular state is analytically tractable because the problem is mapped to a calculation of the entanglement between nearby wires.

\subsection{1D Fermi Liquid Arrays}

\begin{figure}[h!]
  \centering
\includegraphics[width=\linewidth]{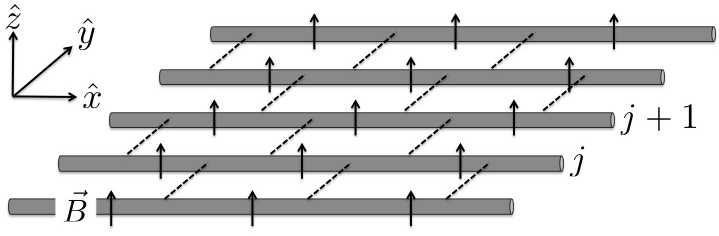}\\
a) An array of wires in a constant background field. Dashed lines indicate interactions between neighboring wires.
\includegraphics[width=\linewidth]{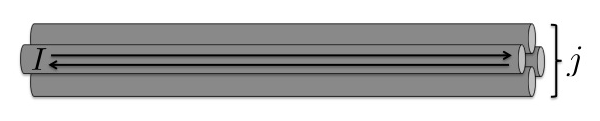}\\
b) A single wire, $j$, with four internal channels. Each internal channel has a right- and left-moving mode, as indicated by the arrows on the outermost channel.
\caption{Schemiatic Illustration of a Coupled Wire array}\label{fig:wirearray}
\end{figure}

The coupled-wire construction begins with ${\cal N}$ parallel wires that each host an $N$ channel 1D spinless Fermi liquid; spin may be incorporated into the channel index if desired.
Each wire is oriented parallel to the $\hat{x}-$direction and an array of  wires is placed in the $xy$ plane, where the wires are separated from one another by a distance $d$ in the $\hat{y}-$direction.
A background magnetic field $B$ is applied in the $\hat{z} = \hat{x} \times \hat{y}$-direction.
This set-up is shown in Fig~\ref{fig:wirearray}.

The 1D free Fermi liquid action for the collection of wires takes the form:
\begin{align}
\label{fermiaction}
S_{{\rm FL}} & = \sum_{j = 1}^{{\cal N}} \int dt dx \sum_{I = 1}^N \Big[\Psi_{I, j}^\dagger i \partial_t \Psi_{I,j} \cr 
& + {1 \over 2 m} \Psi^\dagger_{I,j} (\partial_x - i e A_x)^2 \Psi_{I,j} + \mu_I \Psi^\dagger_{I, j} \Psi_{I,j} \Big],
\end{align}
where the operator $\Psi_{I,j}^\dagger$ creates an electron on wire $j = 1, ..., {\cal N}$ in channel $I = 1, ..., N$.
For simplicity, we take the electron mass $m,$ and a channel-dependent chemical potential $\mu_I,$ to be the same on each wire.
We choose a gauge where the background electromagnetic field, $A_\mu = (0, - B y, 0, 0)$, which is convenient for our choice of wire orientation.
In subsequent formulas, sums over repeated indices will be understood unless otherwise specified.

At low energies, we may linearize the spatial derivatives in the action and restrict our attention to quasiparticle excitations near the Fermi points at $k_F^{(R/L, I, j)}$.
To isolate this low energy physics, we introduce fermionic operators $\psi_{(R/L, I, j)}^\dagger$ that create excitations near the Fermi points:
\begin{align}
\label{slowmode}
\Psi^\dagger_{I,j} = e^{i k_{F}^{(R, I, j)} x} \psi^\dagger_{(R, I, j)} + e^{i k_{F}^{(L, I, j)} x} \psi^\dagger_{(L, I, j)},
\end{align} 
where the $\psi^\dagger_{(R/L, I, j)}$ fields are slowly varying with respect to the scale set by the ``bare" Fermi momenta, $k^{(0)}_{F, I} = \sqrt{2 m \mu_I}$.
The slowly varying assumption allows us to avoid the possibility of double counting the low-energy degrees of freedom, and effectively places a cutoff $\Lambda \ll k^{(0)}_{F, I}$ on the allowed momenta of the fields $\psi^\dagger_{(R/L, I, j)}$.
The subscripts on $\psi^\dagger_{(R/L, I, j)}$ remind us that the associated excitations are purely right-moving or left-moving along wire $j,$ and belong to channel $I$ (in the absence of inter-wire interactions).
The Fermi momenta $k_{F}^{(R/L, I, j)},$ and the low energy action for the excitations about these points, are found by substituting the expansion in Eqn. (\ref{slowmode}) into the Fermi liquid action of Eqn. (\ref{fermiaction}).
We obtain the linearized action:
\begin{align}
\label{freefermion}
S_{{\rm linearized}} & = \int dt dx \Big[\psi_{(R, I, j)}^\dagger i (\partial_t + v_{F, I} \partial_x) \psi_{(R, I, j)} \cr
&  + \psi_{(L, I, j)}^\dagger i (\partial_t - v_{F,I} \partial_x) \psi_{(L, I, j)}\Big],
\end{align}
where the bare Fermi velocities are:
\begin{align}
v_{F,I} = {k_{F,I}^{(0)} \over m},
\end{align}
and the Fermi momenta are given by:
\begin{align}
k^{(R/L, I, j)}_{F} = \pm k_{F,I}^{(0)} + e B j d 
\end{align}\noindent where the second term arises from the minimal coupling to the vector potential. 
About the decoupled wire fixed point, higher-order interactions are irrelevant in the renormalization group (RG) sense and have been suppressed.

The 2D electron density for each channel, $\rho^{(I)}_{2D} = {1 \over \pi d} k_{F,I}^{(0)}$.
The channel index $I$ may be thought of as a layer index in the full 2D system, e.g. two channels can be treated as a bi-layer.
The wire construction allows one to create (different types of) FQH states at filling fraction $\nu = (2\pi \rho_{2D})/(e B)$, where $\rho_{2D} =  \sum_I \rho_{2D}^{(I)}$, which are obtained by suitable inter-wire coupling terms.
When the tunneling couplings are tuned to zero, we have a highly anisotropic 2D Fermi liquid (with an open Fermi surface).
At non-zero coupling, the bulk is gapped, however, there may exist gapless modes living on the boundary of the system consistent with the desired Hall state. 
In addition, the allowed couplings might be restricted by symmetries of the system.

\subsection{1D Luttinger Liquid Arrays}

To proceed, we bosonize the system.
Within each wire $j$ and for each channel $I$, we introduce two (real) bosonic fields $\phi^{R/L}_{I, j}$ associated to the right/left-moving fermion fields:
\begin{align}
\psi_{(R/L, I, j)}^\dagger = {\gamma_{(R/L, I, j)} \over \sqrt{2 \pi \alpha}} e^{\pm i \phi^{R/L}_{I, j}},
\end{align}
where $\alpha$ is a short-distance cutoff.
The Klein factors $\gamma_{(R/L, I, j)}$, which we take to be real, satisfy the Clifford algebra:
\begin{align}
\{\gamma_{(a, I, j)}, \gamma_{(b, J, k)} \} = 2 \delta_{ab} \delta_{I J} \delta_{j k},
\end{align}
where $a, b = R$ or $L$.
In some equations, it will be convenient to substitute, $R = 1$ and $L=-1$ when appropriate.
The Klein factors ensure that the bosonized expressions for fermion operators with different quantum numbers, $(R/L, I, j)$, anti-commute.
The statistics of operators with the same quantum numbers follows from the equal-time anti-commutation relations:
\begin{align}
[\phi_{I,j}^a(x), \phi_{J, k}^b(y)] = - i  \pi a  \delta_{ab} \delta_{I J} \delta_{j k} {\rm sgn}(x-y),
\end{align}
where $a=R/L\equiv \pm 1,$ and ${\rm sgn}(x-y) = \pm 1$ for $x-y >0 $ or $x-y < 0$, respectively.
The bosonic fields satisfy periodicity conditions and are only defined up to integer multiples of $2\pi$:
\begin{align}
\label{intperiod}
\phi^{R/L}_{I, j} \sim \phi^{R/L}_{I, j} + 2 \pi p^{R/L}_{I, j},
\end{align}
where $p^{R/L}_{I, j} \in \mathbb{Z}$.
The periodicity conditions in Eqn. (\ref{intperiod}) determine the allowed exponential, i.e., vertex,  operators that can be constructed from the bosonic fields.
Vertex operators written in terms of integer multiples of the $\phi^{R/L}_{I,j}$  are created from products of the fundamental fermion fields.

From these definitions we find the bosonized form of the free fermion action in Eqn. (\ref{freefermion}) to be:
\begin{align}
\label{freeluttliquid}
S_{LL} = & {1 \over 4 \pi} \int dt dx \Big[\partial_x \phi^R_{I, j} (- \partial_t - v_{F, I} \partial_x)\phi^R_{I, j} \cr
& + \partial_x \phi^L_{I, j} (\partial_t - v_{F, I} \partial_x)\phi^L_{I, j} \Big].
\end{align}
The fermion charge density in the bosonized language is: $\psi^{\dagger}_{(R/L, I, j)} \psi_{(R/L, I, j)} = {1 \over 2\pi} \partial_x \phi^{R/L}_{I, j}$.
Short-ranged density-density (and current-current) interactions within wire $j$ can be incorporated via the interaction matrix, $V^{(j), (a,b)}_{I J}$, after which the action becomes:
\begin{align}
\label{luttliquid}
S_{LL} = & {1 \over 4 \pi} \int dt dx \Big[- \partial_x \phi^R_{I, j} \partial_t \phi^R_{I, j} + \partial_x \phi^L_{I, j} \partial_t \phi^L_{I, j} \cr 
& - V_{IJ}^{(j),(a,b)} \partial_x \phi_{I, j}^a \partial_x \phi_{J, j}^b \Big],
\end{align}
where $a,b=R/L=\pm 1$ and we have absorbed the mode velocities $v_{F,I}$ into the interaction matrices $V_{IJ}^{(j), (a,a)}$.
Stability of the theory requires the $2N \times 2N$ matrix,
\begin{align}
(U^{(j)})_{I J} = 
\begin{pmatrix}
V^{(j), (R,R)}_{I, J} & V^{(j), (R,L)}_{I,J-N} \cr V^{(j), (L,R)}_{I-N, J} & V^{(j), (L,L)}_{I-N, J-N}
\end{pmatrix},
\end{align}
to be positive-definite. The effects of these interactions can be summarized as follows. 
The ``off-diagonal" interactions, mediated by $V^{(j), (R,L)}_{I J} = V^{(j), (L, R)}_{J I}$, between excitations of opposite chirality can renormalize the scaling dimensions of the vertex operators, and may thus affect the RG relevance of certain tunneling interactions.
Density-density interactions between fields of the same chirality do not affect the scaling dimension of fields; rather, they merely renormalize the velocities of the edge excitations.

\subsection{Coupled Luttinger Liquids}

Gapped 2+1D phases, including fractional quantum Hall states, are obtained by coupling the gapless 1+1D wires.
There are two important classes of interactions to consider.

(i) The first type of interaction tunnels electrons and/or holes between nearby wires.
In these interactions, the explicit Fermi momenta dependence from the decomposition in Eqn. (\ref{slowmode}) must be included so that each appearance of $\psi^\dagger_{(R/L, I, j)}$ in any interaction is accompanied by the oscillating phase factor, $\exp(i k_F^{(R/L, I, j)}x)$.

We can represent any such perturbation that couples together modes on nearest-neighbor wires using an integer vector ${\vec{m}_{j, j+1}} = (m^a_{I, j}, m^b_{J, j+1})$ which defines an interaction of the form:
\begin{align}
{\cal O}_{\vec{m}_{j,j+1}} = \Gamma_{\vec{m}_{j, j+1}} \exp\Big(i m_{I, j}^a \phi^a_{I, j} + i m_{J, j+1}^b \phi^b_{J, j+1}\Big) + {\rm h.c.},
\end{align}
where repeated indices denoting the chirality $(a,b)$, and channel $(I,J)$, of the fields are understood to be summed over unless otherwise specified, and the product of Klein, cut-off, and Fermi momenta factors is
\begin{align}
\label{kleinprod}
\Gamma_{\vec{m}_{j,j+1}} = & {\exp \Big(i a k_F^{(a, I, j)} m^a_{I, j} x + i b k_F^{(b, J, j+1)} m^b_{J, j+1} x \Big)} \over (2 \pi \alpha)^{M_{\vec{m}_{j,j+1}}} \cr
& \times \prod_{a, I} \gamma_{(a,I,j)}^{|m^a_{I, j}|} \prod_{b, J} \gamma_{(b ,J ,j+1)}^{|m^b_{J, j+1}|},
\end{align}
where
\begin{align}
M_{\vec{m}_{j,j+1}} = \sum_{a, I} |m^a_{I, j}| + \sum_{b,J} |m^b_{J, j+1}|.
\end{align} 
Interactions that contain $|m^a_{I, j}| > 1$ are understood to be defined using the operator product expansion (OPE).
For instance, when $m^a_{I,j} = 2$,
\begin{align}
\lim_{\epsilon \rightarrow 0} \psi^\dagger_{(a, I, j)}(x+\epsilon) \psi^\dagger_{(a,I,j)}(x) := {1 \over 2 \pi \alpha} e^{2 a  i \phi^a_{I,j}(x)},
\end{align}
where there is no implied sum on $a,$ and we have omitted any constants or (derivative) terms that vanish in the zero-momentum limit.

Using the definition of the charge operator below Eqn. (\ref{freeluttliquid}),  then requiring ${\cal O}_{\vec{m}_{j, j+1}}$ to be charge-conserving requires:
\begin{align}
\sum_{I} \Big[m^R_{I, j} - m^L_{I, j} + m^R_{I, j+1} - m^L_{I, j+1} \Big] = 0. 
\end{align}
Additionally, enforcing translation-invariance along the wires requires the oscillating exponential factor in Eqn. (\ref{kleinprod}) to vanish:
\begin{align}
\sum_{I, a} \Big[a k_F^{(a, I, j)} m^a_{I, j} + a k_F^{(a, I, j+1)} m^a_{I, j+1}\Big] = 0.
\end{align}
Imposing additional symmetries, e.g., time-reversal symmetry, can lead to further constraints on the allowed interactions.

Occasionally, we will study situations where a perturbation to the Hall state breaks either one or both of these symmetries.
Translation-invariance is not a fundamental symmetry of the topological phases considered here, and any physically-realized state inevitably contains impurities.
Further, we may imagine breaking charge-conservation by bringing a superconductor near some section of the sample so that tunneling is mediated by exchange with Cooper pairs.

(ii) The second type of interaction that we must consider couples together fermion densities on separate wires (we already mentioned the effects of intra-wire density interactions above).
We shall only consider short-ranged density-density interactions, and
the form of these interactions is identical to the intra-wire density-density interactions appearing in Eqn. (\ref{luttliquid}), except that we now allow such couplings between modes on different wires.
These interactions can affect the scaling dimensions of tunneling perturbations, and are required to ensure that certain higher-body interactions are relevant.
In general, discrete symmetries can restrict the allowed inter-wire (and intra-wire) density-density interactions.
However, such constraints do not enter in the description of the states considered in this paper.

We note that locality plays an important role in restricting the allowed inter-wire interactions.
We have considered nearest-neighbor interactions above.
However, it will prove necessary to introduce next-nearest-neighbor interactions in one of the examples described below.
A perturbation is local (and, therefore, allowed) if the interaction only involves a finite number of wires in the thermodynamic limit, i.e., when the number of wires comprising the system, ${\cal N} \rightarrow \infty$.

Equipped with these two types of interactions, we may now construct a 2D state by sewing together the Luttinger liquid wires so that the only remaining gapless degrees of freedom live on the boundary of the system.
Intuitively, for chiral systems with only nearest-neighbor couplings, two wires are sewn together if the right-moving modes on one wire and the left-moving modes on the other obtain a gap in the presence of a tunneling operator.
To gap the right-moving modes on wire $j$ and the left-moving modes on wire $j+1$, we need $N$ ``mutually-commuting" integer vectors $\vec{m}_{j, j+1}^{(\beta)}$ that satisfy:\cite{Haldanestability}
\begin{align}
\label{null}
a (m^{(\beta)})^a_{I, j} (m^{(\gamma)})^a_{I, j} +  a (m^{(\beta)})^a_{J, j+1} (m^{(\gamma)})^a_{J, j+1} = 0,
\end{align}
for all $\beta, \gamma = 1, \dots, N$ with sums over $a=\pm 1$ and $I, J = 1, \ldots, N$ implied.
Vectors satisfying Eqn. (\ref{null}) are said to be ``null" and the set of interaction terms they generate can simultaneously pin all of the low-energy boson modes.
Applying the same sewing perturbation between all wires ensures that tunneling operators between distinct wires commute so that all of the low-energy modes, except possibly those on a boundary, become pinned and gapped. 

Since our fundamental particles are electrons the presence of Klein factors can sometimes impede a straightforward analysis in terms of the bosonic fields.
(For a pedagogical discussion of these issues, please see Refs. [\onlinecite{Giamarchibook, SchulzcoupledLLs}]).
We may simultaneously diagonalize the Klein factors if they commute:
\begin{align}
\Big[\Gamma_{\vec{m}^{(\beta)}_{j, j+1}}, \Gamma_{\vec{m}^{(\gamma)}_{k,k+1}}\Big] = 0,
\end{align}
for all inter-wire perturbations.
(The presence of additional Fermi momentum and cutoff factors does not affect the above condition.)
Remarkably, the Klein factors do commute with one another if Eqn. (\ref{null}) is satisfied, as proven in Appendix \ref{kleinappendix}.
With the ability to simultaneously diagonalize the set of Klein factors, we may replace them by $c$-numbers which can be absorbed into the coupling constants.
\footnote{Relative signs or phases that may appear between eigenvalues of the Klein factors, associated with the presence of two or more tunneling interactions, ensure the fermionic statistics of the underlying electron operators is faithfully represented when computing correlation functions.}

\subsubsection{Illustrating the Construction with $\nu=1/m$ Laughlin States}
\label{laughlinstates}

Following previous work, it is helpful to illustrate the rather technical discussion above by working through the coupled wire construction of the $\nu=1/m$ Laughlin state of fermions, an example that will be used throughout the remainder of the article.
In this case, each wire hosts a single channel which is filled to a density such that $2 k_F^{(0)} \nu^{-1} = e B d$.
The Fermi points lie at $k_F^{(R/L, j)} = k_F^{(0)} (2 j \nu^{-1} \pm 1)$.
We take the dominant sewing perturbation connecting each wire together to be defined by the integer vector $({m+1 \over 2}, {m - 1 \over 2}, {m-1 \over 2}, {m+1 \over 2})$:
\begin{widetext}
\begin{align} 
{\cal O}^{1/m}_{j, j+1} = {\gamma_{(R, j)}^{{m+1 \over 2}} \gamma_{(L, j)}^{{m-1 \over 2}} \gamma_{(R, j+1)}^{{m-1 \over 2}} \gamma_{(L, j+1)}^{{m+1 \over 2}} \over (2 \pi \alpha)^{m}}\cos\Big({m+1 \over 2} \phi_j^R + {m -1  \over 2} \phi_j^L +  {m-1 \over 2} \phi_{j+1}^R + {m+1 \over 2} \phi_{j+1}^L\Big),
\end{align}
\end{widetext}
where the oscillating factor vanishes by construction, and we see that this particular perturbation requires $m$ to be odd -- we can only create a Laughlin state of fermions.
We immediately verify that the integer vector satisfies the null condition of Eqn. (\ref{null}) so the perturbation can open a gap and the Klein factors can be ignored. We generally expect additional tunneling interactions to be present.
However, for the moment, let us assume all additional interactions vanish.

To analyze the effects of perturbation by ${\cal O}_{j, j+1}^{1/m}$, we make the field redefinition:
\begin{align}
\label{redefinelaughlin}
\phi_j^R = & {m + 1 \over 2} \varphi_j^R + {1 - m \over 2} \varphi_j^L, \cr
\phi_j^L = & {1-m \over 2} \varphi_j^R + {m+1 \over 2} \varphi_j^L,
\end{align}
in terms of which the above perturbation simplifies:
\begin{align}
{\cal O}^{1/m}_{j,j+1} = {1 \over (2 \pi \alpha)^m} \cos(m \varphi_j^R + m \varphi^L_{j+1}).
\label{laughlinoperator}
\end{align}
This has the form of an electron transfer operator between edges of a $\nu=1/m$ Hall state.
These operators can be made relevant by appropriate tuning of the inter-wire density-density interactions.
The fields $\varphi_1^L$ and $\varphi_{{\cal N}}^R$ are unaffected by these perturbations and so when the wires are sewn together, these two fields remain in the low-energy spectrum as the familiar FQH edge modes.

Strictly speaking, the newly defined fields obey the periodicity conditions:
\begin{align}
\label{newperiod}
\varphi_j^{R/L} \sim \varphi_j^{R/L} + {2 \pi \over m} P^{R/L}_j,
\end{align}
where the integers $P_j^{R} + P_j^L \in m \mathbb{Z}$.
The perturbation ${\cal O}_{j, j+1}^{1/m}$ pins the fields so that $|\langle \exp(m i \varphi_j^R + m i \varphi_{j+1}^L) \rangle| = 1$.
Expansion about a particular vacuum of the cosine requires the integers appearing in Eqn. (\ref{newperiod}) to obey the constraint, $P_j^R = - P_{j+1}^L$, which implies $P_1^L + P_{{\cal N}}^R \in m \mathbb{Z}$.
Thus, the low-energy theory inherits non-local operators of the form:
\begin{align}
\label{wilson}
{\cal O}_{1, {\cal N}} = \exp\Big(i \sum_{j,a} \varphi^a_j\Big) = \exp\Big(i \varphi_1^L + i \varphi_{{\cal N}}^R\Big),
\end{align}
where, in the second equality, we have replaced the bulk fields by the expectation values they obtain after being pinned.
These operators correspond to the transfer of an electron across the system, and represent the anomalous behavior of the two edges.
Within the Chern-Simons description of such states, we may identify the operator in Eqn. (\ref{wilson}) as a Wilson line operator with end points on the two separated boundaries.
We are allowed to consider operators like ${\cal O}_{1,{\cal N}}$ without recourse to the gapped fields in the bulk. 

Additionally, we can separate ${\cal O}_{1,{\cal N}}$ into quasi-independent left and right pieces, which are localized on opposing boundaries. The fields are not completely independent because an edge can exhibit anomalous charge transfer processes which are compensated by the other boundary. However, we shall treat the low-energy fields $\varphi^L_1$ and $\varphi^R_{\cal N}$ living along the boundary (or any boundary created by an entanglement cut) as having an independent $2\pi$-periodicity.
This allows us to concentrate on the low-energy modes near the entanglement cut. 

The final step of the construction lies in writing out the low-energy action, which now only depends on the $\varphi^L_1$ and $\varphi^R_{\cal N}$ fields.
Substituting the redefinition in Eqn. (\ref{redefinelaughlin}) into the action in Eqn. (\ref{luttliquid}), we obtain:
\begin{align}
\label{laughlin}
S_{\nu=1/m} = & {m \over 4 \pi} \int dt dx \Big[\partial_x \varphi^R_{{\cal N}} (- \partial_t - \tilde{v}_{F, {\cal N}} \partial_x)\varphi^R_{{\cal N}} \cr
& + \partial_x \varphi^L_{1} (\partial_t - \tilde{v}_{F,1} \partial_x)\varphi^L_{1}  \Big],
\end{align}
where the $\tilde{v}_{F, J}$ are determined by the original $v_{F, J}$ and any short-ranged density-density interactions. 
Locality prevents any significant coupling between $\varphi_1^L$ and $\varphi_{\cal N}^R$.
We have ignored any gapped excitations obtained after expanding about the vacuum induced by the sewing perturbations.
Thus, we have obtained the low-energy (edge) theory that describes the $\nu=1/m$ Laughlin state, which is fundamentally different from the free-fermion gapless modes in the decoupled wires.

\section{Entanglement Spectrum of Multi-Component Luttinger Liquids} 
\label{sec:entspec}

In this section we will calculate the entanglement spectrum of a multi-component gapped, chiral Luttinger liquid generalizing Refs.~[\onlinecite{lundgrenentanglement, chenentanglement}].
Since we focus on chiral states, our entanglement cut partitions the multi-component fluid into right- and left-movers, however the extension to non-chiral topological phases is trivial. From this choice of spatial cut we compute the reduced density matrix of the right-movers, and then try to interpret the result as the ground state density matrix of a 2D FQH fluid with open boundary.

\subsection{Effective Action at an Entanglement Cut}
\label{sec:effectiveaction}

If we cut the coupled-wire system open between wires $j$ and $j+1$ by removing all tunneling and density-density interactions that sew the wires together, we will find low-energy modes living along the cut.
For the $\nu=1/m$ example reviewed in the previous section, we obtain a non-chiral Luttinger liquid at the cut when both sides of the cut are considered. This system is described by an action identical to Eqn. (\ref{laughlin}) (with the replacement $1 \rightarrow j+1$ and ${\cal N} \rightarrow j$).
Therefore, within the wire construction, we generally expect that the action at the cut is a non-chiral Luttinger liquid that is identical in form to the sum of the two actions describing the gapless chiral modes that exist at each boundary.
This observation provides the crucial simplification for the calculation of the entanglement spectrum of a particular Abelian Hall state. 

There is an important caveat to this logic.
Thus far, we have assumed that different wires interact via the same set of tunneling interactions.
Generally, there is no reason why this must be the case; on the contrary, there are infinitely many tunneling terms that will maintain the bulk gap.
From the decoupled wire fixed point, each of these tunneling interactions is on the same footing, given the ability to tune inter-wire forward scattering interactions.
We will show that the choice of tunneling term can have a dramatic impact on the entanglement spectrum and entropy.

The most general (decoupled) action for the low-energy degrees of freedom, including both sides of the entanglement cut, takes the form:
\begin{align}
\label{usualaction}
S = & {1 \over 4 \pi} \int dt dx \Big[- K^{R}_{IJ} \partial_t \varphi^R_I \partial_x \varphi^R_J - V^R_{IJ} \partial_x \varphi^R_I \partial_x \varphi^R_J  \cr 
& + K^{L}_{IJ} \partial_t \varphi^L_I \partial_x \varphi^L_J - V^L_{IJ} \partial_x \varphi^L_I \partial_x \varphi^L_J\Big],
\end{align}
where $I,J = 1, \ldots, N$.
The right/left-moving bosonic modes are periodically identified,
\begin{align}
\label{diagonalperiod}
\varphi_I^{R/L} \sim \varphi_I^{R/L} + 2 \pi P^{R/L}_I,
\end{align}
with $P^{R/L}_I \in \mathbb{Z}$.
Because we are considering fully chiral states here, we have dropped the wire indices $j, j+1$ of the previous section for convenience.

The right/left-moving modes, $\varphi^{R/L}$, live on opposite sides of the entanglement cut and describe the excitations of a chiral Luttinger liquid on each side of the cut parameterized by the positive-definite, symmetric integer $K$-matrices $K^{R/L}$.
We refer to the action above, for the independently $2\pi$-periodic modes, as being defined by the matrices $K^{R/L}$.
The fundamental fermion excitations are composites of the $\varphi^{R/L}_I$ modes.
Density-density interactions along an edge are parameterized by the positive-definite, symmetric matrices $V^{R/L},$
while density-density interactions across the entanglement cut are given by,
\begin{align}
S_{U} = U_{IJ} \int dt dx \Big[\partial_x \varphi_I^R \partial_x \varphi^L_J \Big],
\end{align}
where $U_{IJ}$ is a symmetric matrix. The importance and effects of these types of terms was discussed above.

Before we made the entanglement cut, many-particle tunneling of local electrons/holes across the cut occurs via some set of interactions parameterized by integer vectors ${\cal M}^{R/L}_{\beta I} = K^{R/L}_{I J} (m^{(\beta)})_J^{R/L}$:
\begin{align}
\label{generaltunneling}
S_{\rm tunneling} = &  {1 \over 4 \pi} \int dt dx \Big[ g_\beta \cos\Big({\cal M}^{a}_{\beta I} \varphi^a_I\Big)\Big],
\end{align}
where the vectors ${\cal M}^{R/L}_{\beta I}$ for $\beta = 1, \ldots, N$ are linearly independent and satisfy the analog of Eqn. (\ref{null}): 
\begin{align}
\label{newnull}
{\cal M}^{R}_{\beta I} (K^R)^{-1}_{I J} {\cal M}^{R}_{\gamma J} - {\cal M}^{L}_{\beta I} (K^L)^{-1}_{I J}{\cal M}^{L}_{\beta J} = 0.
\end{align}
Tunneling interactions that satisfy Eqn. (\ref{newnull}) generate a gap in the inter-wire spectrum for finite $g_\beta$.\cite{Haldanestability}
Note that in each tunneling interaction in Eqn.~(\ref{generaltunneling}), the integer vector ${\cal M}^{R/L}_{\beta I} = K^{R/L}_{I J} (m^{(\beta)})_J^{R/L}$, where the $(m^{(\beta)})_J^{R/L}$ are integral; the factor $K^{R/L}$ ensures that multiples of the ``fundamental'' electrons and holes are tunneling, rather than the emergent fractionally-charged quasiparticles. There could be additional tunneling interactions, however, we only study $N$ independent, gap-generating operators (i.e., those satisfying Eqn. (\ref{newnull})) at a time.
We assume these perturbations are the dominant interactions between two wires, either because they have the leading (relevant) scaling dimensions, or their coupling constants have been taken to be large.

For a general choice of  tunneling interaction, specified by the vectors $({\cal M}^R, {\cal M}^L)$, we must be careful that the inter-wire vacuum that they determine is non-degenerate. 
Any such degeneracy is {\it not} topological and can be lifted by local perturbations between the two wires.
Further, precisely the perturbations that lift the degeneracy can be used to label the degenerate vacua. 
\footnote{Interestingly, the actual values for the corrections to the topological entanglement entropy that we will find appear to be independent of primitivity considerations: the primitive gapping vectors $({\cal M}^R, {\cal M}^L)$ and the non-primitive gapping vectors $(2 {\cal M}^R, 2 {\cal M}^L)$ give identical subleading corrections to the entropy.}
To this end, we require the $N$ $2N$-component vectors $({\cal M}^{R}_{\beta, I} (K^{R}_{IJ})^{-1}, {\cal M}^{L}_{\beta, I} (K^{L}_{IJ})^{-1})$ to be {\it primitive}.
A thorough discussion of the primitivity condition of sets of null vectors can be found in Refs. [\onlinecite{LevinSternclassoffracs, WangLevinweaksymmetry}].
The vectors are primitive if and only if the greatest common divisor of the $N \times N$ minors of the matrix $({\cal M}^{R}_{\beta, I} (K^{R}_{IJ})^{-1}, {\cal M}^{L}_{\beta, I} (K^{L}_{IJ})^{-1})$ is unity.
The greatest common divisor of this matrix is equal to the degeneracy of the inter-wire vacuum.
A simple example of a non-primitive coupling can be seen for a $\nu=1$ integer state. 
At the entanglement cut, the edge theories are described by the K-matrices $K^{R/L}= (1).$ 
A possible gapping term has ${\cal M}^R = {\cal M}^L = 2$, for which it appears the total ground-state degeneracy is $2.$ However, adding additional \emph{local} tunnelling couplings corresponding to the null vector $(1,1)$ would split the additional degeneracy back to its primitive value of $1$.

We remark that generally $K^R$ need not equal $K^L$.
However, from the assumption that the bulk state is gapped, $K^R$ and $K^L$ must be such that there exist perturbations that will induce a gap in the low-energy spectrum of Eqn. (\ref{usualaction}).
In other words, the excitations living along the entanglement cut must be fully ``gappable."  
This is possible exactly when there exist $N$ integer vectors, ${\cal M}^{R/L}_{\beta I}$, satisfying the null criterion in Eqn. (\ref{newnull}).
An example of a gappable edge with $K^R \neq K^L$ is $K^R = (1)$ and $K^L = (9)$, in a system without charge conservation.\cite{levinprotected}

Symmetry considerations may restrict the allowed integer vectors ${\cal M}^{R/L}_{\beta I}$.
In this section, we analyze arbitrary tunneling perturbations, however,
when we consider particular examples in the next section, we will be careful to specify the symmetries that a particular set of tunneling interactions preserve.
For convenience, we also assume conformal symmetry, $V^{R/L} = K^{R/L}$ and $U = 0$, in this section.
This assumption is not essential for the results of our analysis, however, it does affect the quantitative physical description.
For instance, when studying a particular tunneling interaction, it may be necessary to assume its coefficient is finite, rather than its scaling dimension is relevant, in order to ensure that the modes at the entanglement cut obtain a gap.
We show in Appendix~\ref{sec:gappingvectorsrelevant} that any linearly independent set of $N$ vectors satisfying the null criterion of Eqn.~(\ref{newnull}) can be made relevant by tuning $V^{R/L}$ and $U$.

\subsection{Symmetry Breaking}
\label{symmetrybreaking}

We assume the coupling constants $g_\beta > 0$ in Eqn. ({\ref{generaltunneling}) and, without loss of generality, expand the gapped theory about the vacuum:
\begin{align}
\label{vacchoice}
\langle {\cal M}^{R}_{\beta I} \varphi^R_I + {\cal M}^{L}_{\beta I} \varphi^L_I \rangle = 0,
\end{align}
for each $\beta$.
While the cosine interaction is invariant under the independent shifts, $\varphi^{R/L}_{J} \sim \varphi^{R/L}_J + 2 \pi P^{R/L}_J$ with $P^{R/L}_J \in \mathbb{Z}$, the expansion about a particular vacuum breaks this symmetry. 
The vacuum in Eqn. (\ref{vacchoice}) is only invariant under shifts by $P_J^{R/L}$ that satisfy:
\begin{align}
\label{vacrestrict}
{\cal M}^{R}_{\beta I} P^R_I + {\cal M}^{L}_{\beta I} P^L_I = 0.
\end{align}
This shows the spontaneously broken independent shift symmetries of the bosonic fields.

If $({\cal M}^{R})^{-1} {\cal M}^L$ and its inverse are both integer matrices, i.e., $({\cal M}^{R})^{-1} {\cal M}^L \in GL(N, \mathbb{Z})$, then every integer vector $P^L$ uniquely determines another integer vector $P^R$ via Eqn. (\ref{vacrestrict}). 
However, if $({\cal M}^{R})^{-1} {\cal M}^L \notin GL(N,\mathbb{Z})$, then not all integer vectors $P^L$ will yield an integer solution for $P^R$. 
Instead, the allowed $P^{R/L}$ take the restricted form, $P_{J}^{R/L} = v_{JK}^{R/L} z_K$, where $z_K$ is an arbitrary integer vector and $v^{R/L}$ are integer matrices determined via the following algorithm.

First, put the $N\times 2N$ matrix $\begin{pmatrix} {\cal M}^R & {\cal M}^L \end{pmatrix}$ into its Smith normal form: 
\begin{equation} \begin{pmatrix} S & 0_{N\times N} \end{pmatrix} = {\cal U} \begin{pmatrix} {\cal M}^R & {\cal M}^L \end{pmatrix} \begin{pmatrix} v_1 & v_2 \\ v_3 & v_4 \end{pmatrix},\end{equation}
where $S, {\cal U}, 0_{N\times N}$, and $v_i$ are all integer $N\times N$ matrices.
${\cal U}$ is invertible over the integers and so is the matrix composed of the $v_i$.
Utilizing the invertibility of ${\cal U}$, we see that ${\cal M}^R v_2 + {\cal M}^L v_4 = 0$; hence $v^R = v_2$ and $v^L = v_4$ are exactly the matrices we are seeking. Because ${\cal M}^{R/L}$ are non-singular (proven in Appendix~\ref{sec:linearindependence}), this solution is unique up to multiplication by a unimodular matrix.
The restricted solution set $P^{R/L} = v^{R/L}z$ will play an important role in computing the entanglement entropy.

\subsection{Quadratic Approximation}

We {\it approximate} the cosine interactions in $\mathcal{S}_{\rm tunneling}$ by the quadratic mass term:
\begin{align}
\label{cosapprox}
& g_\beta \cos\Big({\cal M}^{a}_{\beta I} \varphi^a_I\Big)  = - {1 \over 2} g_\beta \Big({\cal M}^{a}_{\beta I} \varphi^a_I\Big) \Big({\cal M}^{b}_{\beta J} \varphi^b_J\Big) + \ldots,
\end{align}
where $\ldots$ represents a constant shift and higher-order interactions.
This approximation is justified when the interactions in Eqn. (\ref{generaltunneling}) generate a spectral gap, i.e., when the $N$ independent vectors ${\cal M}^{R/L}_{\beta I}$ satisfy the null condition in Eqn. (\ref{newnull}).
It is not reliable when there are competing tunneling interactions.  We will see below that this approximation drastically simplifies some of the calculations. Thus, we expect that dealing with competing interactions will be a challenging problem. Although we will not consider it further here, it would be interesting to consider the behavior of the entanglement spectrum in such circumstances.

\subsection{Lattice Structure}
\label{lattices}

It is convenient to make a field redefinition in order to simplify the analysis of the action in Eqns. (\ref{usualaction}) and (\ref{cosapprox}) and to illuminate the underlying physics.\cite{Read90, WenZee92}
We take the ``square-root" of $K^{R/L}$ by introducing basis vectors $e^{R/L}_I$ satisfying:
\begin{align}
(e^{R/L}_I)_i (e^{R/L}_J)_i = & K^{R/L}_{IJ},
\end{align}
where $i = 1, \dots, N$.
These vectors define integer lattices $\Lambda^{R/L}= \lbrace n_I e_I^{R/L} | n_I \in \mathbb{Z} \rbrace $, of which the $K^{R/L}$ are the so-called Gram matrices with unit cell volume equal to $|\det K^{R/L} |$.
The dual basis (reciprocal basis) is defined by the vectors $(f_I^{R/L})_i = (K^{R/L})^{-1}_{IJ} (e_J^{R/L})_i$, which satisfy:
\begin{align}
(f_I^{R/L})_i (f_J^{R/L})_i = & (K^{R/L})^{-1}_{IJ}, \cr
(f_I^{R/L})_i (e_J^{R/L})_i = & \delta_{IJ}, \cr
(e_I^{R/L})_i (f_I^{R/L})_j = & \delta_{i j}.
\end{align}
The choice of basis (and, therefore, dual basis) for each chirality is unique up to $SO(N)$ rotations, $(e^{R/L}_{I})_i \rightarrow (O^{R/L})_{i j} (e_I^{R/L})_j$ with $O^{R/L} \in SO(N)$. We will exploit this freedom momentarily. 
The physical properties are also invariant under certain redefinitions: $(e^{R/L}_{I})_i \rightarrow W_{IJ}^{R/L}(e_J^{R/L})_i$ for $W^{R/L} \in GL(N,\mathbb{Z})$, which amounts to a field redefinition in Eqn. (\ref{usualaction}), $\varphi^{R/L}_I \rightarrow (W^{R/L})^{-1}_{IJ} \varphi^{R/L}_{J}$.

We put this formalism to work by defining the fields $X_i^{R/L}$:
\begin{align}
\varphi^{R/L}_I = (f_I^{R/L})_i X^{R/L}_i,
\label{Xdefinition}
\end{align}
which have the virtue of diagonalizing the action in Eqn. (\ref{usualaction}):
\begin{align}
\label{luttaction}
S = & {1 \over 4 \pi} \int dt dx \Big[- \partial_t X^R_i \partial_x X^R_i - \partial_x X^R_i \partial_x X^R_i \cr
& + \partial_t X^L_i \partial_x X^L_i - \partial_x X^L_i \partial_x X^L_i 
\end{align}
We show in Appendix~\ref{fieldredef} that we can use the freedom in the choice of the $e^{R/L}_I$ to choose a basis such that not only is Eqn.~(\ref{luttaction}) diagonal, but that the cosine terms, expanded to quadratic order as in Eqn.~(\ref{cosapprox}), take the form
\begin{equation} \label{tunndiag}
S_{\rm tunneling} = - \frac{1}{4\pi} \int dt dx \lambda_i (X_i^R + X_i^L)(X_i^R + X_i^L) + \ldots, \end{equation}
where the $\dots$ represent a constant shift and the $\lambda_i$ are positive eigenvalues of the matrix with $ij$th entry, $(f^R_I)_i {\cal M}^{R}_{\beta I} g_\beta {\cal M}^{R}_{\beta J} (f^R_J)_j$; with the correct choice of basis, these eigenvalues are invariant under the replacement $R \rightarrow L$, as shown in Appendix~\ref{fieldredef}. Because this basis choice diagonalizes the mass matrix, it will depend on the gapping vectors and the coupling constants $g_\beta$.

The new fields are periodic up to elements in the lattices $\Lambda^{R/L}$:
$X^{R/L}_i \sim X^{R/L}_i + 2 \pi P^{R/L}_I (e_I^{R/L})_i$.
However, the expansion about a particular minimum of the cosine tunneling operators,
Eqn. (\ref{vacrestrict}), restricts the allowed lattice points by constraining $P^{R/L}_J = v^{R/L}_{JK} z_K$. When this restriction is included, the periodicity condition on the fields is,
\begin{align}
\label{periodrestriction}
X^{R/L}_i \sim X^{R/L}_i + 2 \pi  (e_I^{R/L})_i v^{R/L}_{IJ} z_J.
\end{align}
where $z_J \in\mathbb{Z}$. This defines the {\it restricted lattice},
\begin{align} \Lambda' = \lbrace z_J v_{IJ}^{R}(e_I^{R})_i | z_J \in \mathbb{Z}^N \rbrace\label{definerestrictedlattice} \end{align} 
and an ``effective K matrix'', 
\begin{align} 
\label{effectiveK}
K_{IJ}^{\rm eff} = v^{R}_{MI}(e_M^{R})_j (e_K^{R})_j v^{R}_ {KJ}= ((v^R)^T K^R v^R)_{IJ}.
\end{align}
Both $\Lambda'$ and $K^{\rm eff}$ are unchanged up to basis transformations under the substitution $R\rightarrow L$, as shown in Appendix~\ref{fieldredef}. 
They will directly enter our calculation of the entanglement spectrum and entanglement entropy. The invariance under $R\rightarrow L$ reflects the fact that the spectrum and entropy are independent of whether we trace over the right or left side of the system. 

Recall that physically the lattices $\Lambda^{R/L}$ determine the form of fundamental quasiparticle creation/annihilation operators, $\exp\Big(i n_I (e_I^{R/L})_i X^{R/L}_i\Big)$.
On the other hand, `fractional' quasiparticle creation/annihilation operators, $\exp\Big(i n'_I (f_I^{R/L})_a X^{R/L}_a \Big)$, are formed by elements of the dual lattice. Since there is a connection between the lattice and the fundamental quasi-particle excitations, the restriction of the lattice $\Lambda'$ implies that there is  a reduced set of \emph{local} quasi-particles that are allowed to tunnel.

\subsection{Quantization and Diagonalization of the Hamiltonian}

We now canonically quantize the Hamiltonian associated to the action of Eqns. (\ref{luttaction}) and (\ref{tunndiag}) in order to compute the ground state wave function of the gapped system obtained at finite $\lambda_i$. Placing the system on a (spatial) circle of circumference $\ell$, we perform a mode-decomposition of the right- and left-moving fields at time $t=0$:
\begin{align}
\label{modeexpansion}
X^R_i = & X^R_{i,0} + {2 \pi N_{i}^R \over \ell} x + \sum_{n > 0} \Big({\alpha_{i,n} \over \sqrt{|n|}} e^{{2 \pi i n x \over \ell}} + {\alpha^\dagger_{i,n} \over \sqrt{|n|}} e^{- {2 \pi i n x \over \ell}}\Big), \cr 
X_i^L = & X^L_{i,0} + {2 \pi N_{i}^L \over \ell} x + \sum_{n < 0} \Big({\alpha_{i,n} \over \sqrt{|n|}} e^{{2 \pi i n x \over \ell}} + {\alpha^\dagger_{i,n} \over \sqrt{|n|}} e^{- {2 \pi i n x \over \ell}}\Big),
\end{align}
where the sum on $n$ ranges over the positive/negative integers for the right/left-moving fields.
We refer to the $X^{R/L}_{i,0}$ and $N^{R/L}_i$ operators in Eqn. (\ref{modeexpansion}) as the zero mode operators and the $\alpha_{i,n}$ as the oscillator mode operators.
To preserve the spatial periodicity under $x \rightarrow x + \ell$, as well as the periodicity of the fields required by Eqn.~(\ref{periodrestriction}),
eigenstates of the zero mode operators $N_i^{R/L}$ have eigenvalues 
\begin{equation}
n^{R/L}_i = (e_I^{R/L})_i v^{R/L}_{IJ} z_J,
\label{zeromodeeigenvalues}
\end{equation}
for any integers $z_J$. 
From the analysis in the previous section, we see that these eigenvalues lie in the lattice $\Lambda'$.
The modes in the field expansion obey the algebra:
\begin{align}
\label{algebra}
[X_{i,0}^R, N_j^R] & = - i \delta_{ij}, \cr
[X_{i,0}^L, N_j^L] & = i \delta_{ij}, \cr 
[\alpha_{i,m}, \alpha_{j,n}^\dagger] & = \delta_{ij} \delta_{m n}
\end{align}
and all other commutators vanish.

Acting upon the ground state, the Hamiltonian associated to the action of Eqns. (\ref{luttaction}) and (\ref{tunndiag}) takes the decoupled form:
\begin{align}
H = \sum_{i=1}^N \Big[ H_i^{\rm zero} + H_i^{\rm osc}\Big],
\end{align}
where
\begin{align}
H^{\rm zero}_i & = {\pi \over 2 \ell} \Big[\Big(N^R_i - N^L_i\Big)^2 +  \ell^2 \lambda_i \Big(X_{i,0}^R + X_{i,0}^L\Big)^2 \Big], \cr
H^{\rm osc}_i & = {\pi \over 2 \ell} \sum_{n \in \mathbb{Z} - \{0\}} \Big[4 |n| \alpha_{i, n}^\dagger \alpha_{i,n} \cr
& +  {\ell^2 \lambda_i \over |n|} \Big(\alpha_{i,n} \alpha_{i, - n} + \alpha_{i,n} \alpha^\dagger_{i, n} \cr
& + \alpha^\dagger_{i,n} \alpha_{i,n} + \alpha^\dagger_{i,n} \alpha^\dagger_{i,-n}\Big) + 2 |n|\Big].
\label{Hexpandedtilde}
\end{align}
The zero modes and oscillator modes decouple because of the quadratic approximation.
Eigenstates of the zero mode part of the Hamiltonian represent distinct sectors of the theory which we will label by their $N^{R/L}_i$ eigenvalues. 
The sectors that enter in the description of the wave function are constrained by Eqn. (\ref{vacrestrict}) so that our expansion about a particular vacuum remains consistent.
The oscillator modes represent excitations on top of each zero mode sector.
Note that we have used the fact that the ground state is a linear combination of states that are annihilated by the operator $N^R_i + N^L_i$, a fact that follows from Eqn. (\ref{vacrestrict}) after performing the field redefinitions in the previous section. Additionally, we can heuristically understand this by noting that configurations where $N^R_i+N^L_i$ was non-vanishing would be energetically costly since the $X_i$ fields would have appreciable spatial dependence which in-turn deforms the field away from its vacuum configuration\cite{lundgrenentanglement}. 

Finally, we must now determine the excitation spectrum for each $H^{\rm osc}_i$.
We complete the diagonalization of the oscillator part of the Hamiltonian by performing the Bogoliubov transformation:
\begin{align}
\label{bogo}
\begin{pmatrix}
\alpha_{i,n} \cr \alpha^\dagger_{i, -n}
\end{pmatrix}
=
\begin{pmatrix}
\cosh(\theta_{i,n}) & \sinh(\theta_{i,n}) \cr \sinh(\theta_{i,n}) & \cosh(\theta_{i,n})
\end{pmatrix}
\begin{pmatrix}
\beta_{i,n} \cr \beta^\dagger_{i, -n},
\end{pmatrix}
\end{align}
where 
\begin{align}
\cosh(2\theta_{i,n}) &= \frac{ |n| + {\ell^2 \lambda_i \over 2 |n|}}{\sqrt{|n|^2 + \ell^2 \lambda_i}}, \cr
\sinh(2\theta_{i,n}) &= - \frac{{\ell^2 \lambda_i \over 2 |n|}}{\sqrt{|n|^2 + \ell^2 \lambda_i}}.
\end{align}
Rewriting the oscillator Hamiltonian in terms of the $\beta_{i,n}$ modes, we obtain:
\begin{align}
H^{\rm osc}_i = \sum_{n \in \mathbb{Z} - \{0\}} E_{i,n} \Big(\beta_{i,n}^\dagger \beta_{i,n} + {1 \over 2}\Big),
\end{align}
where $E_{i,n} = \frac{2 \pi}{\ell}\sqrt{|n|^2 + \ell^2 \lambda_i}$.
The ground state of $H^{\rm osc}_i$ is annihilated by the operators $\beta_{i,n}$.
Note that the small $|n|$ regime is precisely the same as the $\ell \rightarrow \infty$ regime which will be of interest momentarily.

\subsection{Reduced Density Matrices}
\label{sec:reduceddensitymatrices}

We now find the reduced density matrix, $\rho^R$ for the ground state of $H = \sum_i (H_i^{\rm zero} + H_i^{\rm osc})$ after tracing out the left-moving degrees of freedom.
To the order at which we are working, i.e., the quadratic approximation, the ground state is a tensor product of the zero mode and oscillator mode degrees of freedom.
Therefore, the reduced density matrix takes the separable form $\rho^R = \rho^R_{\rm zero} \otimes \rho^R_{\rm osc}$. This allows us to calculate them independently.

\subsubsection{Zero Mode Reduced Density Matrix}

To calculate the ground state in the zero mode sector, we note that $H^{\rm zero}_i$ is equivalent to the harmonic oscillator Hamiltonian after identifying $m_i = (\pi \ell \lambda_i)^{-1}, \omega_i = 2 \pi \sqrt{\lambda_i}, X_i = (N^R_i - N^L_i)/2$, and ${\cal P}_i = X^R_{i,0} + X^L_{i,0}$:
\begin{align}
H^{\rm zero}_i = {1 \over 2} \Big({{\cal P}_i^2 \over m_i} + m_i \omega_i^2 X_i^2\Big),
\end{align}
 since $\Big[{N^R_i - N^L_i \over 2}, X^R_{j,0} + X^L_{j,0}\Big] = [X_i, {\cal P}_j] =  i \delta_{ij}$.
We take the continuum approximation which is valid in the limit $\ell \rightarrow \infty$.
Therefore, the un-normalized ground state is given by:
\begin{align}
|\psi^{\rm zero}_0 \rangle =  \sum_{n_i^R, n_i^L \in \Lambda'} e^{- \sum_{i=1}^N {1 \over 4 \ell \sqrt{\lambda_i}} (n^R_i - n^L_i)^2} |n^R_i, n^L_i \rangle.
\end{align}
Imposing the constraint $(N^R_i + N^L_i) |\psi_0^{\rm zero} \rangle = 0$ gives:
\begin{align}
|\psi^{\rm zero}_0 \rangle =  \sum_{n_i \in \Lambda'} e^{- \sum_{i=1}^N {n_i^2 \over \ell \sqrt{\lambda_i}}} |n_i, - n_i \rangle.
\end{align}
The un-normalized reduced density matrix for the right-moving sector is thus:
\begin{align}
\rho_{\rm zero}^R & = {\rm Tr}_L \Big(|\psi_0^{\rm zero} \rangle \langle \psi_0^{\rm zero}| \Big) \cr 
& =  \sum_{n_i \in \Lambda'} e^{- \sum_{i=1}^N {2 n^2_i \over \ell \sqrt{\lambda_i}}} |n_i \rangle \langle n_i |.
\end{align}
Retracing our steps, it is evident that $\rho^R_{\rm zero} = \rho^L_{\rm zero}$.

\subsubsection{Oscillator Reduced Density Matrix}

We use the method of Peschel\cite{peschelreduced}  to calculate the reduced density matrix for the right-moving oscillator modes.
Using the Bogoliubov transformation in Eqn. (\ref{bogo}) we compute:
\begin{align}
\label{firstway}
\langle \alpha_{i,n}^\dagger \alpha_{i,n} \rangle = \langle \beta_{i,-n} \beta^\dagger_{i,-n} \sinh^2(\theta_{i,n}) \rangle = \sinh^2(\theta_{i,n}).
\end{align}
The expectation value is taken in the ground state of $H^{\rm osc}_i$.

Alternatively, we can write $\rho_{\rm osc}^R$ formally as
\begin{align}
\rho^R_{\rm osc} = {e^{- H_e^{\rm osc}} \over Z_e^{\rm osc}}.
\end{align}
Because we are working at quadratic order, the entanglement Hamiltonian takes the form
\begin{align}
H_e^{\rm osc} = \sum_i \sum_{n>0} \omega_{i,n} \Big(\alpha^\dagger_{i,n} \alpha_{i,n} + {1 \over 2}\Big),
\end{align}
where the dispersion $\omega_{i,n}$ is to be determined. Thus,
\begin{align}
Z_e^{\rm osc} = {\rm Tr_R}(e^{- H_e^{\rm osc}})= \prod_i\prod_{n>0}\frac{1}{2}{\rm csch} \left( \frac{\omega_{i,n}}{2}\right),
\end{align}
where the trace is performed over the right-moving sector.
The correlator in Eqn. (\ref{firstway}) can be rewritten as:
\begin{align}
\label{secondway}
{\rm Tr_R}\left(\alpha^\dagger_{i,n} \alpha_{i,n} \rho^R_{{\rm osc}}\right)  &= - \partial_{\omega_{i,n}} \log\left( Z_e^{\rm osc} \right)-\frac{1}{2} \cr 
&= \frac{1}{2}{\rm coth}\left( \frac{\omega_{i,n}}{2} \right) -\frac{1}{2}.
\end{align}
Equating Eqns. (\ref{firstway}) and (\ref{secondway}) yields the dispersion
\begin{align}
\omega_{i,n} = \log\Big({\cosh(2 \theta_{i,n}) + 1 \over \cosh(2 \theta_{i,n}) - 1} \Big).
\end{align}
In the $\ell \rightarrow \infty$ (low energy) limit, we find
\begin{align}
\omega_{i,n} = {4 |n| \over \ell \sqrt{\lambda_i}} - {2 |n|^3 \over 3 (\ell^2 \lambda_i)^{3/2}} + {\cal O}({1 \over \ell^5}) .
\end{align}

\subsection{Entanglement Spectrum}
\label{sec:TEE}

Multiplying the un-normalized zero mode and oscillator mode density matrices together, we find the un-normalized reduced density matrix in the $\ell \rightarrow \infty$ limit:
\begin{align}
\label{reducedgeneral}
\rho^R = \rho^R_{{\rm zero}} \otimes \rho^R_{\rm osc} & = \Big(\sum_{n_i \in \Lambda'} e^{- \sum_{i=1}^N {2 n_i^2 \over \ell \sqrt{\lambda_i}}} |n_i \rangle \langle n_i| \Big) \cr
& \otimes \Big(e^{- \sum_{i=1}^N \sum_{n>0} {4 n \over \ell \sqrt{\lambda}_i} (\alpha^\dagger_{i,n} \alpha_{i,n} + {1 \over 2})} \Big). \cr
\end{align}
This immediately yields the entanglement Hamiltonian:
\begin{align}
\label{entham}
H_e^R \equiv & - {\rm log}(\rho^R) \cr
= &  \sum_{i=1}^N {2\over \ell \sqrt{\lambda}_i} \Big((N^R_i)^2 + 2 \sum_{n>0} (n \alpha^\dagger_{i,n} \alpha_{i,n} + {n \over 2}) \Big)
\end{align}
\noindent where we have restored the \emph{operator} $N_{i}^{R}.$ Normalizing  the density matrix merely shifts the zero spectrum by a constant that can be absorbed into the regularization of the spectrum; furthermore it makes no contribution to the entanglement entropy since that involves a derivative.
The Hamiltonian breaks up into sectors labelled by the eigenvalues of $N^R_a$ in $\Lambda'$. 
Each oscillator excitation carries ``entanglement energy" $4n /(\ell \sqrt{\lambda_i})$.
We see that we have benefitted from the the decoupling of the zero modes and the oscillator modes, which is a consequence of the quadratic approximation made in Eqn.~(\ref{cosapprox}). We observe that this decoupling is present in both the entanglement spectrum and the edge spectrum.

We can immediately see the important result that the entanglement spectrum depends strongly upon the interactions along the entanglement cut. 
First, the factors $\lambda_i$ depend on the coupling constants $g_\beta$.
This represents a dependence upon non-universal parameters such as the UV cutoff and is not of intrinsic interest.
Second, and more importantly, the eigenvalues of $N^R_i$ are constrained to lie in the \emph{restricted} lattice $\Lambda'$, which is determined by the null vectors ${\cal M}^{R/L}_{\beta I}$. 
This is a dependence upon universal parameters which are determined by the phase structure of the edge excitations.
Thus, both the allowed tunneling terms, labelled by ${\cal M}^{R/L}_{\beta I}$, which are a universal feature of the topological phase, and their coupling constants, $g_\beta$, determine the entanglement spectrum. This dependence leads to a remarkable correction to the entanglement entropy below.

\subsection{Entanglement Entropy}

We can now compute the entanglement entropy, which defined as the thermodynamic entropy of the density operator  in Eqn. (\ref{reducedgeneral}). We define the partition functions as a function of ``temperature,'' $T=1/\beta$, which we have until now set to 1: 
\begin{align}
Z_e(T) = {\rm Tr}_R ( e^{-\beta H_e^R} ) 
& = Z^{\rm zero}_e Z^{\rm osc}_e,
\end{align}
where
\begin{equation}
Z_e^{\rm zero}(\tau)  =\prod_i \sum_{n_i \in \Lambda'} e^{\pi i \tau \sum_i {1 \over \sqrt{\lambda}_i} n_i^2}= \sum_{z_I \in \mathbb{Z}^N} e^{\pi i \tau z_I \Omega_{IJ} z_J} 
\label{zeropart}
\end{equation}
and
\begin{equation}
Z_e^{\rm osc}(\{\tau_i\}) = \prod_{i} e^{-{\pi i \tau_i \over 12}}\prod_{n>0}  \frac{1}{1-e^{2 \pi i \tau_i n}},
\label{oscpart} \end{equation}
where have used Zeta function regularization to compute $\sum_{n>0} n = - 1/12$ and defined $\tau = {2 i \over \pi \ell T}$, $\tau_i = {2 i \over \pi \ell \sqrt{\lambda_i} T}$ and 
\begin{align}
\label{omegadef}
\Omega_{LK} = v^R_{I L}( {e}^R_I)_i{1 \over \sqrt{\lambda_i}}(e^R_J)_i  v^R_{JK}.
\end{align}
$\Omega$ is dictated by the form of the eigenvalues in Eqn.~(\ref{zeromodeeigenvalues}).

We are interested in the entanglement entropy, $S_e$, in the $\ell \rightarrow \infty$ limit. When $S_e$ is expanded in powers of $\ell$, the term of order $\ell^0$ is the topological entanglement entropy, which is sensitive to universal features of the topological phase. 
We have introduced $\tau$, $\tau_i$ and $\Omega$ in order to write the partition functions in terms of Riemann $\theta$ and Dedekind $\eta$ functions, whose modular properties we will exploit to compute the leading terms of $S_e$ in the $\ell\rightarrow \infty$ ($\tau,\tau_i \rightarrow 0$) limit. To this end:
\begin{align}
Z_e^{\rm zero}(\tau) &= \theta(0 | \tau\Omega) = \frac{\theta (0|-\tau^{-1}\Omega^{-1})}{\sqrt{ {\rm det}(-i\tau\Omega)}},\cr
Z_e^{\rm osc}(\{\tau_i\}) &= \prod_i \frac{1}{\eta(\tau_i)} = \prod_i \frac{\sqrt{-i\tau_i}}{\eta(-\tau_i^{-1})},
\end{align}
where the first equality in each line follows from the definitions of the partition functions, and the second equalities utilize a modular transformation.
Thus,
\begin{align}
Z^{\rm zero}_{e} &= (\det(- i \tau \Omega))^{-{1 \over 2}} \sum_{m_I \in \mathbb{Z}^N} e^{- i \pi \tau^{-1} m_I (\Omega^{-1})_{IJ} m_J} \nonumber\\
&\xrightarrow{\ell \rightarrow\infty}  \Big[\det\left({2\over \pi \ell T} \Omega\right)\Big]^{- {1 \over 2}} + {\cal O}(e^{- c \ell}),
\end{align}
for some positive constant $c$.
Similarly,
\begin{align}
Z^{\rm osc}_e \xrightarrow{\ell \rightarrow \infty}  \prod_i \Big({2 \over \pi \ell \sqrt{\lambda_i} T} \Big)^{1/2} e^{{\pi^2 \ell T \sqrt{\lambda_i} \over 24}}.
\end{align}
Utilizing the definition Eqn. (\ref{omegadef}), we find ${\rm det}(\Omega) = {\rm det}^2(v^R){\rm det}(K^R)\prod_i \frac{1}{\sqrt{\lambda_i}}$.
Thus,
\begin{align}
Z_e(\ell \rightarrow \infty) =\frac{ \prod_i e^{{\pi^2 \ell T \sqrt{\lambda_i} \over 24}}}{ |{\rm det}(v^R)| \left( {\rm det}(K^R)\right)^{1/2}}.
\end{align}
The entanglement entropy can now be computed using:
\begin{align}
\label{entropygeneral}
S_e & = {\partial \Big( T \log(Z_e(\ell \rightarrow \infty))\Big) \over \partial T}\Big|_{T = 1} \cr
& = {\pi^2 \ell \over 12} \sum_{i} (\sqrt{\lambda_i}) - {1 \over 2} \log |\det(K^R)| - \log | \det(v^R) |, \cr
& = \alpha \ell - {1 \over 2} \log|\det{(K^{\rm eff})}|,
\end{align}
where $K^{\rm eff} = (v^R)^T K^R v^R = (v^L)^T K^L v^L$and subleading terms in $\ell$ have been suppressed.
We give the leading corrections to Eqn. (\ref{entropygeneral}) in Appendix \ref{finitecorrections}.

Eqn. (\ref{entropygeneral}) is the primary result of this paper. The first term in Eqn. (\ref{entropygeneral}) is the leading non-universal area law term,  
which depends upon the coupling constants $g_\beta$ (and, ultimately, the cutoff) through its dependence upon the $\lambda_i$.
The sub-leading constant in Eqn. (\ref{entropygeneral}) contains the topological entanglement entropy, $\gamma_{K^R} = \log \sqrt{\det |K^R|}$ and a
 constant correction, $\log | \det(v^R)| > 0$, that is sensitive to the dominant interactions, encoded in the matrix $v^R$, occurring near the entanglement cut.
These interactions depend upon both $K^{R/L}$ and the null vectors ${\cal M}^{R/L}_{\beta I},$ but not on any non-universal features. Additionally, since our entanglement cut has a smooth boundary this term does not arise from properties of the boundary geometry itself.
Thus, we find that we must view the parameterization of possible interactions as additional universal data characterizing a topological phase, as this data encodes information about interfaces between two phases $K^R$ and $K^L.$

We note that the correction to the entropy due to interactions serves to lower the entanglement between two regions separated by an entanglement cut.
This is natural as the term is controlled by the ${\cal M}^{R/L}_{\beta I}$ and, ultimately, $K^{R/L},$ which determine how quasiparticles tunnel from one region to another.
The choice of tunneling interaction, ${\cal M}^{R/L}$, is not unique and may restrict the allowed \emph{local} quasiparticles that may tunnel.
This restriction and the total constant sub-leading term in Eqn. (\ref{entropygeneral}) may be incorporated into an effective $K^{\rm eff}$-matrix that determines the types of quasiparticles that can tunnel between the right-moving and left-moving edges. This is the same $K^{\rm eff}$ that we have already seen in Eqn. (\ref{effectiveK}).
In fact, the tunneling interactions that give rise to non-zero corrections {\it simulate} the effects of a strip of topological fluid defined by the $K^{\rm eff}$-matrix through their restriction on the form of the allowed fundamental quasiparticle tunneling operators across the interface -- see Fig. \ref{fig:effectivephase}. This result has at least two interesting implications: (i) if we  turn on one set  of gapping interactions in a small region of our wire array then an entanglement cut in that region can have a different TEE than in other regions where a different set of gapping interactions is used (ii) we can have two gapped phases with the same edge theories but different choices of gapping interactions throughout the bulk and these phases can be distinguished by the correction to their TEE -- we might say these two phases are homologous.

In the next three sections, we apply our results to a number of interesting examples to illustrate how the entanglement spectrum depends upon the interactions.

\section{Examples Part I: Tunneling Dependence}
\label{tunsec}

Here we consider examples where $K^R = K^L$, i.e. the system is symmetric across the entanglement cut, but where interactions near the entanglement cut, or throughout the bulk, are such that some local quasiparticles cannot move freely across the cut. This limitation contributes a constant correction to the entanglement entropy, as described in the previous section.

\subsection{$\nu=4$}
\label{fournugap}

A $\nu=4$ state can be constructed as four copies of the integer quantum Hall state at $\nu=1$, each built as described in Sec. \ref{laughlinstates}, with the parameter $m=1$.
Within this construction, modes at the interface between any two nearest-neighbor wires $j$ and  $j+1$ become gapped in the presence of the single-particle backscattering term, $(\psi^L_{I, j+1})^\dagger \psi^R_{I, j} + {\rm h.c.} \sim \cos(\varphi_{I, j+1}^L + \varphi_{I, j}^R)$, where $I=1, \ldots, 4$ labels the layer of the $\nu=4$ state.
This tunneling term preserves charge and translations along the interface. Since this is a free-fermion topological phase without topological order we would not expect any contribution to the topological entanglement entropy. 

Now imagine that a different set of tunneling operators are dominant along the entanglement cut. 
In particular, consider the set of tunneling operators, $\cos\Big(({\cal M}^R_{(4)})_{\beta I} \varphi^R_{I,j} + ({\cal M}^L_{(4)})_{\beta I} \varphi^L_{I,j+1}  \Big)$, defined by the integer vectors:
\begin{align}
\label{nu4vecs}
({\cal M}_{(4)}^R)_{1 I} = & (1, 0, 1, 0), \quad &({\cal M}_{(4)}^L)_{1 I} =& (0, 1, 0,1), \cr
({\cal M}_{(4)}^R)_{2 I} = & (0, 1, 0, 1), \quad &({\cal M}_{(4)}^L)_{2 I} =& (1, 0, 1, 0), \cr
({\cal M}_{(4)}^R)_{3 I} = & (0, 0, 1, 1), \quad &({\cal M}_{(4)}^L)_{3 I} =& (1, 1, 0, 0), \cr
({\cal M}_{(4)}^R)_{4 I} = & (0, 0, 1,-1), \quad &({\cal M}_{(4)}^L)_{4 I} =& (0, 0, -1, 1).
\end{align}
These operators tunnel a pair of right-moving electrons into a pair of left-moving electrons and are {\it marginal} about the free fermion fixed point. 
These tunneling terms conserve both charge and translations along the interface and are primitive.
Because the rows of $({\cal M}^R_{(4)}, {\cal M}^L_{(4)})$ satisfy the null condition in Eqn. (\ref{newnull}) (with $K^{R/L} = \mathbb{I}_4$), and are linearly independent, they may be made relevant simultaneously by tuning inter-wire density-density interactions, as explained in Appendix~\ref{sec:gappingvectorsrelevant}.

However, the matrix $({\cal M}^R_{(4)})_{I \beta}^{-1} ({\cal M}^L_{(4)})_{\beta J}$ is {\it not} in ${\rm GL}(4, \mathbb{Z})$ because its entries are half-integral (although it does have unit determinant.)
Therefore, we expect a constant correction to the entanglement entropy, as given by Eqn. (\ref{entropygeneral}).
To determine this correction, we follow the prescription of Sec~\ref{symmetrybreaking} and write the $4 \times 8$ matrix, $({\cal M}_{(4)}^R, {\cal M}^L_{(4)})$, in terms of its Smith normal form:
\begin{align}
{\cal U}_{(4)} ({\cal M}_{(4)}^R, {\cal M}_{(4)}^L) {\cal V}_{(4)} =  {\cal S}_{(4)},
\end{align}
where
\begin{align}
{\cal U}_{(4)} = & \left(
\begin{array}{cccc}
 1 & 0 & 0 & 0 \\
 0 & 1 & 0 & 0 \\
 0 & 0 & 1 & 0 \\
 0 & 0 & 1 & -1 \\
\end{array}
\right), \quad {\cal S}_{(4)} = \left(
\begin{array}{cccccccc}
 1 & 0 & 0 & 0 & 0 & 0 & 0 & 0 \\
 0 & 1 & 0 & 0 & 0 & 0 & 0 & 0 \\
 0 & 0 & 1 & 0 & 0 & 0 & 0 & 0 \\
 0 & 0 & 0 & 1 & 0 & 0 & 0 & 0 \\
\end{array}
\right), \cr
{\cal V}_{(4)} = & \left(
\begin{array}{cccccccc}
 1 & 0 & -1 & 1 & -1 & -1 & -1 & 0 \\
 0 & 1 & 0 & -1 & 1 & 1 & 0 & -1 \\
 0 & 0 & 1 & -1 & 1 & 0 & 1 & -1 \\
 0 & 0 & 0 & 0 & 1 & 0 & 0 & 0 \\
 0 & 0 & 0 & 1 & -2 & -1 & -1 & 1 \\
 0 & 0 & 0 & 0 & 0 & 1 & 0 & 0 \\
 0 & 0 & 0 & 0 & 0 & 0 & 1 & 0 \\
 0 & 0 & 0 & 0 & 0 & 0 & 0 & 1 \\
\end{array}
\right).
\end{align}
We identify
\begin{align}
v_{(4)}^R = \left(
\begin{array}{cccc}
 -1 & -1 & -1 & 0 \\
 1 & 1 & 0 & -1 \\
 1 & 0 & 1 & -1 \\
 1 & 0 & 0 & 0 \\
\end{array}
\right), \quad 
v_{(4)}^L = \left(
\begin{array}{cccc}
 -2 & -1 & -1 & 1 \\
 0 & 1 & 0 & 0 \\
 0 & 0 & 1 & 0 \\
 0 & 0 & 0 & 1 \\
\end{array}
\right),
\end{align}
and find $|\det(v_{(4)}^R)| = |\det(v_{(4)}^L)| = 2$. The effective $K$-matrix is
\begin{align}
K^{\rm eff}=(v^{R}_{(4)})^{T}v^{R}_{(4)}=(v^{L}_{(4)})^{T}v^{L}_{(4)}\nonumber\\
=\left(\begin{array}{cccc}
 4 & -2 & -2 & 2 \\
-2 & 2 & 1 & -1 \\
 -2 & 1 & 2 & -1 \\
 2 & -1 & -1 & 2 \\
\end{array}\right).
\end{align}
Therefore, using Eqn. (\ref{entropygeneral}), we compute the entanglement entropy,
\begin{align}
S_e(\nu = 4, {\cal M}^a_{(4)}) = \alpha \ell - \log(2)
\end{align}
where $\alpha$ is a non-universal constant, and we have suppressed corrections that vanish in the limit $\ell \rightarrow \infty$.
We see that there is a $\log(2)$ correction to the entropy which differs from the vanishing topological entanglement entropy that we would have computed if the single-electron tunneling terms had dominated over the terms described by Eqn. (\ref{nu4vecs}).
Furthermore, $\log(2)$ is the minimal non-zero correction, because the entries of ${\cal V}_{(4)}$ are necessarily integral.
Larger corrections are generally possible, however, they would require higher-body interactions across the cut.

While inter-wire tunneling terms defined by the vectors ${\cal M}^a_{(4)}$ in Eqn. (\ref{nu4vecs}) that are turned on near the entanglement cut can generate such a correction, the region in which these interactions dominate does not need to be limited to the region around the entanglement cut. 
Instead, we could have sewn {\it all} wires together using these interactions to create a kind of alternate $\nu=4$ phase. 
Near the UV fixed point, consisting of a collection of decoupled wires, these tunneling interactions can be made arbitrarily relevant by tuning the (exactly) marginal inter-wire density-density interactions. 
If the IR fixed point is chosen by the most relevant tunneling interactions, this construction for the $\nu = 4$ state is on equal footing with the usual state, in the sense that which state occurs in a given system is determined by the microscopic density-density couplings. 
The inter-wire interactions in Eqn. (\ref{nu4vecs}) are marginal about the free fermion fixed point.

It is interesting to ask whether or not the 2D state defined by the vectors ${\cal M}^a_{(4)}$ is a truly a distinct integer state from the ``conventional" $\nu=4$ state where nearest-neighbor wires are sewn together via the single-particle backscattering term, $(\psi^L_{I, j+1})^\dagger \psi^R_{I,j} + {\rm h.c.}$
Both phases have the same electrical and thermal Hall conductance since the edge mode structure is identical, however, the unconventional bulk state has a different constant sub-leading term in its entanglement entropy compared to  the ``conventional" $\nu=4$ state.
On a torus, the ground state defined by the vectors ${\cal M}^a_{(4)}$ is non-degenerate.
In addition, the state defined by the vectors ${\cal M}^a_{(4)}$ has different local bulk excitations: in the strict limit where {\it only} the tunneling terms of ${\cal M}^a_{(4)}$ are present, single-electron tunneling across wires is not allowed, rather, only electron-pairs can tunnel. In this sense, gapped single-electron excitations that are confined along the wire directions.
We note that perturbation by single-particle hopping terms allows electrons to be transported between wires.
Due to the bulk gap, the state determined by the inter-wire interactions in Eqn. (\ref{nu4vecs}) is robust to negligibly small perturbations by such single-particle tunneling.

We note that on the finite cylinder, the edge structure of the 2D state defined by the vectors ${\cal M}^a_{(4)}$ is simply the four-channel free chiral Fermi liquid.
This is not the edge structure that might be expected from a phase defined by the $K$-matrix equal to $K^{\rm eff}=(v^R_{(4)})^T v^R_{(4)}$, which happens to define a bosonic topological phase as can be seen from the even-integer entires on the diagonal. Instead, since the interface between both bulk phases of the $\nu=4$ system can be gapped, we should imagine that we capped off each end of the state defined by the ${\cal M}^a_{(4)}$ with a  strip of `conventional' $\nu=4$ fluid.

\subsection{$\nu=2$}
\label{twonugap}

The integer quantum Hall state at $\nu = 2$ admits a similar construction, and corresponding correction to the entanglement entropy, if we violate {\it both} charge conservation and  translation invariance along the entanglement cut.
A proof that charge conservation must be broken in order for the $\nu=2$ state to admit a constant correction is given in Appendix \ref{nu2charge}.
As an example for how the constant correction can occur, consider the nearest-neighbor tunneling operators defined by the primitive integer vectors: 
\begin{align}
({\cal M}^R_{(2)})_{1 I} = & (3, -1), \quad ({\cal M}^L_{(2)})_{1 I} = (3,1), \cr
({\cal M}^R_{(2)})_{2 I} = & (2, 1), \quad ({\cal M}^L_{(2)})_{2 I} = (1,2).
\end{align}
These vectors are null (satisfy Eqn. (\ref{newnull})) and $({\cal M}_{(2)}^R)^{-1} {\cal M}^L_{(2)}$ is not in ${\rm GL}(2, \mathbb{Z})$. 

To calculate the constant correction, we compute the Smith normal form:
\begin{align}
{\cal U}_{(2)} ({\cal M}_{(2)}^R, {\cal M}_{(2)}^L) {\cal V}_{(2)} =  {\cal S}_{(2)},
\end{align}
where 
\begin{align}
{\cal U}_{(2)} = & \left(
\begin{array}{cc}
 -1 & 0 \\
 1 & 1 \\
\end{array}
\right), \quad {\cal S}_{(2)} = \left(
\begin{array}{cccc}
 1 & 0 & 0 & 0 \\
 0 & 1 & 0 & 0 \\
\end{array}
\right), \cr
{\cal V}_{(2)} = & \left(
\begin{array}{cccc}
 0 & 0 & 0 & 1 \\
 1 & 2 & -5 & -7 \\
 0 & 1 & -3 & -5 \\
 0 & -1 & 4 & 5 \\
\end{array}
\right).
\end{align}
We see:
\begin{align}
v^R_{(2)} = \left(
\begin{array}{cc}
 0 & 1 \\
 -5 & -7 \\
\end{array}
\right), \quad v^L_{(2)} = \left(
\begin{array}{cc}
 -3 & -5 \\
 4 & 5 \\
\end{array}
\right),
\end{align}
and find $|\det(v^R_{(2)})| = |\det(v^L_{(2)})| = 5$. The effective K-matrix is 
\begin{align}
K^{\rm eff}=\left(\begin{array}{cc}
25 & 35\\
35& 50\\
\end{array}\right).
\end{align}
Thus, the entanglement entropy is given by,
\begin{align}
S_e(\nu=2, {\cal M}^a_{(2)}) = \alpha \ell - \log(5).
\end{align}
While we have no proof, brute force searching suggests the $\log(5)$ correction to be the minimal possible non-trivial value for a constant correction to the $\nu=2$ entanglement entropy. 
This provides another example of a free-fermion state whose bulk can be modified to generate a non-vanishing contribution to the topological entanglement entropy, albeit in this case we must break extra symmetries compared to the $\nu=4$ case.

\subsection{$\nu = 4/3$}
\label{fracgapvec}

Let us now consider a fractionalized bulk state.
We will show that the $\nu=4/3$ state admits a constant correction to its entanglement entropy, although our construction violates translation invariance along the cut.
We construct the bilayer $\nu=4/3$ state from one layer of $\nu = 1$ and one layer of $\nu = 1/3$, where within each layer, the wires are sewn together with the tunneling operator defined in Eqn. (\ref{laughlinoperator}) with $m=1$ and $m=3$, respectively.

As explained in Sec. \ref{laughlinstates}, if the tunneling operators are removed between two nearest-neighbor wires, we obtain a non-chiral Luttinger liquid whose action is defined by the matrices, $K^{R/L} = \begin{pmatrix} 1 & 0 \cr 0 & 3 \end{pmatrix}$, and $2\pi$-periodic fields, $\varphi^{R/L}_{1,2}$.
With respect to these Luttinger liquid variables, the tunneling operators that we have removed, take the form: 
\begin{align}
{\cal O}^{(1)} = & \cos(\varphi_1^L + \varphi^R_1), \cr
{\cal O}^{(1/3)} = & \cos(3 \varphi^L_2 + 3 \varphi^R_2).
\end{align}

Now suppose a different set of nearest-neighbor tunneling interactions are enabled to become strong along an entanglement cut or throughout the bulk.
With respect to the Luttinger liquid variables, we consider the tunneling operators, $\tilde{\cal O}^{(4/3)}_\beta = \cos\Big(({\cal M}_{({4 \over 3})}^a)_{\beta I} \varphi^a_I\Big)$ defined by the integer vectors:
\begin{align}
({\cal M}^R_{({4 \over 3})})_{1 I} = & (2, 0), \quad & ({\cal M}^L_{({4 \over 3})})_{1 I} = (1,3), \cr
({\cal M}^R_{({4 \over 3})})_{2 I} = & (1, 3), \quad & ({\cal M}^L_{({4 \over 3})})_{2 I} = (2,0).
\end{align}
These integer vectors are null (they satisfy Eqn. ({\ref{newnull})), and the corresponding tunneling operators conserve charge.
However, $({\cal M}^R_{({4 \over 3})})^{-1} {\cal M}^L_{({4 \over 3})}$ is not in ${\rm GL}(2, \mathbb{Z}),$
and thus, we expect a constant correction to the entanglement entropy.

To find this correction, we compute the Smith normal form:
\begin{align}
{\cal U}_{({4 \over 3})} ({\cal M}_{({4 \over 3})}^R, {\cal M}_{({4 \over 3})}^L) {\cal V}_{({4 \over 3})} =  {\cal S}_{({4 \over 3})},
\end{align}
where
\begin{align}
{\cal U}_{({4 \over 3})} = \left(
\begin{array}{cc}
 1 & 0 \\
 -2 & -1 \\
\end{array}
\right), \quad 
{\cal S}_{({4 \over 3})} = \left(
\begin{array}{cccc}
 1 & 0 & 0 & 0 \\
 0 & 3 & 0 & 0 \\
\end{array}
\right), \cr
{\cal V}_{({4 \over 3})} = \left(
\begin{array}{cccc}
 0 & 0 & 1 & 0 \\
 0 & 1 & 1 & 2 \\
 1 & 0 & -2 & -3 \\
 0 & 0 & 0 & 1 \\
\end{array}
\right).
\end{align}
We observe:
\begin{align}
v^R_{({4 \over 3})} = \left(
\begin{array}{cc}
 1 & 0 \\
 1 & 2 \\
\end{array}
\right), \quad v^L_{({4 \over 3})} = \left(
\begin{array}{cc}
 -2 & -3 \\
 0 & 1 \\
\end{array}
\right),
\end{align}
and find $|\det(v^R_{({4 \over 3})})| = |\det(v^L_{({4 \over 3})})| = 2$. The effective $K$-matrix is
\begin{align}
K^{\rm eff}=\left(\begin{array}{cc}
4 & 6\\
6& 12\\
\end{array}\right).\end{align}
Using Eqn. (\ref{entropygeneral}), we obtain the entanglement entropy:
\begin{align}
S_e(\nu={4 \over 3}, {\cal M}^a_{({4 \over 3})}) = \alpha \ell - {1 \over 2} \log(3) -  \log(2).
\end{align}
We thus find that the original TEE  $\log(\sqrt{3})$ is corrected by a $\log(2)$ contribution due to the modified interactions along the entanglement cut.

All of these examples illustrate the key feature that starting with a conventional wire construction of a quantum Hall phase one can shift the TEE by either modifying the tunneling terms at the entanglement cut itself, or creating a fully modified bulk phase which can still be attached to the same edge theory. The implications of these results are surprising because it, for example, illustrates that we can have identical edge theories attached to bulk phases with differing TEE contributions. These bulk phases appear to differ by the allowed local excitations that can tunnel between neighboring wires -- in certain strict limits -- and this constraint on the local tunneling processes reduces the total entanglement entropy.

\section{Examples Part Deux: Stable Edge Phases}
\label{parttwo}

In the previous section, we showed how the entanglement spectrum and entropy depends on the choice of tunneling interactions across the interface, but always with $K^R = K^L$.
In this section, we consider heterojunctions where this is not the case.
In particular, we consider the case where $K^R$ and $K^L$ are not exactly equivalent, but are only stably equivalent, i.e. there exists an invertible, integral matrix, $W$, such that  $W^T (K^R \oplus \sigma_z)W = K^L \oplus\sigma_z$ and there does not exist any such $W$ when the $\sigma_z$ factors are removed.
\footnote{It is possible to be more general by replacing each $\sigma_z$ by unimodular matrices, $U_R, U_L$, i.e., matrices with unit determinant, such that there exists an invertible, integer matrix $W$ such that $W^T(K^R \oplus U_R) W = K^L \oplus U_L$.
We shall not explore this possibility here.}
Examples, and a more detailed discussion, of stable equivalence can be found in Refs. [\onlinecite{PlamadealaE8}] and  [\onlinecite{generalstableequivalence}].
In the two examples we discuss, we will first explain how each phase can be built from the wire construction, although the constructions are by no means unique, and then consider the entanglement spectrum and entropy.

Notice that Eqn.~(\ref{newnull}) implies that whenever $(M^R)^{-1}M^L$ is in ${\rm GL}(N,Z)$ that $K^R$ and $K^L$ are equivalent up to a ${\rm GL}(N,Z)$ transformation. 
However, when $K^R$ and $K^L$ are not \emph{exactly} equivalent,  $(M^R)^{-1}M^L \notin GL(N,Z)$ and we expect a correction to the constant sub-leading term in the entanglement entropy. Our examples verify this fact.
For these statements, we have assumed $|\det(K^R)| = |\det(K^L)|,$
but this assumption will be relaxed in Sec. \ref{laughlininterface} where we consider interfaces between general Laughlin states.

\subsection{$\nu=8$}
\label{intreconstruct}

Our first example occurs at filling $\nu = 8$.
We will consider an interface between a free-fermion $\nu=8$ integer quantum Hall state and its stably equivalent partner the $E_{8}$ bosonic integer quantum Hall state. We will begin by first explicitly constructing the stably equivalent partner to the usual $\nu = 8$ state.

\subsubsection{$\nu=8$ Edge Reconstruction}

Here we describe how modifying the couplings on wires $j$ and $j-1$ will reconstruct the usual $\nu = 8$ edge (i.e., 8 layers of $\nu = 1$) on wire $j$ into its stably equivalent partner. 
First, we remove the nearest-neighbor backscattering terms, 
\begin{equation}\mathcal{O}^{(8)}_{I;j-1,j} =  \cos(\varphi_{I, j}^L + \varphi_{I, j-1}^R) \sim (\psi^L_{I, j})^\dagger \psi^R_{I, j-1} + {\rm h.c.},\label{I8interactions}\end{equation}
where $I=1, \ldots, 8$ is the layer index, as well as the analogous term between wires $j$ and $j+1$. 
Then the free fermion Luttinger liquid on wire $j$ is defined by matrices $K_j^{R/L} = \mathbb{I}_8$.
Uniqueness of unimodular lattices with signature $(M, N)$, with $M, N > 0$ implies the matrix equality,
\begin{align}
\label{integeruniqueness}
\mathbb{I}_8 \oplus (- \mathbb{I}_1) =W_{(8,1)}^T \left( K_{\rm E_8} \oplus (- \mathbb{I}_1) \right) W_{(8,1)}
\end{align}
where the ${\rm E_8}$ Cartan matrix, $K_{E_8}$ is given in Eqn.~(\ref{KE8})
and $W_{(8,1)}$ is some matrix in ${\rm GL}(9, \mathbb{Z})$ (the form of $W_{(8,1)}$ and an explanation of how to find it are given in Appendix~\ref{sec:I8toE8}).
Because the diagonal entries of $K_{\rm E_8}$ are even, all vertex operators that create local particles have integer spin and hence describe bosonic quasiparticles.
Thus, the equality in Eqn. (\ref{integeruniqueness}) is somewhat remarkable: it implies the (non-chiral) eight-channel free \emph{fermion} Luttinger liquid may be decomposed into a left-moving free Fermi liquid and a ``bosonic" right-moving sector, exactly.

We implement this decomposition on wire $j$ by the field redefinition,
\begin{align}
\label{integerchange}
\begin{pmatrix} \varphi^R \cr \varphi^L \end{pmatrix}_{I,j}= \left( W^{-1}_{(8,8)} \right)_{I J} \begin{pmatrix}\tilde{\varphi}^R \cr \tilde{\varphi}^L \end{pmatrix}_{J,j},
\end{align}
where $W_{(8,8)} = W_{(8,1)} \oplus \mathbb{I}_7 \in {\rm GL}(16,\mathbb{Z})$ has no effect on the left-moving fields, $\varphi^L_{I,j}$, for $I=2, \ldots, 8$.
Using the identity in Eqn. (\ref{integeruniqueness}), the Luttinger liquid action for the new fields, $\tilde{\varphi}_{I, j}^{R/L}$, is defined with respect to the matrices, $\tilde{K}^R_j = K_{\rm E_8}$ and $\tilde{K}^L_j = \mathbb{I}_8$.

We sew wires $j$ and $j-1$ back together by adding tunneling interactions that gap out the left-moving modes on wire $j$ and the right-moving modes on wire $j-1$,
\begin{align}
\label{integeroptrans}
\tilde{{\cal O}}_{I; j-1, j}^{(8)} = \cos(\tilde{\varphi}^L_{I, j} + \varphi^R_{I, j-1}).
\end{align}
These operators are charge conserving (the total charge density operator on wire $j$ is ${1 \over 2\pi} t^a_I \partial_x \tilde{\varphi}^a_{I,j}$, where $t^R_{1,j}=-2, t^R_{4,j}=2,$ and otherwise $t^R_{I,j}=0$; additionally $t^L_{I,j}=1$).
Since $\tilde{\varphi}^L_{I,j} = \varphi^L_{I,j}$ for $I=2,...,8$, the interactions are identical to Eqn.~(\ref{I8interactions}) when $I\neq 1$. However, $\tilde{{\cal O}}_{I=1; j-1, j}^{(8)}$ differs from ${\cal O}_{I=1; j-1, j}^{(8)}$ and breaks translation invariance by carrying $4k_F^{(0)}$ units of extra momentum. Although it is a single-particle backscattering term in the new variables, it is a many-body interaction in the original fields, 
\begin{widetext}
\begin{align}
\tilde{{\cal O}}^{(8)}_{1, j-1, j} = \cos\Big(\varphi^R_{1, j} + \varphi^R_{2,j} + \varphi^R_{3,j} + \varphi^R_{4,j} + \varphi^R_{5,j} - \varphi^R_{6,j} - \varphi^R_{7,j} - \varphi^R_{8,j} + 3 \varphi^L_{1,j} + \varphi^R_{1, j-1}\Big).
\end{align}
\end{widetext}
The argument of the cosine does {\it not} commute with the argument of ${\cal O}_{I=1; j-1, j}^{(8)}$.
This implies that there will be a line of critical excitations {\it if} both $\tilde{{\cal O}}^{(8)}_{1, j-1, j}$ and  ${\cal O}_{I=1; j-1, j}^{(8)}$ were to be present in the action and be of equal dominance.
We expect this zero crossing can be avoided, however.

When the $\tilde{{\cal O}}_{I; j-1, j}^{(8)}$ interactions dominate over the interactions in Eqn.~(\ref{I8interactions}), the gapless right-moving modes on wire $j$ enter a chiral ``bosonic" Luttinger liquid phase defined by the matrix $\tilde{K}^R_j = K_{\rm E_8}$.
Thus, we have explicitly shown how altering the interactions near the entanglement cut between wires $j$ and $j+1$ can change the $K$ matrix on one side of the cut.

We can compute the entanglement spectrum for a cut between wires $j$ and $j+1$ following the method described in Sec.~\ref{sec:TEE}, once we specify whether the low-energy modes on wire $j+1$ are eight fermionic modes or eight bosonic modes; the construction for the latter is analogous to that of this section. There are three possible cases: the asymmetric case, where $K_{j+1}^L = \mathbb{I}_8$, $K_j^R = K_{\rm E_8}$; the symmetric-${\rm E}_8$ case, where $\tilde{K}_j^R = \tilde{K}_{j+1}^L =  K_{\rm E_8}$; and the symmetric-$\mathbb{I}_8$ case, where $ K^R_j = K^L_{j+1} = \mathbb{I}_8$. We will consider them separately.

\subsubsection{Symmetric Interface Entanglement Spectra}
\label{symmint}

It is simplest to begin by considering the two possible symmetric interfaces.
We drop the wire labels, $j$ and $j+1$: the right-moving modes belong to wire $j$ and the left-moving modes belong to wire $j+1$.
We first sew together the symmetric-$\mathbb{I}_8$ interface, corresponding to an un-perturbed $\nu=8$ state, by adding single-particle backscattering terms, $\cos(({\cal M}^R_{(8)})_{\beta I}\varphi^R_I + ({\cal M}^L_{(8)})_{\beta I}\varphi^L_I)$, where
$({\cal M}^R_{(8)})_{\beta I} = ({\cal M}^L_{(8)})_{\beta I} = \delta_{\beta I}.$
There is no constant correction to the topological entanglement entropy because $({\cal M}^R_{(8)})^{-1} {\cal M}^L_{(8)} = \mathbb{I}_8$.
Since this state has no fractionalization, the topological entanglement entropy vanishes.

Likewise, when expressed in terms of the modes $\tilde{\varphi}^{R/L}$, the symmetric-${\rm E_8}$ interface can be sewn together using the operators  $\cos(({\cal M}^R_{(8)})_{\beta I}\tilde{\varphi}^R_I + ({\cal M}^L_{(8)})_{\beta I}\tilde{\varphi}^L_I)$.
Therefore, there is no constant correction to the topological entanglement, which again vanishes because $\det{(\tilde{K}_{\rm E_8})} = 1$.

Since the universal part of the entropy does not distinguish the phases we can consider the entanglement spectrum. The entanglement Hamiltonian takes the form given in Eqn. (\ref{entham}).
The only difference between the spectrum for the symmetric-$\mathbb{I}_8$ and symmetric-${\rm E_8}$ interfaces lies in the eigenvalues of the zero mode operators.
These eigenvalues reflect the underlying lattice of local quasiparticles: for the symmetric-$\mathbb{I}_8$ interface, the eigenvalues are defined by lattice vectors in $\mathbb{I}_8$ which square to arbitrary integers; for the symmetric-${\rm E_8}$ interface, the eigenvalues are defined by lattice vectors in ${\rm E_8}$ which square to even integers.
A basis for the ${\rm E_8}$ lattice is given in Appendix~\ref{sec:I8toE8}.

For the special, fine-tuned values of the parameters, $\lambda_a = \lambda_b$ for all $a,b$ that enter Eqn. (\ref{tunndiag}) this difference in quasiparticle lattice structure will be  manifest in the spectrum. However, such values are very fine-tuned and their generic dependence upon the coupling constants will make it difficult to clearly distinguish the two interfaces.
For this reason, it is useful to compare the response of the spectrum to the insertion of $\pi$-flux through the cylinder.
The spectrum of the symmetric-$\mathbb{I}_8$ interface shifts after $\pi$-flux is inserted, while the symmetric-${\rm E_8}$ interface is invariant under the $\pi$-flux insertion.

\subsubsection{Asymmetric Interface}
\label{asymmint}

We now consider the asymmetric interface, gapped by the tunneling interactions, $\cos\Big(({\cal M}^R_{(8)/(\tilde{8})})_{\beta I} \tilde{\varphi}^{L}_{I} + ({\cal M}^L_{(8)/(\tilde{8})})_{\beta I} \varphi^L_{I}\Big)$ , defined by the integer vectors:
\begin{widetext}
\begin{align}
({\cal M}^R_{(8)/(\tilde{8})})_{1 I} = & \begin{pmatrix} 0^7 & 1\end{pmatrix}, \quad & ({\cal M}^L_{(8)/(\tilde{8})})_{1 I} = & \begin{pmatrix} 2 & 0^7\end{pmatrix}, \cr
({\cal M}^R_{(8)/(\tilde{8})})_{2 I} = & \begin{pmatrix} -1 & 0^5 & 1 & -1\end{pmatrix}, \quad & ({\cal M}^L_{(8)/(\tilde{8})})_{2 I} = & \begin{pmatrix} -1 & 1 & 0^6\end{pmatrix}, \cr
({\cal M}^R_{(8)/(\tilde{8})})_{3 I} = & \begin{pmatrix} -1 & 1 & 0^4 & -1 & 0 \end{pmatrix}, \quad & ({\cal M}^L_{(8)/(\tilde{8})})_{3 I} = & \begin{pmatrix} -1 & 0 & 1 & 0^5\end{pmatrix}, \cr
({\cal M}^R_{(8)/(\tilde{8})})_{4 I} = & \begin{pmatrix} 0 & -1 & 1 & 0^5\end{pmatrix}, \quad & ({\cal M}^L_{(8)/(\tilde{8})})_{4 I} = & \begin{pmatrix} -1 & 0^2 & 1 & 0^4\end{pmatrix}, \cr
({\cal M}^R_{(8)/(\tilde{8})})_{5 I} = & \begin{pmatrix} 0^2 & -1 & 1 & 0^4\end{pmatrix}, \quad & ({\cal M}^L_{(8)/(\tilde{8})})_{5 I} = & \begin{pmatrix} -1 & 0^3 & 1 & 0^3\end{pmatrix}, \cr
({\cal M}^R_{(8)/(\tilde{8})})_{6 I} = & \begin{pmatrix} 0^3 & 1 & -1 & 0^3\end{pmatrix}, \quad & ({\cal M}^L_{(8)/(\tilde{8})})_{6 I} = & \begin{pmatrix} 1 & 0^4 & 1 & 0^2\end{pmatrix}, \cr
({\cal M}^R_{(8)/(\tilde{8})})_{7 I} = & \begin{pmatrix} 0^4 & 1 & -1 & 0^2\end{pmatrix}, \quad & ({\cal M}^L_{(8)/(\tilde{8})})_{7 I} = & \begin{pmatrix} 1 & 0^5 & 1 & 0\end{pmatrix}, \cr
({\cal M}^R_{(8)/(\tilde{8})})_{8 I} = & \begin{pmatrix} 0^5 & 1 & 0^2\end{pmatrix}, \quad & ({\cal M}^L_{(8)/(\tilde{8})})_{8 I} = & \begin{pmatrix} 1 & 0^6 & 1\end{pmatrix}
\end{align}
\end{widetext}\noindent where $0^n$ represents $n$ consecutive zero entries in the vector.
If re-fermionized in the symmetric-$\mathbb{I}_8$ basis, these gap-generating interactions take the form of charge-conserving single-particle backscattering terms, $(\psi^L_I)^\dagger \psi^R_I + {\rm h.c.}$
They preserve charge conservation, but not translation invariance along the interface.
Because $({\cal M}^R_{(8)/(\tilde{8})})^{-1} {\cal M}^L_{(8)/(\tilde{8})} \notin {\rm GL}(8, \mathbb{Z})$, we expect a constant correction to the entanglement entropy.

To find this correction, we compute the Smith normal form:
\begin{align}
\label{asymsmith}
{\cal U}_{(8)/(\tilde{8})} ({\cal M}_{(8)/(\tilde{8})}^R, {\cal M}_{(8)/(\tilde{8})}^L) {\cal V}_{(8)/(\tilde{8})} =  {\cal S}_{(8)/(\tilde{8})}.
\end{align}
Because of the large size of the matrices appearing in Eqn. (\ref{asymsmith}), we have relegated the explicit forms of ${\cal U}_{(8)/(\tilde{8})}, {\cal V}_{(8)/(\tilde{8})}$, and ${\cal S}_{(8)/(\tilde{8})}$ to Appendix \ref{largematrices}.
However, we may decompose ${\cal V}_{(8)/(\tilde{8})}$ to find:
\begin{align}
v^R_{(8)/(\tilde{8})} = & \left(
\begin{array}{cccccccc}
 0 & 1 & 0 & 0 & 0 & 0 & 0 & 0 \\
 -5 & 0 & 0 & 1 & 1 & -1 & -1 & -1 \\
 -4 & 0 & 0 & 0 & 1 & -1 & -1 & -1 \\
 -3 & 0 & 0 & 0 & 0 & -1 & -1 & -1 \\
 -2 & 0 & 0 & 0 & 0 & 0 & -1 & -1 \\
 -1 & 0 & 0 & 0 & 0 & 0 & 0 & -1 \\
 -6 & -1 & 1 & 1 & 1 & -1 & -1 & -1 \\
 -2 & 0 & 0 & 0 & 0 & 0 & 0 & 0 \\
\end{array}
\right), \cr
v^L_{(8)/(\tilde{8})} = & \left(
\begin{array}{cccccccc}
 1 & 0 & 0 & 0 & 0 & 0 & 0 & 0 \\
 5 & 2 & -1 & -1 & -1 & 1 & 1 & 1 \\
 0 & 0 & 1 & 0 & 0 & 0 & 0 & 0 \\
 0 & 0 & 0 & 1 & 0 & 0 & 0 & 0 \\
 0 & 0 & 0 & 0 & 1 & 0 & 0 & 0 \\
 0 & 0 & 0 & 0 & 0 & 1 & 0 & 0 \\
 0 & 0 & 0 & 0 & 0 & 0 & 1 & 0 \\
 0 & 0 & 0 & 0 & 0 & 0 & 0 & 1 \\
\end{array}
\right),
\end{align}
from which we find $|\det{(v^R_{(8)/(\tilde{8})})}| = |\det{(v^L_{(8)/(\tilde{8})})}| = 2$.
Thus, the entanglement entropy across the asymmetric-$\mathbb{I}_8/{\rm E_8}$ interface is:
\begin{align}
S_e(\mathbb{I}_8/{\rm E_8}, {\cal M}^a_{(8)/(\tilde{8})}) = \alpha \ell -  \log(2).
\end{align}

We generally expect there to always be a constant correction to the entanglement entropy at such asymmetric interfaces.
The reason is that neither the $\mathbb{I}_8$ lattice nor the ${\rm E_8}$ lattice contains the other, i.e., there is no way to perfectly embed one in the other.
Recall that the cosine interactions are invariant under independent shifts of the right/left-moving fields by elements in their respective lattices, however, the expansion about a particular vacuum breaks this shift symmetry down to one where any shift of a right-moving field is accompanied by a compensating shift of a left-moving field.
When an arbitrary shift of a right-moving field can be canceled by a left-moving one, i.e., when the matrix $({\cal M}^R)^{-1} {\cal M}^L \in GL(N, \mathbb{Z})$ for some $N$, there is no constant correction to the entropy.
However, if $({\cal M}^R)^{-1} {\cal M}^L \notin GL(N, \mathbb{Z})$, then there is a restriction on the allowed shifts down to a common sub-lattice. 
When neither of the two lattices contain the other, there is necessarily a restriction and a constant correction to the entropy is expected.

\subsection{$\nu = 8/15$}
\label{fracsectstable}

Our second example occurs at filling $\nu=8/15$. 
Again, we imagine making an entanglement cut between wires $j$ and $j+1,$
however, the construction for this stable-equivalence requires that we modify nearest-neighbor {\it and} next-nearest-neighbor interactions in order to access a novel chiral edge phase on wire $j$.

\subsubsection{$\nu=8/15$ Edge Reconstruction}
\label{symmconstruct}

We construct the bilayer $\nu=8/15$ state using the technique described in Sec. \ref{laughlinstates}, where the first layer is in the $\nu=1/3$ state and the second layer is in the $\nu=1/5$ state.
Within each layer, wires are filled to the appropriate density and are sewn together with the tunneling operator defined in Eqn. (\ref{laughlinoperator}) with $m=3$ and $m=5$, respectively.

To construct the stably equivalent phase for gapless right-moving modes on wire $j$, we will alter the interactions on wires $j+1, j, j-1$, and $j-2$. We begin by removing all interactions that couple these wires together.
Once decoupled, these wires are in a 1D Fermi liquid phase.
It will be useful to describe excitations on wire $j$ using a ``Luttinger liquid basis" of field variables and the excitations on wire $j-1$ using a ``free fermion basis" of variables.
For instance, it is convenient to describe the excitations on wire $j$ in terms of the variables appropriate to the decoupled $\nu = 1/3 + 1/5$ state, $\varphi^{R/L}_{I,j}$, which are described by the action~(\ref{usualaction}) with $K^{R/L}_j = \begin{pmatrix} 3 & 0 \cr 0 & 5\end{pmatrix}$ and obey the periodicity condition given by Eqn. (\ref{newperiod}) where $m=3$ or $5$. On wire $j-1$, we consider the original electron fields, $\phi^{R/L}_{I,j+1}$, which are described by the action~(\ref{usualaction}) with $K^{R/L}_{j-1} = \pm \mathbb{I}_2$.

We will exploit the following identity:
\begin{align}
\label{fracrelation}
(W^{({8 \over 15})})^T \begin{pmatrix} 3 & 0 & 0 & 0\cr 0 & 5 & 0 & 0 \cr 0 & 0 & 1 & 0 \cr 0 & 0 & 0 & -1  \end{pmatrix} W^{({8 \over 15})} = \begin{pmatrix} 2 & 1 & 0 & 0 \cr 1 & 8 & 0 & 0 \cr 0 & 0 & 1 & 0 \cr 0 & 0 & 0 & -1 \end{pmatrix},
\end{align}
where,
\begin{align}
W^{({8 \over 15})} =
\label{fracw}
 \left(
\begin{array}{cccc}
 -1 & -3 & 0 & -1 \\
 0 & -3 & 0 & -1 \\
 0 & 0 & 1 & 0 \\
 1 & 8 & 0 & 3 \\
\end{array}
\right) \in {\rm GL}(4, \mathbb{Z}).
\end{align}
Eqn. (\ref{fracrelation}) motivates us to make the change of variables,
\begin{align}
\label{fracchange}
\begin{pmatrix}
\varphi^R_{1, j} \cr \varphi^R_{2, j} \cr \phi^R_{1, j-1} \cr \phi^L_{1, j-1}
\end{pmatrix}
\equiv W^{({8 \over 15})} 
\begin{pmatrix}
\tilde{\varphi}^R_{1, j} \cr \tilde{\varphi}^R_{2, j} \cr \tilde{\phi}^R_{1, j-1} \cr \tilde{\phi}^L_{1, j-1}
\end{pmatrix}.
\end{align}
We observe that $\phi^R_{1, j-1} = \tilde{\phi}^R_{1, j-1}$.
The action in terms of the fields $\tilde{\varphi}^R_{I,j}$ is Eqn.~(\ref{usualaction}) with $\tilde{K}^R_j = \begin{pmatrix} 2 & 1 \cr 1 & 8\end{pmatrix}$ while the action for the $\tilde{\phi}^{R/L}_{1,j-1}$ is the same as for the $\phi^{R/L}_{1,j-1}$ fields.
On wire $j-1$, we then switch to the $\varphi_{j-1}^{R/L}$ fields, defined via Eqn.~(\ref{redefinelaughlin}).

To sew the wires $j, j-1, j-2$ back together, we couple the left-movers on wire $j$ to the right-movers on wire $j-1$, and left-movers on wire $j-1$ to the right-movers on $j-2,$ by adding the tunneling interactions in Eqn. (\ref{laughlinoperator}) within each layer. The remaining low-energy modes are $\tilde{\varphi}^R_{I, j}$. 
\footnote{The careful reader might be worried about the periodicity conditions on the fields. Inverting $W^{({8 \over 15})}$ shows that the fields $\tilde{\phi}^{R/L}_{I, j-1}$ are independently $2\pi$-periodic. Thus, they can be used to define  $\tilde{\varphi}^{R/L}_{I, j-1}$ via Eqn.~(\ref{redefinelaughlin}), which have the periodicity of Eqn.~(\ref{newperiod}). When the couplings between wires $j-2$ and $j-1$ and between $j-1$ and $j$ are implemented, the remaining gapless fields on wire $j$ have $2\pi$ periodicity.}
Thus, we have provided an explicit construction for the interactions that must dominate in order that the low-energy modes on wire $j$ at the entanglement cut between $j$ and $j+1$ are described by $\tilde{K}^R_j = \begin{pmatrix} 2 & 1 \cr 1 & 8\end{pmatrix}$.

\subsubsection{Symmetric Interfaces}
\label{symmfrac}

To calculate the entanglement spectrum with respect to an entanglement cut between wires $j$ and $j+1$, we must specify the phase of the low-energy modes on both sides of the interface.
As in the $\nu = 8$ case, there are three situations to consider: the symmetric interface where $K^R_j = K^L_{j+1}  = \begin{pmatrix} 3 & 0 \cr 0 & 5 \end{pmatrix}$, the symmetric interface where $K^R_j = K^L_{j+1}  = \begin{pmatrix} 2 & 1 \cr 1 & 8 \end{pmatrix}$, and the asymmetric interface.

We compute the spectrum following Sec~\ref{sec:TEE}, beginning with the symmetric interfaces.
In what follows, we drop the wire label: right-moving modes belong to wire $j$ and the left-moving modes belong to wire $j+1$.
We sew together the symmetric-$\begin{pmatrix} 3 & 0 \cr 0 & 5\end{pmatrix}$ interface, corresponding to an un-perturbed $\nu=8/15$ state, by adding the backscattering terms, $\cos ( ({\cal M}^R_{({8 \over 15}), s1})_{\beta I}\varphi_I^R + ({\cal M}^L_{({8 \over 15}), s1})_{\beta I}\varphi_I^L)$, where $({\cal M}^{R/L}_{({8 \over 15}), s1}) = \begin{pmatrix} 3 & 0 \\ 0 & 5 \end{pmatrix}$.
There is no constant correction to the topological entanglement entropy because $({\cal M}^R_{({8 \over 15})})^{-1} {\cal M}^L_{({8 \over 15})} = \mathbb{I}_2$.
The topological entanglement entropy $\gamma_{{8 \over 15}} = \log(\sqrt{15})$ which is the expected conventional value.

Likewise, when expressed in terms of the modes $\tilde{\varphi}^{R/L}$, the symmetric-$\begin{pmatrix} 2 & 1 \cr 1 & 8\end{pmatrix}$ interface can be sewn together using tunneling operators 
$\cos ( ({\cal M}^R_{({8 \over 15}), s2})_{\beta I}\varphi_I^R + ({\cal M}^L_{({8 \over 15}), s2})_{\beta I}\varphi_I^L)$, where ${\cal M}^L_{({8 \over 15}), s2} = \begin{pmatrix} 2 & 1 \\ 1 & 8 \end{pmatrix}$.
Again, there is no constant correction to the topological entanglement, which is given by  $\gamma_{{8 \over 15}} = \log(\sqrt{15})$.

The entanglement Hamiltonian for both cases takes the form given in Eqn. (\ref{entham}).
The only difference between the spectrum for the symmetric-$\begin{pmatrix} 3 & 0 \cr 0 & 5\end{pmatrix}$ and symmetric-$\begin{pmatrix} 2 & 1 \cr 1 & 8\end{pmatrix}$ interfaces lies in the eigenvalues of the zero mode operators.
These eigenvalues reflect the underlying lattice: for the symmetric-$\begin{pmatrix} 3 & 0 \cr 0 & 5\end{pmatrix}$ interface, the eigenvalues are defined by lattice vectors in $\sqrt{3} \mathbb{Z} \oplus \sqrt{5} \mathbb{Z}$ which square to arbitrary integer multiples of $3$ and $5$; for the symmetric-$\begin{pmatrix} 2 & 1 \cr 1 & 8\end{pmatrix}$ interface, the eigenvalues may be defined by integral linear combinations of the vectors: $\begin{pmatrix} \sqrt{2} & 0 \end{pmatrix}$, $\begin{pmatrix} 1/\sqrt{2} & \sqrt{15/2} \end{pmatrix}$.
The square of elements in the lattice defined by these two basis vectors is always even.

As mentioned above for the fine-tuned values of the parameters, $\lambda_a = \lambda_b$ for all $a,b$, these differences are manifest in the spectrum.
However, for generic values  a clear identification of the two interfaces will be difficult. 
For this reason, it is useful to compare the response of the spectrum to the insertion of $\pi$-flux through the cylinder.
The spectrum of the symmetric-$\begin{pmatrix} 3 & 0 \cr 0 & 5\end{pmatrix}$ interface shifts after $\pi$-flux is inserted, while the symmetric-$\begin{pmatrix} 2 & 1 \cr 1 & 8\end{pmatrix}$ interface is invariant under the $\pi$-flux insertion.

\subsubsection{Asymmetric Interface}
\label{asymmfrac}

A set of operators that gap out the modes at an asymmetric-$\begin{pmatrix} 3 & 0 \cr 0 & 5\end{pmatrix}/\begin{pmatrix} 2 & 1 \cr 1 & 8\end{pmatrix}$ interface are
$\cos ( ({\cal M}^R_{({8 \over 15}), {\cal A}})_{\beta I}\varphi_I^R + ({\cal M}^L_{({8 \over 15}), {\cal A}})_{\beta I}\varphi_I^L)$ where ${\cal M}^R_{({8 \over 15}), {\cal A}} = \begin{pmatrix} 5 & -5 \\ 1 & -7\end{pmatrix}$ and ${\cal M}^L_{({8 \over 15}), {\cal A}} = \begin{pmatrix} 0 & 10 \\ -3 & 5\end{pmatrix}$.
These interactions conserve charge, but break translation invariance along the interface.
Notice $({\cal M}^R_{({8 \over 15}), A})^{-1} {\cal M}^L_{({8 \over 15}), A} \notin {\rm GL}(2, \mathbb{Z})$ as its entries are half-integral. 
Therefore, we expect a constant correction to the entanglement entropy.

To find this correction, we compute the Smith normal form:
\begin{align}
\label{asymsmith}
{\cal U}_{({8 \over 15}), A} ({\cal M}_{({8 \over 15}), A}^R, {\cal M}_{({8 \over 15}), A}^L) {\cal V}_{({8 \over 15}), A} =  {\cal S}_{({8 \over 15}), A},
\end{align}
where
\begin{align}
{\cal U}_{({8 \over 15}), A} = & \left(
\begin{array}{cc}
 0 & 1 \\
 1 & -5 \\
\end{array}
\right), \quad {\cal S}_{({8 \over 15}), A} = \left(
\begin{array}{cccc}
 1 & 0 & 0 & 0 \\
 0 & 15 & 0 & 0 \\
\end{array}
\right), \cr
{\cal V}_{({8 \over 15}), A} = & \left(
\begin{array}{cccc}
 1 & 3 & 1 & -2 \\
 0 & 0 & 1 & 0 \\
 0 & 1 & -2 & 1 \\
 0 & 0 & 0 & 1 \\
\end{array}
\right).
\end{align}
We see that $|\det(v^R_{({8 \over 15}), A})| = |\det(v^L_{({8 \over 15}), A})| = 2$. We can also calculate an effective K-matrix in this case via
\begin{align}
K^{\rm eff}=(v^{R})^{T}K^Rv^{R}=(v^{L})^{T}K^Rv^{L}\nonumber\\
=\left(\begin{array}{cc}
8& -6\\
-6& 12\\
\end{array}\right).
\end{align}
Therefore, the entanglement entropy across the asymmetric-$\begin{pmatrix} 3 & 0 \cr 0 & 5\end{pmatrix}/\begin{pmatrix} 2 & 1 \cr 1 & 8\end{pmatrix}$ interface is:
\begin{align}
S_e({\cal M}^a_{({8 \over 15}), A}) = \alpha \ell - {1 \over 2} \log(15) -  \log(2).
\end{align}
Again, we generally expect such a correction across asymmetric interfaces when the underlying lattices do not contain one another.

\section{Examples Part III: Interfaces between distinct Laughlin States}
\label{laughlininterface}

Our final set of examples are perhaps the simplest.
However, the interfaces we consider provide interesting features.
We consider an entanglement cut through gapped interfaces that separate distinct Abelian states.
The only relation that these states have to one another is that they may share a gapped interface. 

\subsection{Laughlin Interfaces}

First, we consider an entanglement cut made between the two distinct Laughlin states at filling fractions, $\nu=1/k^R$ and $\nu' = 1/k^L$. An interface between identical, single-component Laughlin states does not provide any interesting entanglement corrections.

The filling fractions cannot arbitrary: we require the modes living at the interface to have an instability to a gap-generating perturbation that transfers some number of electrons between the edges. Such a perturbation must take the form $\cos({\cal M}_{(k)}^R \varphi^R + {\cal M}_{(k)}^L \varphi^L)$, where ${\cal M}_{(k)}^{R/L}$ is a multiple of $k^{R/L}$ and $\varphi^{R/L}$ are the edge modes for the state at $\nu=1/k^R$ and $\nu'=1/k^L$, respectively.
To generate a gap, the integers, ${\cal M}_{(k)}^{R/L}$, must satisfy:
\begin{align}
\label{gapcondition}
{({\cal M}_{(k)}^R)^2 \over k^R} = {({\cal M}_{(k)}^L)^2 \over k^L}.
\end{align}
Thus, defining $g = {\rm gcd}[k^R, k^L]$, $k^{R/L} = g(\tilde{k}^{R/L})^2$, for some integers $\tilde{k}^{R/L}$.
In the remainder, we assume ${\cal M}_{(k)}^{R/L}$ and $k^{R/L}$ are related by Eqn. (\ref{gapcondition}).

If $k^R \neq k^L$, any such perturbation will necessarily violate charge conservation, in addition to translation invariance along the cut. Thus, the above interaction must be mediated by (some number of) Cooper pairs from a proximity-coupled superconductor and a periodic background potential.

The entanglement entropy is immediate using our algorithm.
We compute the Smith normal form:
\begin{align}
\label{asymsmith}
{\cal U}_{(k)} ({\cal M}_{(k)}^R, {\cal M}_{(k)}^L) {\cal V}_{(k)} =  {\cal S}_{(k)},
\end{align}
where
\begin{align}
{\cal U}_{(k)} = & \left(
\begin{array}{cc}
 1
\end{array}
\right), \quad {\cal S}_{(k)} = \left(
\begin{array}{cccc}
 {\rm gcd}[{\cal M}_{(k)}^{R/L}] & 0 \\
\end{array}
\right), \cr
{\cal V}_{(k)} = & \left(
\begin{array}{cccc}
 v_{1} & -\tilde{k}^L \\
 v_{2} & \tilde{k}^R  \\
\end{array}
\right),
\end{align}
where ${\rm gcd}[{\cal M}_{(k)}^{R/L}]$ is the greatest common divisor of ${\cal M}^R_{(k)}$ and ${\cal M}^L_{(k)}$ and
$v_1$ and $v_2$ are integers satisfying $v_1\tilde{k}^R+ v_2\tilde{k^L}= 1$. The effective K-matrix is $K^{\rm eff}=k^R k^L/g.$
Thus,
\begin{align}
S_e({\cal M}_{(k)}^{R/L}, {k^{R/L}}) = \alpha \ell - {1 \over 2}\left(  \log(k^R) + \log(k^L) - \log(g)\right),
\end{align}
which is invariant under interchanging $R \leftrightarrow L$.
When $k^R = k^L$, $g=k^L$ and the constant subleading term reduces to the expected result, $\log(\sqrt{k^R})$.

\subsection{Symmetry-Preserving Interfaces}

In this subsection, we break from our pattern and study interfaces between non-chiral states.
In general, such states present a new opportunity for how an interface may be gapped: edge modes on opposite sides of the interface need no longer interact; they may in principle obtain a spectral gap via intra-wire interactions.
For example, suppose $K^R=K^L=\sigma_z$. 
Then, in principle, we could choose the gapping vectors $({{1,1}, {0,0}})$ and $({{0,0}, {1,1}})$. 
These are valid gapping terms, but the ${\cal M}^{R/L}$ matrices are not invertible and so our methods do not apply when such terms are added.
However, in many cases symmetry preservation can preclude such interactions. 
Indeed, we will consider two examples where symmetry preservation ensures that a completely gapped interface can only occur through the interaction between edge modes on opposite sides of the entanglement cut.
These examples provide a simple illustration for how our methods may be applied to non-chiral states.

\subsubsection{Topological Insulator and Strong-Pairing Insulator Interface}

Our first example was the topic of a recent paper by Wang and Levin.\cite{WangLevinweaksymmetry}.
We consider the interface between a fermionic topological insulator and a strong pairing insulator.
Charge-conservation and time-reversal invariance prevents the edge modes of the fermionic topological insulator from obtaining a gap when bordering the topologically trivial vacuum.
However, these edge modes may obtain a gap in a symmetry-preserving manner when they border the strong pairing insulator.

The interface can be described by the Luttinger liquid action:
\begin{align}
S_{\rm SPI/TI} = & \int dt dx\Big[(K_{\rm SPI/TI})_{IJ} \partial_t \varphi_I \partial_x \varphi_J - V_{IJ} \partial_x \varphi_I \partial_x \varphi_J \cr
+ & g_\beta \cos(({\cal M}^R_{{\rm SPI/TI}}, {\cal M}^L_{{\rm SPI/TI}})_{\beta I} \varphi_I) \Big],
\end{align}
where
\begin{align}
& K_{\rm SPI/TI} = \begin{pmatrix} -8 & 0 & 0 & 0 \cr 0 & 8 & 0 & 0 \cr 0 & 0 & 1 & 0 \cr 0 & 0 & 0 & -1 \end{pmatrix}, \quad V = v \mathbb{I}, \cr
& ({\cal M}^R_{{\rm SPI/TI}}, {\cal M}^L_{{\rm SPI/TI}}) = \begin{pmatrix} 0 & 8 & 1 & 3  \cr 8 & 0 & 3 & 1\end{pmatrix},
\end{align}
and some positive constant $v$.
We note that the $R/L$ label on ${\cal M}^{R/L}_{\rm SPI/TI}$ refer to modes living on the right/left side of the entanglement cut.
We take the strong-pairing insulator to live on the right side of the cut and its edge excitations to be described by the modes $\varphi_{1,2}$.
The topological insulator edge modes live on the left side of the cut and are described by $\varphi_{3,4}$.
The interface is gapped in a time-reversal invariant and charge-conserving way by the cosine interactions, parameterized by $g_\beta$.

To compute the entanglement entropy obtained for an entanglement cut across such an interface, we put $({\cal M}^R_{{\rm TI/SPI}}, {\cal M}^L_{{\rm SPI/TI}})$ into its Smith normal form:
\begin{align}
\label{asymsmith}
{\cal U}_{{\rm SPI/TI}} ({\cal M}_{{\rm SPI/TI}}^R, {\cal M}_{{\rm SPI/TI}}^L) {\cal V}_{{\rm SPI/TI}} =  {\cal S}_{{\rm SPI/TI}},
\end{align}
where
\begin{align}
{\cal U}_{{\rm SPI/TI}} = & \left(
\begin{array}{cc}
 1 & 0 \\
 -3 & 1 \\
\end{array}
\right), \quad {\cal S}_{{\rm SPI/TI}} = \left(
\begin{array}{cccc}
 1 & 0 & 0 & 0 \\
 0 & 8 & 0 & 0 \\
\end{array}
\right), \cr
{\cal V}_{{\rm SPI/TI}} = &\left(
\begin{array}{cccc}
 0 & 1 & 3 & 1 \\
 0 & 0 & 1 & 0 \\
 1 & 0 & -8 & -3 \\
 0 & 0 & 0 & 1 \\
\end{array}
\right).
\end{align}
We find $|\det(v^R_{{\rm SPI/TI}})| = 1$ and $|\det(v^L_{{\rm SPI/TI}})| = 8$. The effective K-matrix is
\begin{align}
K^{\rm eff}=-\left(\begin{array}{cc}
64& 24\\
24& 8\\
\end{array}\right).
\end{align}

Thus, the entanglement entropy is,
\begin{align}
S_e({\cal M}_{{\rm SPI/TI}}^{R/L}) = \alpha \ell - {1 \over 2} \log(64).
\end{align}
Again, this result is symmetric under tracing out the strong-pairing or topological insulator edge degrees of freedom.

\subsubsection{Topological Insulator and Doubled Semion Interface}

In our second example, we consider the interface between the topological insulator and the doubled-semion state which was recently studied by Lu and Lee.\cite{LuLeesymmetryedges}
Again, the preservation of $U(1) \ltimes Z_2$ symmetry prevents the edge modes of the topological insulator from obtaining a spectral gap when bordering the trivial vacuum; however, the interface between the topological insulator and doubled-semion state may obtain a symmetry-preserving gap.

The interface can be described by the Luttinger liquid action:
\begin{align}
S_{\rm TI/DS} = & \int dt dx\Big[(K_{\rm TI/DS})_{IJ} \partial_t \varphi_I \partial_x \varphi_J - V_{IJ} \partial_x \varphi_I \partial_x \varphi_J \cr
+ & g_\beta \cos(({\cal M}^R_{{\rm TI/DS}}, {\cal M}^L_{{\rm TI/DS}})_{\beta I} \phi_I) \Big],
\end{align}
where
\begin{align}
& K_{\rm TI/DS} = \begin{pmatrix} 0 & 2 & 0 & 0 \cr 2 & 0 & 0 & 0 \cr 0 & 0 & 1 & 0 \cr 0 & 0 & 0 & -1 \end{pmatrix}, \quad V = v \mathbb{I}_4, \cr
& ({\cal M}^R_{{\rm TI/DS}}, {\cal M}^L_{{\rm TI/DS}}) = \begin{pmatrix} 0 & 2 & 1 & 1  \cr 2 & 0 & 1 & - 1\end{pmatrix},
\end{align}
for some positive constant $v$.
As before, the $R/L$ label on ${\cal M}^{R/L}_{{\rm TI/DS}}$ refer to modes living on the right/left side of the entanglement cut.
We take the doubled-semion state to live on the right side of the cut and its edge excitations to be described by the modes $\varphi_{1,2}$.
The topological insulator edge modes live on the left side of the cut and are described by $\varphi_{3,4}$ fields.
The interface is gapped in a time-reversal invariant and charge-conserving way by the cosine interactions, parameterized by $g_\beta$.

To compute the entanglement entropy obtained for an entanglement cut across such an interface, we put $({\cal M}^R_{{\rm TI/DS}}, {\cal M}^L_{{\rm TI/DS}})$ into its Smith normal form:
\begin{align}
\label{asymsmith}
{\cal U}_{{\rm TI/DS}} ({\cal M}_{{\rm TI/DS}}^R, {\cal M}_{{\rm TI/DS}}^L) {\cal V}_{{\rm TI/DS}} =  {\cal S}_{{\rm TI/DS}},
\end{align}
where
\begin{align}
{\cal U}_{{\rm TI/DS}} = & \left(
\begin{array}{cc}
 1 & 0 \\
 1 & -1 \\
\end{array}
\right), \quad {\cal S}_{{\rm TI/DS}} = \left(
\begin{array}{cccc}
 1 & 0 & 0 & 0 \\
 0 & 2 & 0 & 0 \\
\end{array}
\right), \cr
{\cal V}_{{\rm TI/DS}} = & \left(
\begin{array}{cccc}
 0 & 0 & 1 & 0 \\
 0 & 1 & 1 & -1 \\
 1 & -2 & -2 & 1 \\
 0 & 0 & 0 & 1 \\
\end{array}
\right).
\end{align}
We find $|\det(v^R_{{\rm TI/DS}})| = 1$ and $|\det(v^L_{{\rm TI/DS}})| = 2$. The effective K-matrix is
\begin{align}
K^{\rm eff}=\left(\begin{array}{cc}
4& -2\\
-2& 0\\
\end{array}\right).
\end{align}
Thus, the entanglement entropy is,
\begin{align}
S_e({\cal M}_{{\rm TI/DS}}^{R/L}) = \alpha \ell - {1 \over 2} \log(4).
\end{align}
Again, this result is symmetric under tracing out the doubled-semion or topological insulator edge degrees of freedom.

\section{Flux Insertion}

In Sec. \ref{parttwo}, we saw examples for which entanglement cuts gave identical entanglement entropies, even while the edge modes appearing at the entanglement cuts were in very different phases.
This is not surprising as the topological entanglement entropy measures the modular S-matrix of the topological phase and the different edge phases we considered have identical S-matrices.
Unfortunately, the identification of the distinct edge phases through the entanglement spectrum is also difficult. We also mentioned in that section that flux insertion was a possible tool for distinguishing the phases. We will discuss that in more detail here. 

In this section, we describe how flux insertion through the cylinder can help differentiate the possible edge phases occurring at the entanglement cut.
When the topological phase is put on a cylinder, flux insertion through the hole of the cylinder can generally be used to access different ground state sectors corresponding to the different bulk quasiparticles.
Given a particular flux sector, the entanglement spectrum reflects the spin of the operator (and the topological spin of the bulk quasiparticle created by the operator) in the edge theory associated to the bulk (quasiparticle) sector.

Extracting the actual value of the spin again appears difficult at generic points in moduli space for arbitrary states.
However, there is a coarser way in which distinct edge phases may be differentiated.
Recall that the edge phases we considered were either fermionic or bosonic, as the fermionic phase admit operators of half-integer spin while the boson phase only contain operators with integer spin.
It is known that $\pi$-flux has an important effect on the fermionic phase while it has no effect on the bosonic phase: the boundary conditions of the fields in the fermionic phase are modified by the presence of the $\pi$-flux, while the fields comprising the bosonic phase are unaffected.
The modified boundary conditions result in a shift of the ``ground state energy" in the entanglement spectrum of the fermionic phase from which the two phases may be distinguished.

\subsection{Generalities}

We thread flux $\Phi$ through the cylinder by turning on a constant vector potential $A_x = \Phi/\ell$ so that:
\begin{align}
\Phi = \int_0^\ell dx A_x.
\end{align}
We are interested in the effect of the threaded flux on the edge modes occurring at an entanglement cut. 

First, consider free fermions living on the two sides of the entanglement cut.
We may gauge away a constant vector potential at the cost of modifying the boundary conditions of the fermions:
\begin{align}
\label{bcs}
A_\mu \rightarrow A_\mu + \partial_\mu \lambda, \quad \psi \rightarrow e^{- i \lambda} \psi,
\end{align}
where $\lambda = (\Phi x)/\ell$.
Only when we thread $2\pi$ flux does the system remain invariant.
$\pi$-flux interchanges periodic and anti-periodic boundary conditions on the fermions.
Bosonizing the free fermion, $(\psi^{R/L})^\dagger \sim e^{\pm i \varphi^{R/L}}$, we see that anti-periodic boundary conditions correspond to the boundary conditions:
\begin{align}
\label{pifluxfermion}
\varphi^{R/L}(\ell) = \varphi^{R/L}(0) + 2 \pi \Big(P^{R/L} + {1 \over 2} \Big),
\end{align}
for $P^{R/L} \in \mathbb{Z}$.
Thus, $\pi$-flux merely shifts of the origin of the underlying lattice so that the shortest lattice vector has length-squared equal to $1/4$.

The generalization to a single-channel non-chiral Luttinger liquid at level $k$ living at the entanglement cut is immediate.
The electron operator takes the form, $(\psi^{R/L})^\dagger \sim e^{\pm i k \varphi^{R/L}}$.
If flux $\Phi$ is threaded, $\varphi^{R/L}$ obeys the boundary condition:
\begin{align}
\varphi^{R/L}(\ell) = \varphi(0) + 2 \pi \Big(P^{R/L} + {\Phi \over 2 \pi k}\Big),
\end{align}
 for $P^{R/L} \in \mathbb{Z}$.
Thus, there are $k$ different sectors, each corresponding to a value of flux, $\Phi = 2 \pi m$, with $m = 0, \dots, k-1$ which are accessed through $2\pi$-flux insertion.
However, for our needs, we require only $\pi$-flux insertion.

\subsection{Flux Insertion: Symmetric Interfaces of Integer States}

We now show how the entanglement spectrum of the symmetric-$\mathbb{I}_8$ interface differs from that of the symmetric-${\rm E_8}$ interface when $\pi$-flux threads the cylinder.

First, consider the effect of $\pi$-flux on the symmetric-$\mathbb{I}_8$ interface: fermions with periodic boundary conditions are transformed into fermions with anti-periodic boundary conditions.
In bosonized language, where $(\psi^{R/L}_I)^\dagger = \exp(\pm i \varphi^{R/L}_I)$ for $I = 1, \ldots, 8$, the bosons satisfy the boundary conditions:
\begin{align}
\label{ffbcs}
\varphi_I^{R/L}(\ell) = \varphi^{R/L}_I(0) + 2\pi \Big(P^{R/L}_I + {1 \over 2}\Big),
\end{align}
for $P_I^{R/L} \in \mathbb{Z}$.

The matrix $W_{(1,8)} \in {\rm GL}(9, \mathbb{Z})$, relates the $\mathbb{I}_8 \oplus (-\mathbb{I}_1)$ lattice to the $E_8 \oplus (-\mathbb{I}_1)$ lattice via Eqn.~(\ref{integeruniqueness})
and determines the boundary conditions on the edge modes of the ${\rm E_8}$ phase:
\begin{align}
\tilde{\varphi}^R_I(\ell) = \tilde{\varphi}^R_I(0) + 2 \pi (\tilde{P}^{R}_I + \tilde{S}^R_I),
\end{align}
where $\tilde{P}^R_I \in \mathbb{Z}$ and $\tilde{S}^R_I = \frac{1}{2} \sum_J \left( W_{(1,8)}\right)_{IJ} = \begin{pmatrix}
 -7 & -12 & -9 & -6 & -4 & -2 & -8 & -4
\end{pmatrix}$.
Since $\tilde{S}^R_I \in \mathbb{Z}$, it can be absorbed by the arbitrary integer $\tilde{P}^R_I$.
Thus, unlike the fields, $\varphi^R_I$, the fields $\tilde{\varphi}^R_I$ are unaffected by the $\pi$-flux.
This makes sense because all the vertex operators constructed from the $\tilde{\varphi}^R_I$ are bosonic and hence should not acquire a phase upon encircling $\pi$-flux.

We would like to compute the reduced density matrix for the symmetric-$\mathbb{I}_8$ interface when $\pi$-flux threads the cylinder following Sec. \ref{sec:entspec}.
We assume a tunneling interaction of the form $\cos(\varphi^R_I + \varphi^L_I)$ across the entanglement cut.
The only novelty arising from the flux insertion is that the eigenvalues of the zero mode operators take the form $N^{R/L}_I + 1/2$.
We find the entanglement Hamiltonian:
\begin{align}
H^R_e(\pi) = \sum_{a=1}^8 {2 \over \ell \sqrt{\lambda_a}} \Big((N^R_a + {1 \over 2})^2 + 2 \sum_{n>0} (n \alpha^\dagger_{a,n} \alpha_{a,n} + {n \over 2}) \Big).
\end{align}
Crucially, there is a finite difference in the ground state energy of the entanglement Hamiltonian with and without $\pi$-flux present:
\begin{align}
E_0(\pi) - E_0(0) = {1 \over 2 \ell} \sum_{a=1}^8 {1 \over \sqrt{\lambda_a}} \geq 0.
\label{fluxenergydifference}
\end{align}
This is a finite-size effect which vanishes as $\ell \rightarrow \infty$.
The constant sub-leading term in the entanglement entropy is unaffected by this shift.
We remark that this does not necessarily hold true for non-Abelian states, however.

Since the periodicity conditions on the $\tilde{\varphi}^{R/L}_I$ fields are unchanged by the $\pi$-flux insertion, the entanglement Hamiltonian of the ${\rm E}_8$ edge is invariant, and there is no change in energy analogous to Eqn.~(\ref{fluxenergydifference}). Thus, whether or not the energy difference in Eqn.~(\ref{fluxenergydifference}) is finite gives us a (coarse) prescription to distinguish the two symmetric interfaces possible at the entanglement cut.

\subsection{Flux Insertion: Symmetric Interfaces of Fractional States}

We now show how $\pi$-flux insertion can be used to distinguish the symmetric-$\begin{pmatrix} 3 & 0 \cr 0 & 5\end{pmatrix}$ and symmetric-$\begin{pmatrix} 2 & 1 \cr 1 & 8\end{pmatrix}$ interfaces.
Again, our strategy is to observe whether or not there is a finite energy shift in the ground state energy of the entanglement Hamiltonian under $\pi$-flux insertion.

First, consider the effect of $\pi$-flux on the symmetric-$\begin{pmatrix} 3 & 0 \cr 0 & 5\end{pmatrix}$ interface.
The fundamental fermions with periodic boundary conditions are transformed into fermions with anti-periodic boundary conditions.
In bosonized language, $(\psi^{R/L}_I)^{\dagger} = \exp(\pm i k^{R/L}_I \varphi^{R/L}_I)$ for $k^{R/L}_1 =  3$ and $k^{R/L}_2 = 5.$ Hence, we may take the bosons to satisfy the ``shifted" boundary conditions:
\begin{align}
\label{fracbcs}
\varphi_I^{R/L}(\ell) = \varphi^{R/L}_I(0) + 2\pi \Big(P^{R/L}_I + {1 \over 2}\Big),
\end{align}
for $P_I^{R/L} \in \mathbb{Z}$.

The matrix $W^{({8 \over 15})} \in {\rm GL}(4, \mathbb{Z})$ in Eqn. (\ref{fracw}) allows us to determine the boundary conditions induced on the fields, $\tilde{\varphi}^{R/L}_I$ describing the symmetric-$\begin{pmatrix} 2 & 1 \cr 1 & 8\end{pmatrix}$ interface:
\begin{align}
\tilde{\varphi}^{R/L}_I(\ell) = \tilde{\varphi}^{R/L}_I(0) + 2 \pi (\tilde{P}^{R/L}_I + \tilde{S}^{R/L}_I),
\end{align}
where $\tilde{P}^{R/L}_I \in \mathbb{Z}$ and the integer vector $\tilde{S}^{R/L}_I = \begin{pmatrix}
 0 & -2
\end{pmatrix}$.
We may absorb the integer shift $\tilde{S}^{R/L}_I$ into the arbitrary integers $\tilde{N}^{R/L}_I$.
Thus, the fields $\tilde{\varphi}^{R/L}_I$ are unaffected by the $\pi$-flux.

The calculation of the reduced density matrix for the symmetric-$\begin{pmatrix} 3 & 0 \cr 0 & 5\end{pmatrix}$ interface when $\pi$-flux threads the cylinder follows the approach described in Sec. \ref{sec:entspec}.
We assume a tunneling interaction of the form, $\cos(k^R_I \varphi^R_I + k^L_I \varphi^L_I)$, across the entanglement cut.
Again it is only the eigenvalues of the zero mode operators that shift to $N^{R/L}_I + 1/2$.
For this case as well there is a nonzero difference in energy of the ground state of the entanglement Hamiltonian with and without the $\pi$-flux present:
\begin{align}
E_0(\pi) - E_0(0) = {1 \over 2 \ell} \sum_{a=1}^2 {1 \over \sqrt{\lambda_a}} \geq 0.
\end{align}
The constant sub-leading term in the entanglement entropy is unaffected by this shift. 

Because the $\tilde{\varphi}^{R/L}_I$ fields are invariant under the $\pi$-flux insertion, the symmetric-$\begin{pmatrix} 2 & 1\\1 & 8\end{pmatrix}$ edge sees no such shift.

\section{Summary}

In this paper, we have studied how distinct edge phases of a given Hall state manifest themselves in the entanglement spectrum and entanglement entropy.
Surprisingly, we have found a universal constant correction to the topological entanglement entropy that is reflective of the distinct ways in which the edge modes appearing at an entanglement cut can be gapped.
In addition, we have observed how the distinct edge phases affect the entire entanglement spectrum.

There are a number of directions for future work.
These range from straightforward extensions to more speculative possibilities:

\noindent
$\bullet$ We have concentrated on fully chiral Abelian topological states of fermions in 2+1D. 
We expect our results and methods to readily generalize to bosonic states, non-chiral symmetry protected states, and non-Abelian states, as well as topologically ordered states in other dimensions.

\noindent
$\bullet$ Our states were placed on a cylinder and we studied an entanglement cut running around the cylinder. 
It would be of interest to study possible corrections in other geometries, similar to Refs. [\onlinecite{fradkintopologiesent, groverbraiding}], and to understand whether the corrections we find survive the prescriptions in Refs. [\onlinecite{LevinWenentropy, KitaevPreskillentropy}] that isolate the constant sub-leading term in the entanglement entropy. 

\noindent
$\bullet$
We found a dependence of the entanglement spectrum on the actual coupling constants parameterizing the interactions across the entanglement cut.
It would be of interest to better understand how universal information in the entanglement spectrum might be extracted.

\noindent
$\bullet$ We have studied the situation in which an independent set of sewing perturbations were present at the entanglement cut.
It would be interesting to understand how the entanglement spectrum behaves when different sets of sewing perturbations ``compete" with one another. 
On a related note, it is of interest to understand if the different gapped phases occurring at an entanglement cut for, say, two independent sets of sewing perturbations were connected to one another without a closing of the gap.

\noindent
$\bullet$
A confirmation of our results through numerical experiments using model wave functions would be of great interest. Additionally, it would be illustrative to find a prescription in which our correction could be studied in the Chern-Simons formulation of the topological phase. We expect that this is an interesting, but non-trivial problem.

\noindent
$\bullet$ 
The tunneling interactions that resulted in constant corrections to the entanglement entropy were found by requiring the matrix $({\cal M}^R)^{-1} {\cal M}^L$ -- formed by the integer vectors defining the interaction -- to not lie in ${\rm GL}(N, \mathbb{Z})$.
A more systematic understanding for such occurrences is desirable.
In particular, does a given state admit only a finite number of possible constant corrections to its entropy?
While it does not appear to be possible to obtain arbitrary corrections, without any symmetry constraints, it appears that there is no restriction on the number of different corrections, i.e., there are a countable number of different corrections to the entropy.

\noindent
$\bullet$
In Sec. \ref{fournugap}, we studied constant corrections to the entanglement entropy in a $\nu=4$ example resulting from novel tunneling interactions across the entanglement cut.
We remarked that it is possible to create a 2D state using such interactions without changing the free Fermi liquid edge structure; however, such a state only allows electrons to tunnel in pairs so there is a type of ``confinement" of single-particle excitations.
A better understanding of the resulting 2D state and analogous constructions at other fillings would be of interest.

\noindent
$\bullet$
Chiral edge phase transitions were studied in Refs. [\onlinecite{PlamadealaE8, generalstableequivalence}].
Given an edge phase defined by a particular $K$-matrix, these transitions proceeded by the interaction of these modes with the edge modes arising from a strip of $\nu=1$ fluid.\cite{chamonwenreconstruction}
Our work here suggests a generalization of this procedure: we imagine attaching a (narrow, but finite) `strip' of a different (possibly fractional) fluid and allowing the different edge excitations to interact with one another.
The one requirement is that these two fluids be able to share a gapped boundary.

\acknowledgments

It is a pleasure to thank L. Balents, A. Chandran, M. Cheng, D. Clarke, M. Fisher, E. Fradkin, M. Freedman, T. Grover, C. Nayak, M. Stoudenmire, and Z. Wang for helpful discussions and encouragement.
J.C. acknowledges the support of the National Science Foundation Graduate Research Fellowship under Grant No. DGE1144085.
T.L.H acknowledges the support of National Science Foundation CAREER Grant No. DMR-1351895 and the ICMT at UIUC.
M.M. is partially supported by a grant from the Templeton Foundation on Emergence and Entanglement .
M.M. acknowledges the support and hospitality of Microsoft Station Q, where this work was initiated, and the Kavli Institute for Theoretical Physics at the University of California Santa Barbara, where this work was completed.
This research was supported in part by the National Science Foundation under Grant No. NSF PHY11-25915.

\appendix

\section{Klein Factors}
\label{kleinappendix}

The wire construction provides us with a set of chiral fermion creation operators, $\psi^\dagger_{I,j, a}$, where $I$ is a channel index for wire $j,$ and $a=R$ or $L$.
We introduce a collective index $i$ to replace both the channel $I$ and wire $j$ indices.
Therefore, we denote $\psi^\dagger_{I,j, a}$ by $\psi^\dagger_{i, a}$ in the remainder of this Appendix.
The $\psi^\dagger_{i,a}$ are bosonized according to
\begin{equation} \psi^\dagger_{i,R/L} \sim \gamma_{i,R/L} e^{\pm i\phi_{I}^{R/L}}. \end{equation}

The $\gamma_{i,a}$ operators are Klein factors: they satisfy anti-commutation relations,
\begin{equation} \lbrace \gamma_{i, a}, \gamma_{j, b} \rbrace = 2 \delta_{i j} \delta_{ab},\end{equation}
and ensure that the fermion operators with different $i,a$ obey the correct commutation relations. The commutation relations for fermion operators with the same $i,a$ are ensured by the commutation relations of the $\phi_{i}^{a}$ in Eqn. (\ref{appeq:commutator}). 

Here we show that given a set of fermion-parity-conserving operators $\mathcal{O}_{\vec{m}}$ that will gap (some of) the degrees of freedom in the wires, the products of Klein factors that enter in each $\mathcal{O}_{\vec{m}}$ will mutually commute, provided that the bosonic parts of each $\mathcal{O}_{\vec{m}}$ also commute. 

To be concrete, the operators are products of fermion annihilation and creation operators, which we express schematically as
\begin{equation} \mathcal{O}_{\vec{m}} = \prod_{I,a} \psi_{i,a}^{m_{i,a}}, \end{equation}
where $\vec{m}$ is a vector of integers; when $m_{i,a}$ is negative $\psi_{i,a}^\dagger$ instead of $\psi_{i,a}$ should enter the product. Thus, each operator can be decomposed as 
\begin{equation} \mathcal{O}_{\vec{m}} = \Gamma_{\vec{m}} e^{i\vec{m}\cdot \vec{\phi}},\end{equation}
where the notation $\vec{m}\cdot \vec{\phi}$ is shorthand for $\sum_{i,a} m_{i,a}\phi_{i}^a$ and 
\begin{equation} \Gamma_{\vec{m}} = \prod_{i,a} \gamma_{i,a}. \end{equation}
We show that if the operators are chosen such that the set of $e^{i \vec{m}\cdot \vec{\phi}}$ mutually commute, then the $\Gamma_{\vec{m}}$ will also mutually commute. Thus, one can work in a basis of eigenvectors of all the $\Gamma_{\vec{m}}$ and the Klein factors effectively become classical variables.

\subsection{Proof that the Klein factors commute}
The bosonic operators obey the equal-time commutation relation
\begin{equation} \left[ \phi_{i}^a(x) , \phi_{j}^b(x') \right] = - a\pi i \delta_{ab}\delta_{ij}\text{sgn}(x-x'), \label{appeq:commutator} \end{equation}
with $a = R/L = \pm 1$ from which it follows,
\begin{align} 
e^{i\vec{m}\cdot \vec{\phi}(x)} e^{i\vec{m'}\cdot \vec{\phi}(x')} 
&= e^{i\vec{m'}\cdot \vec{\phi}(x')} e^{i\vec{m}\cdot \vec{\phi}(x)} e^{[i\vec{m}\cdot \vec{\phi}(x), i\vec{m'}\cdot \vec{\phi}(x')]} \nonumber\\
&=e^{i\vec{m'}\cdot \vec{\phi}(x')} e^{i\vec{m}\cdot \vec{\phi}(x)} e^{- i\pi \sum_{i,a} m_{i,a}m'_{i,a}}.
\end{align}
Hence, the condition that the bosonic operators commute is exactly the condition 
\begin{equation} \sum_{i,a} m_{i,a}m'_{i,a}=0 \mod 2, \label{eq:bosoncondition} \end{equation}
for all $\vec{m},\vec{m}'$.
Now consider the Klein factors. Each $\Gamma_{\vec{m}}$ is a product of an even number of $\gamma$-operators, since the $\mathcal{O}_{\vec{m}}$ conserve fermion parity. Furthermore, since $\gamma_{i,a}^2 = 1$, each $\Gamma_{\vec{m}}$ can be written so that no $\gamma$-operator appears more than once in the product. Thus, the commutation relations between the $\Gamma_{\vec{m}}$ are described by Sec.~\ref{sec:gammacommutator}, which yields 
\begin{equation} \left[ \Gamma_{\vec{m}}, \Gamma_{\vec{m'}} \right] =0 \iff \sum_{i,a} m_{i,a} m'_{i,a}=0 \mod 2. \end{equation}
From Eqn. (\ref{eq:bosoncondition}) we see that 
\begin{equation} \left[ \Gamma_{\vec{m}}, \Gamma_{\vec{m'}} \right] =0 \iff \left[ e^{i\vec{m}\cdot \vec{\phi}(x)}, e^{i\vec{m'}\cdot \vec{\phi}(x')}\right] =0. \end{equation}
Thus, whenever the bosonic parts of the operators $\mathcal{O}_{\vec{m}}$ commute, the contributions from the Klein factors do as well, and we can omit them from our consideration.

\subsection{Commutation relations for products of $\gamma$ matrices}
\label{sec:gammacommutator}
Given the operators $\Gamma^a = \gamma_{a_1}\gamma_{a_2}...\gamma_{a_{2n}}$ and $\Gamma^b = \gamma_{b_1}\gamma_{b_2}...\gamma_{b_{2n'}}$, where all $a_i$ are distinct and all $b_i$ are distinct, and collectively refer to the $i,a$-labels, we now show that $[ \Gamma^a, \Gamma^b ] = 0$ exactly when the set $I = \lbrace a_i\rbrace \cap \lbrace b_i \rbrace$ contains an even number of elements. First, label the elements of $I$, $\gamma_{i_1}, ... , \gamma_{i_N}$. Then define
\begin{align}
\tilde{\Gamma}^a &= \prod_{j=1}^N \gamma_{i_j}\prod_{k=1}^{2n-N}\gamma_{\tilde{a}_k},\nonumber\\
\tilde{\Gamma}^b &= \prod_{k=1}^{2n'-N}\gamma_{\tilde{b}_k}\prod_{j=N}^1 \gamma_{i_j},
\end{align}
where $\lbrace \tilde{a}_k \rbrace \cup I = \lbrace a_i \rbrace $ and $\lbrace \tilde{b}_k \rbrace \cup I = \lbrace b_i \rbrace $. $\tilde{\Gamma}^{a,b}$ differ from $\Gamma^{a,b}$ by at most a sign. We have manufactured the slightly strange-looking ordering of the $\gamma$ matrices so that 
\begin{equation} \tilde{\Gamma}^b \tilde{\Gamma}^a = \prod_{k=1}^{2n'-N}\gamma_{\tilde{b}_k}\prod_{j=1}^{2n-N}\gamma_{\tilde{a}_j} \label{eq:productba}.\end{equation}
Now since $\tilde{a}_k \neq i_j$ and $\tilde{b}_k \neq i_j$ for any $k$ and $j$, we can rewrite 
\begin{align}
\tilde{\Gamma}^a &= (-1)^{N(2n-N)}\prod_{k=1}^{2n-N}\gamma_{\tilde{a}_k}\prod_{j=1}^N \gamma_{i_j},\nonumber\\
\tilde{\Gamma}^b &= (-1)^{N(2n'-N)}\prod_{j=N}^1 \gamma_{i_j}\prod_{k=1}^{2n'-N}\gamma_{\tilde{b}_k}.
\end{align}
Thus,
\begin{equation} \tilde{\Gamma}^a\tilde{\Gamma}^b = \prod_{k=1}^{2n-N}\gamma_{\tilde{a}_k}\prod_{j=1}^{2n'-N}\gamma_{\tilde{b}_j}, \end{equation}
where we have used $(-1)^{N(2n-N)}(-1)^{N(2n'-N)}  = 1$. Since $\tilde{a}_k \neq \tilde{b}_j$ for any $k$ and $j$ (or else such a pair would have been in $I$), we can rewrite 
\begin{equation} \tilde{\Gamma}^a\tilde{\Gamma}^b = (-1)^{(2n-N)(2n'-N)}\prod_{j=1}^{2n'-N}\gamma_{\tilde{b}_j} \prod_{k=1}^{2n-N}\gamma_{\tilde{a}_k}, \end{equation}
when $N$ is even, the right-hand-side is identical to that of Eqn. (\ref{eq:productba}). Hence, when $N$ is even, $\tilde{\Gamma}^a$ and $\tilde{\Gamma}^b$ commute, which implies that $\Gamma^a$ and $\Gamma^b$ also commute. 

Notice that it is not necessary for $\Gamma^a$ and $\Gamma^b$ to both be a product of an even number of $\gamma$ operators, as long as one of them is such a product. On the other hand, if both $\Gamma^a$ and $\Gamma^b$ are a product of an odd number of $\gamma$ operators, then they will commute exactly when $N$ is odd.

\section{Relevance of gapping vectors}
\label{sec:gappingvectorsrelevant}
Here we show that any set of linearly independent gapping terms satisfying the null condition of Eqn.~(\ref{newnull}) can be made arbitrarily relevant by appropriately choosing the interactions between the right- and left- moving fields.

The action at the entanglement cut takes the form
\begin{equation} S= \frac{1}{4\pi} \int dt dx \left( \eta_{ij} \partial_t X_i \partial_x X_j - V_{ij} \partial_x X_i \partial_x X_j \right)\label{generalnonchiralaction} \end{equation}
where  $X_{i\leq N} = X^R_i$ and $X_{i>N}=X^L_{i-N}$, $X^{R/L}$ were defined in Eqn.~(\ref{Xdefinition}), and
\begin{equation} \eta = \begin{pmatrix} - \mathbb{I}_N & 0 \\ 0 & \mathbb{I}_N \end{pmatrix}.\end{equation} 
Eqn.~(\ref{generalnonchiralaction}) is similar to Eqn.~(\ref{luttaction}) but without any assumptions on the form of interactions; instead, interactions are lumped into the $2N\times 2N$ matrix $V$. 
Tunneling terms take the form
\begin{equation} S_{\rm tunneling} = \frac{1}{4\pi} \int dt dx \left[ g_\beta \cos\left( R_{\beta i} X_i + L_{\beta i}X_{i+N}\right)\right],
\end{equation}
where $R$ and $L$ are defined in Eqn.~(\ref{defineRL}) and satisfy $RR^T = LL^T$, from the null condition~(\ref{newnull}).

We assume the ansatz $V = e^{-A}$, where
\begin{equation} A=\begin{pmatrix} 0 & \alpha R^{-1}L \\ \alpha (R^{-1}L)^T  & 0 \end{pmatrix}, \end{equation}
and $\alpha$ is a tuning parameter. The linear independence of gapping vectors assures us that $R$ is invertible, as shown in Appendix~\ref{sec:linearindependence}. Then $A^{2n} = \alpha^{2n}\mathbb{I}$, which gives
\begin{align} V^{-1} = e^A = \mathbb{I} \sum_{k=0}^\infty  \frac{\alpha^{2k}}{(2k)!} + A \sum_{k=0}^\infty  \frac{\alpha^{2k}}{(2k+1)!}.
\end{align}
This yields the scaling dimension, $\Delta_\beta$, of $\cos\left( R_{\beta i} X_i + L_{\beta i}X_{i+N}\right)$:
\begin{align}
\Delta_\beta &= \frac{1}{2}\left( \begin{pmatrix} R &L \end{pmatrix}V^{-1} \begin{pmatrix} R & L \end{pmatrix}^T \right)_{\beta\beta}
=e^\alpha \left( RR^T \right)_{\beta\beta}.
\end{align}
By tuning $\alpha$ to be large and negative, we can manufacture a set of interactions, contained in $V$, which make the scaling dimensions, $\Delta_\beta$, arbitrarily relevant.

\section{Linear independence of rows of ${\cal M}^{R/L}$}
\label{sec:linearindependence}

To completely gap out the entanglement edge, it is necessary that the $2N$-component vectors formed by the rows of $\begin{pmatrix} {\cal M}^R & {\cal M}^L \end{pmatrix}$ are linearly independent. However, this does not necessarily imply that the rows of ${\cal M}^R$ are linearly independent. Here we show that Eqn. (\ref{newnull}) requires this to be the case. First, notice that Eqn. (\ref{newnull}) implies that the inner product of any linear combination of the rows of $\begin{pmatrix} {\cal M}^R & {\cal M}^L \end{pmatrix}$ with another linear combination is zero:
\begin{align}& \left( a_\alpha {\cal M}^R_{\alpha I}\right) (K^R)^{-1}_{IJ} \left( b_\beta {\cal M}^R_{\beta J} \right) 
-  \left( a_\alpha {\cal M}^L_{\alpha I}\right) (K^L)^{-1}_{IJ} \left( b_\beta {\cal M}^L_{\beta J} \right)  \nonumber\\
&= a_\alpha b_\beta \left( {\cal M}^R_{\alpha I}(K^R)^{-1}_{IJ} {\cal M}^R_{\beta J} - {\cal M}^L_{\alpha I}(K^L)^{-1}_{IJ} {\cal M}^L_{\beta J} \right) = 0.
\label{minnerproduct}
\end{align}
Now we prove the rows of ${\cal M}^R$ are linearly independent by contradiction: suppose that for some $\beta_0$, ${\cal M}^R_{\beta_0 I} = \sum_{\beta \neq \beta_0} b_\beta {\cal M}^R_{\beta I}$, for some coefficients $b_\beta$. Then by Eqn. (\ref{minnerproduct}),

\begin{align} &\left( {\cal M}^R_{\beta_0 I} - \sum_{\alpha \neq \beta_0} b_\alpha {\cal M}^R_{\alpha I}\right)  (K^R)^{-1}_{IJ}  \left( {\cal M}^R_{\beta_0 J} - \sum_{\beta \neq \beta_0} b_\beta {\cal M}^R_{\beta J}\right)  \nonumber\\
&=  \left( {\cal M}^L_{\beta_0 I} - \sum_{\alpha \neq \beta_0} b_\alpha {\cal M}^L_{\alpha I}\right)  (K^L)^{-1}_{IJ}  \left( {\cal M}^L_{\beta_0 J} - \sum_{\beta \neq \beta_0} b_\beta {\cal M}^L_{\beta J}\right) \nonumber\\
& = 0.
\end{align}
The chirality of $K^L$ requires ${\cal M}^L_{\beta_0 J} = \sum_{\beta \neq \beta_0} b_\beta {\cal M}^L_{\beta J}$. But this contradicts the fact that the $2N$-component rows of $\begin{pmatrix} {\cal M}^R & {\cal M}^L \end{pmatrix}$ are linearly independent. Hence, we conclude that not only are the $2N$-component vectors linearly independent, but the $N$-component rows of ${\cal M}^{R,L}$ are as well.

Note that this implies that the matrix $g_\beta (f^{R/L}_I)_i{\cal M}^{R/L}_{\beta I}{\cal M}^{R/L}_{\beta J} (f^{R/L}_J)_j$ is positive-definite: for any $N$-component vector $z$, define $y_\beta = \sqrt{g_\beta} {\cal M}^{R/L}_{\beta J} (f^{R/L}_J)_i z_i$ where there is no sum implied on $\beta$. Since the ${\cal M}^{R/L}_{\beta J} (f^{R/L}_J)_i$ are non-singular, $y$ is not equal to zero. Hence, for any $z$, $z_i g_\beta (f^{R/L}_I)_i{\cal M}^{R/L}_{\beta I}{\cal M}^{R/L}_{\beta J} (f^{R/L}_J)_j z_j= y_\beta y_\beta >0$.

\section{Field Redefinitions}
\label{fieldredef}

Here we show that starting with the action in Eqn.~(\ref{luttaction}) and the quadratic mass terms of Eqn.~(\ref{cosapprox}) rewritten in terms of the fields $X_i^{R/L}$, namely,
\begin{equation} 
-\frac{g_\beta}{2}\left( {\cal M}^a_{\beta I}(f_I^a)_i X_i^a\right)\left( {\cal M}^b_{\beta J}(f_J^b)_jX_j^b\right),
\label{mass1} \end{equation}
for some choice of dual basis, $(f_I^{R/L})_i$,
that we can do an orthogonal transformation on the fields such that the quadratic term takes the diagonal form of Eqn.~(\ref{tunndiag}), while Eqn.~(\ref{luttaction}) remains invariant. 

We first introduce the matrices:
\begin{align}
R_{\beta i} = {\cal M}^R_{\beta I}(f_I^R)_i, \quad L_{\beta i} ={\cal M}^L_{\beta I}(f_I^L)_i,
\label{defineRL}
\end{align}
which we (``QR'') decompose as,
\begin{align}
\label{QRdecomp}
R_{\beta i} = T^R_{\beta a} Q^R_{a i}, \quad L_{\beta i} = T^L_{\beta a} Q^L_{a i},
\end{align}
where $Q^{R/L}$ are orthogonal and $T^{R/L}$ are lower-triangular matrices. Since $R$ and $L$ are non-singular (proven in Appendix~\ref{sec:linearindependence}), this decomposition is unique if we choose the diagonal elements of $T^{R/L}$ to be positive.
From Eqn. (\ref{newnull}), we obtain
$T^R_{\beta a} T^R_{\gamma a} = T^L_{\beta a} T^L_{\gamma a}.$
The uniqueness of the Cholesky decomposition then implies $T^R = T^L \equiv T$.

We now independently rotate the right/left-moving fields to define the fields $\tilde{X}_a^{R/L}$:
\begin{align}
\label{newredef}
X^{R/L}_i = Q^{R/L}_{a i} \tilde{X}^{R/L}_{a}.
\end{align}
Eqn.~(\ref{luttaction}) is invariant under these orthogonal rotations and the cosine terms of Eqn.~(\ref{mass1}) are rewritten as
\begin{equation}  - {g_\beta \over 2} \Big(T_{\beta a} (\tilde{X}^R_a + \tilde{X}^L_a) \Big) \Big(T_{\beta b} (\tilde{X}^R_b + \tilde{X}^L_b) \Big). 
\label{mass2} \end{equation}

Finally, we can diagonalize the mass matrix, ${\cal M}_{a b} = \sum_\beta g_\beta T_{\beta a} T_{\beta b}$ by writing,
\begin{align}
\tilde{X}^{R/L}_a = O_{a b} {\cal X}^{R/L}_b,
\end{align}
where $O_{ab} \in SO(N)$ satisfies:
\begin{align}
{1 \over 2} O_{ac} {\cal M}_{a b} O_{b d} = \lambda_c \delta_{cd}. 
\end{align}
Because $g_\beta T_{\beta a} T_{\beta b}$ is positive, $\lambda_a > 0$.
Thus, the quadratic terms in Eqn.~(\ref{mass2}) now take the diagonal form of Eqn.~(\ref{tunndiag}) with $X_i^{R/L}  \rightarrow {\cal X}_i^{R/L}$.

Returning to the original variables, we find $\varphi_I^{R/L} = (f_I^{R/L})_i Q^{R/L}_{ai} O_{ab}{\cal X}_b^{R/L}$. Thus, the field redefinitions that we have implemented amount to an orthogonal rotation of the dual basis, $(f_I^{R/L})_i \rightarrow ({\cal F}_I^{R/L})_i = (f_I^{R/L})_a Q^{R/L}_{ba} O_{bi}$, or, equivalently, $(e_I^{R/L})_i \rightarrow ({\mathcal E}_I)_i^{R/L}= (e_I^{R/L})_a Q^{R/L}_{ba} O_{bi}$. This transformation is in $SO(N)$ if the matrices $Q^{R/L}$ are in $SO(N)$; if they are not in $SO(N)$, they can be made so by multiplying one row of ${\cal M}^{R/L}$ by a factor of $-1$, which does not change the physics.

Thus, the $({\cal E}_I^{R/L})_i$ basis is the ``correct'' choice of basis. Using this basis, the restricted lattices defined in Eqn.~(\ref{definerestrictedlattice}) are identical: starting from the results of Sec.~\ref{symmetrybreaking},
we see that 
\begin{equation} {\cal M}^R v^R = {\cal M}^L v^L, \label{relateMv}\end{equation}
where we have dropped the indices to reduce clutter. 
Furthermore, by definition, ${\cal M}^R {\cal F}^R = {\cal M}^R f^R (Q^R)^T O = {\cal M}^R f^R (Q^R)^T Q^L (e^L)^T {\cal F}^L = {\cal M}^L {\cal F}^L$, 
where in the last equality we have utilized Eqns.~(\ref{defineRL}) and (\ref{QRdecomp}) and $T^R = T^L$. 
Combined with Eqn.~(\ref{relateMv}), this yields $(v^R)^{-1} {\cal F}^R = (v^L)^{-1}  {\cal F}^L$. Inverting this equation yields, 
$({\cal E}^R)^T v^R = ({\cal E}^L)^T v^L$. Thus, when the ``correct'' basis choice is used, the restricted lattices defined in Eqn.~(\ref{definerestrictedlattice}) are identical.

\section{Finite Size Corrections to the Entanglement Entropy}
\label{finitecorrections}

We have found,
\begin{equation} Z_e(T)=\frac{ \left( {\rm det}\left(-i\tau\Omega\right) \right)^{-1/2} \sum_{m_I\in \mathbb{Z}^N}e^{-i\pi\tau^{-1}m_I\Omega^{-1}_{IJ}m_j}}{\prod_i \frac{1}{\sqrt{-i\tau_i}}e^{-i\pi /12\tau_i}\prod_{n>0}\left(1-e^{-2\pi i n/\tau_i}\right) } \end{equation}
where the $T$ dependence is implicit in $\tau = \frac{2i}{\pi \ell T}, \tau_i = \frac{2i}{\pi \ell\sqrt{\lambda_i}T}$. Using Eqn. (\ref{entropygeneral}) we find:
\begin{widetext}
\begin{align} 
S_e &= \partial_T \left( T \left( -\frac{1}{2} \ln {\rm det}\left(-i\tau\Omega\right)  + \ln \left( \sum_{m_I\in \mathbb{Z}^N}e^{-i\pi\tau^{-1}m_I\Omega^{-1}_{IJ}m_j} \right) \right. \right. \nonumber \cr
&  \left.\left.  + \sum_i \left( {i\pi \tau_i \over 12} + \frac{1}{2} \ln\left( -i\tau_i\right)  - \sum_{n>0} \ln(1-e^{-2\pi i n/ \tau_i} ) \right) \right)\right) \Bigg|_{T=1} \nonumber\\
&= S_e(\ell \rightarrow \infty) + \partial_T \left( T \ln \left( \sum_{m_I\in \mathbb{Z}^N}e^{-\frac{1}{2} \pi^2 \ell T m_I\Omega^{-1}_{IJ}m_j} \right) \right. \left. - \sum_i \sum_{n>0} T\ln (1-e^{-\pi^2 n \ell\sqrt{\lambda_i}T}) \right) \Bigg|_{T=1} \nonumber\\
&=  S_e(\ell \rightarrow \infty) + S_{zero} + S_{osc}.
\end{align} 
\end{widetext}
where
\begin{align} S_{{\rm zero}} &=  \ln \sum_{m}e^{-\frac{1}{2} \pi^2 \ell  m\Omega^{-1}m} \cr
&  + \frac{\sum_{m} -\frac{1}{2}\pi^2 \ell m\Omega^{-1} m e^{-\frac{1}{2} \pi^2 \ell  m\Omega^{-1}m} }{\sum_{m}e^{-\frac{1}{2} \pi^2 \ell  m\Omega^{-1}m} } \nonumber\\
&\approx \sum_{m_0} \left(1 - \frac{1}{2}\pi^2 \ell m\Omega^{-1} m \right)e^{-\frac{1}{2}\pi^2 \ell m\Omega^{-1} m}
 \end{align}
and
\begin{align} S_{\rm osc} &=  - \sum_{i, n>0} \left( \ln (1-e^{-\pi^2 n \ell\sqrt{\lambda_i}})  + \frac{\pi^2 n \ell \sqrt{\lambda_i} e^{-\pi^2 n \ell\sqrt{\lambda_i} } }{1-e^{-\pi^2 n\ell \sqrt{\lambda_i}}}\right) \nonumber\\
&\approx  \sum_{i_0} \left( 1 - \pi^2 \ell\sqrt{\lambda_i} \right) e^{-\pi^2 \ell \sqrt{\lambda_i}},
\end{align}
where we have expanded in the large $\ell$ limit. The sum over $m_0$ is a sum over the (possibly multiple) $m$ which minimize $m\Omega^{-1} m$ and, similarly, the sum over $i_0$ is over the (possibly multiple) $i$ that have the smallest value of $\lambda_i$.


\section{$\mathbb{I}_8 \oplus (- \mathbb{I}_1) = E_8 + (- \mathbb{I}_1)$}
\label{sec:I8toE8}

In this appendix, we provide the bases, Cartan matrices, and basis transformation for the equivalent lattices, $\mathbb{I}_8 \oplus (- \mathbb{I}_1) = E_8 \oplus (- \mathbb{I}_1)$.
We also describe how the basis transformation relating these two lattices is found. 
These two lattices are equivalent because unimodular, signature $(8,1)$ lattices are unique up to $SO(8,1)$ transformations.

As a basis for the $E_8 \oplus (- \mathbb{I}_1)$ lattice, we choose:
\begin{align}
(e_I)_a & = (x_I)_a - (x_{I+1})_a,\ I = 1, ..., 6, \cr
(e_7)_a & = - (x_1)_a - (x_2)_a, \cr
(e_8)_a & = {1 \over 2} \Big((x_1)_a + \dots + (x_8)_a\Big), \cr
(e_9)_a & = (x_9)_a,
\end{align}
where $a = 1, ..., 9$ and $(x_I)_a$ is the unit vector with a ``1" in the I-th entry and zeros otherwise.
Using the inner product, $\eta_{ab} = {\rm diag}(1^8, -1)$, the above basis has the Cartan matrix, $(K_{E_8 \oplus (- \mathbb{I}_1)})_{IJ} = (e_I)_a \eta_{ab} (e_J)_b$,
\begin{align}
K_{E_8 \oplus (- \mathbb{I}_1)} =
\left(
\begin{array}{ccccccccc}
 2 & -1 & 0 & 0 & 0 & 0 & 0 & 0 & 0 \\
 -1 & 2 & -1 & 0 & 0 & 0 & -1 & 0 & 0 \\
 0 & -1 & 2 & -1 & 0 & 0 & 0 & 0 & 0 \\
 0 & 0 & -1 & 2 & -1 & 0 & 0 & 0 & 0 \\
 0 & 0 & 0 & -1 & 2 & -1 & 0 & 0 & 0 \\
 0 & 0 & 0 & 0 & -1 & 2 & 0 & 0 & 0 \\
 0 & -1 & 0 & 0 & 0 & 0 & 2 & -1 & 0 \\
 0 & 0 & 0 & 0 & 0 & 0 & -1 & 2 & 0 \\
 0 & 0 & 0 & 0 & 0 & 0 & 0 & 0 & -1 \\
\end{array}
\right).
\label{KE8}
\end{align}

As a basis for $\mathbb{I}_8 \oplus (- \mathbb{I}_1)$, we choose:
\begin{align}
(e'_I)_a = (x_I)_a.
\end{align}
The associated Cartan matrix is,
\begin{align}
K_{\mathbb{I}_8 \oplus (- \mathbb{I}_1)} =
\left(
\begin{array}{ccccccccc}
 1 & 0 & 0 & 0 & 0 & 0 & 0 & 0 & 0 \\
 0 & 1 & 0 & 0 & 0 & 0 & 0 & 0 & 0 \\
 0 & 0 & 1 & 0 & 0 & 0 & 0 & 0 & 0 \\
 0 & 0 & 0 & 1 & 0 & 0 & 0 & 0 & 0 \\
 0 & 0 & 0 & 0 & 1 & 0 & 0 & 0 & 0 \\
 0 & 0 & 0 & 0 & 0 & 1 & 0 & 0 & 0 \\
 0 & 0 & 0 & 0 & 0 & 0 & 1 & 0 & 0 \\
 0 & 0 & 0 & 0 & 0 & 0 & 0 & 1 & 0 \\
 0 & 0 & 0 & 0 & 0 & 0 & 0 & 0 & -1 \\
\end{array}
\right).
\end{align}

Because these two lattices are equivalent, there exists a matrix $W_{(8,1)} \in GL(9, \mathbb{Z})$ satisfying:
\begin{align}
(W^T_{(8,1)})_{IJ} (K_{E_8 \oplus (- \mathbb{I}_1)})_{JK} (W_{(8,1)})_{KL} = (K_{\mathbb{I}_8 \oplus (- \mathbb{I}_1)})_{IL}.
\end{align}
The explicit form of such a $W_{(8,1)}$ is as follows:
\begin{align}
\label{WI8toE8}
W_{(8,1)} = 
\left(
\begin{array}{ccccccccc}
 -3 & -3 & -3 & -3 & -3 & 3 & 3 & 3 & -8 \\
 -5 & -5 & -5 & -6 & -6 & 6 & 6 & 6 & -15 \\
 -4 & -4 & -4 & -4 & -5 & 5 & 5 & 5 & -12 \\
 -3 & -3 & -3 & -3 & -3 & 4 & 4 & 4 & -9 \\
 -2 & -2 & -2 & -2 & -2 & 2 & 3 & 3 & -6 \\
 -1 & -1 & -1 & -1 & -1 & 1 & 1 & 2 & -3 \\
 -3 & -3 & -4 & -4 & -4 & 4 & 4 & 4 & -10 \\
 -1 & -2 & -2 & -2 & -2 & 2 & 2 & 2 & -5 \\
 1 & 1 & 1 & 1 & 1 & -1 & -1 & -1 & 3 \\
\end{array}
\right).
\end{align}

We now explain how $W_{(8,1)}$ is obtained.
A Dynkin diagram geometrically expresses the content contained in the Cartan matrix.
Each basis element of a given lattice is represented by a dot or node.
A shaded node denotes a basis element with length-squared equal to $+2$ while an open node denotes a basis element of length-squared equal to $+1$.
A single line between two nodes signifies that the associated basis elements have an inner product equal to $-1$.
If no line is drawn between nodes, the associated basis elements are orthogonal.
Thus, an equivalent way of asking for the $W_{(8,1)}$ transformation is to seek basis transformations of $e_I$ and $e'_J$ such that the associated Dynkin diagrams are the same. 
The extended Dynkin diagram for $E_8$ is given in Fig. \ref{fig:dynkin}.
`Extended' refers to an extra node (compared with the $E_8$ Dynkin diagram) at the long-end of the diagram marked with an open circle.

\begin{figure}[h!]
  \centering
\includegraphics[width=\linewidth]{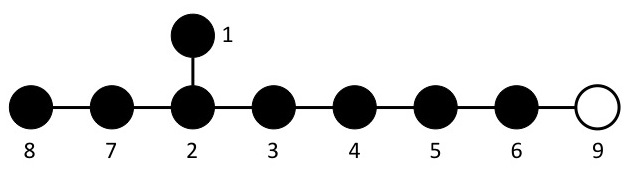}
\caption{Extended Dynkin diagram for $E_8$. Numbers indicate the particular basis vector to which the nearby circle refers.}\label{fig:dynkin}
\end{figure}

A basis is given by:
\begin{align}
(\tilde{e}_I)_a & = (x_I)_a - (x_{I+1})_a,\ I = 1, ..., 6, \cr
(\tilde{e}_7)_a & = - (x_1)_a - (x_2)_a, \cr
(\tilde{e}_8)_a & = {1 \over 2} ((x_1)_a + \dots + (x_8)_a), \cr
(\tilde{e}_9)_a & = (x_7)_a - (x_8)_a + (x_9)_a,
\end{align}
which has the Cartan matrix:
\begin{align}
K_{\tilde{E}_8} = \left(
\begin{array}{ccccccccc}
 2 & -1 & 0 & 0 & 0 & 0 & 0 & 0 & 0 \\
 -1 & 2 & -1 & 0 & 0 & 0 & -1 & 0 & 0 \\
 0 & -1 & 2 & -1 & 0 & 0 & 0 & 0 & 0 \\
 0 & 0 & -1 & 2 & -1 & 0 & 0 & 0 & 0 \\
 0 & 0 & 0 & -1 & 2 & -1 & 0 & 0 & 0 \\
 0 & 0 & 0 & 0 & -1 & 2 & 0 & 0 & -1 \\
 0 & -1 & 0 & 0 & 0 & 0 & 2 & -1 & 0 \\
 0 & 0 & 0 & 0 & 0 & 0 & -1 & 2 & 0 \\
 0 & 0 & 0 & 0 & 0 & -1 & 0 & 0 & 1 \\
\end{array}
\right).
\end{align}
The difference $(x_7)_a - (x_8)_a$ between $\tilde{e}_9$ and $e_9$ is an element of the $E_8$ lattice:
\begin{align}
\label{linear}
x_7 - x_8 & =  -3 e_1 - 6 e_2 - 5 e_3 - 4 e_4  \cr 
& - 3 e_5 - 2 e_6 - 4 e_7 - 2 e_8,
\end{align}
which implies the existence of the basis transformation:
\begin{align}
\tilde{e}_I = M_{IJ} e_J, 
\end{align}
with 
\begin{align}
M_{IJ} =
\left(
\begin{array}{ccccccccc}
 1 & 0 & 0 & 0 & 0 & 0 & 0 & 0 & 0 \\
 0 & 1 & 0 & 0 & 0 & 0 & 0 & 0 & 0 \\
 0 & 0 & 1 & 0 & 0 & 0 & 0 & 0 & 0 \\
 0 & 0 & 0 & 1 & 0 & 0 & 0 & 0 & 0 \\
 0 & 0 & 0 & 0 & 1 & 0 & 0 & 0 & 0 \\
 0 & 0 & 0 & 0 & 0 & 1 & 0 & 0 & 0 \\
 0 & 0 & 0 & 0 & 0 & 0 & 1 & 0 & 0 \\
 0 & 0 & 0 & 0 & 0 & 0 & 0 & 1 & 0 \\
 -3 & -6 & -5 & -4 & -3 & -2 & -4 & -2 & 1 \\
\end{array}
\right).
\end{align}

There exists a similar basis redefinition of the $\mathbb{I}_8 \oplus (-\mathbb{I}_1)$ whose Cartan matrix is represented by the extended $E_8$ Dynkin diagram:
\begin{align}
e''_1 & = - x_1 - x_2 - x_3 + x_9, \cr
e''_i & = x_{i+1} - x_{i+2}, i = 2, ..., 6, \cr
e''_7 & = x_2 - x_3, \cr
e''_8 & = x_1 - x_2, \cr 
e''_9 & = x_8, 
\end{align}
where 
\begin{align}
e''_I = C_{I J} N_{J K} e'_K,
\end{align}
with
\begin{align}
N_{JK} = 
\left(
\begin{array}{ccccccccc}
 -1 & -1 & -1 & 0 & 0 & 0 & 0 & 0 & 1 \\
 0 & 0 & 1 & -1 & 0 & 0 & 0 & 0 & 0 \\
 0 & 0 & 0 & 1 & -1 & 0 & 0 & 0 & 0 \\
 0 & 0 & 0 & 0 & 1 & -1 & 0 & 0 & 0 \\
 0 & 0 & 0 & 0 & 0 & 1 & -1 & 0 & 0 \\
 0 & 0 & 0 & 0 & 0 & 0 & 1 & -1 & 0 \\
 0 & 1 & -1 & 0 & 0 & 0 & 0 & 0 & 0 \\
 1 & -1 & 0 & 0 & 0 & 0 & 0 & 0 & 0 \\
 0 & 0 & 0 & 0 & 0 & 0 & 0 & 1 & 0 \\
\end{array}
\right),
\end{align}
and $C_{IJ} = \mathbb{I}_5 \oplus (- \mathbb{I}_3) \oplus \mathbb{I}_1$.
The matrix $C$ is introduced to preserve charge conservation in the example considered in the main text, but is not essential for relating the two lattices.

Thus, we have the relations:
\begin{align}
C_{I L} N_{L K} (e'_K)_a \eta_{ab} C_{J M} N_{ML} (e'_L)_b & = (e''_I)_a \eta_{ab} (e''_J)_b \cr
& = (\tilde{e}_I)_a \eta_{ab} (\tilde{e}_J)_b \cr
& = M_{IK} (e_K)_a \eta_{ab} M_{JL} (e_L)_b, \cr
\end{align}
from which we read off,
\begin{align}
W_{(8,1)} = M^T (N^T)^{-1} C.
\end{align}

\section{Charge Conservation and $\nu=2$ Tunneling Interactions}
\label{nu2charge}

In this Appendix, we show that a constant correction to the entanglement entropy in the $\nu=2$ state requires charge conservation to be violated. 
Consider the tunneling operators defined by the integer vectors $({\cal M}^R_{1 I} = \begin{pmatrix} a & b \end{pmatrix}, {\cal M}^L_{1 I} = \begin{pmatrix} c & d\end{pmatrix})$ and $({\cal M}^R_{2 I} = \begin{pmatrix} a' & b' \end{pmatrix}, {\cal M}^L_{2 I} = \begin{pmatrix} c' & d'\end{pmatrix})$.
Nullity and charge conservation requires the first vector, $({\cal M}^R_{1 I}, {\cal M}^L_{1 I})$ to satisfy:
\begin{align}
a^2 + b^2 = & c^2 + d^2,\cr
a + b = & c + d.
\end{align}
Squaring the second equation and subtracting from the first yields: $ab = cd$.
Suppose $c\neq 0$, then $d = ab/c$.
Substituting back into the null condition allows us to find: $c^2 = 1/2(a^2 + b^2 \pm |a^2 - b^2|$.
If $c=\pm a$, then $d = \pm b$ and the charge conservation constraint fixes $a=c, d=b$.
Alternatively, if $c= \pm b$, then $d = \pm a$, and charge conservation requires $a = d$ and $b = a$.
On the other hand, if $c=0$, we reach the same conclusions.
Thus, a charge-conserving null vector must take one of two forms: $\begin{pmatrix}a & b & a & b \end{pmatrix}$ or $\begin{pmatrix}a & b & b & a \end{pmatrix}.$
The same analysis implies the ``primed" charge-conserving null vector to take an analogous form.
Requiring the the null vectors to be orthogonal to one another requires, $a a' + b b' - c c' - d d' = 0$, implies $({\cal M}^R)^{-1} {\cal M}^L$ to either be the identity or the Pauli $\sigma_x$ matrix.
Therefore, we must break charge conservation if we are to generate a constant correction to the entanglement entropy for the $\nu=2$ state.

\section{Matrices of Sec. \ref{asymmint}}
\label{largematrices}

In this Appendix, we provide explicit forms for the matrices ${\cal U}_{(8)/(\tilde{8})}, {\cal V}_{(8)/(\tilde{8})}$, and ${\cal S}_{(8)/(\tilde{8})}$ used in Sec. \ref{asymmint}.
We find:
\begin{widetext}
\begin{align}
{\cal U}_{(8)/(\tilde{8})} = & \left(
\begin{array}{cccccccc}
 1 & 0 & 0 & 0 & 0 & 0 & 0 & 0 \\
 1 & 1 & 0 & 0 & 0 & 0 & 0 & 0 \\
 1 & 1 & 1 & 0 & 0 & 0 & 0 & 0 \\
 1 & 1 & 1 & 1 & 0 & 0 & 0 & 0 \\
 1 & 1 & 1 & 1 & 1 & 0 & 0 & 0 \\
 1 & 1 & 1 & 1 & 1 & -1 & 0 & 0 \\
 1 & 1 & 1 & 1 & 1 & -1 & -1 & 0 \\
 1 & 1 & 1 & 1 & 1 & -1 & -1 & -1 \\
\end{array}
\right), \cr
{\cal V}_{(8)/(\tilde{8})} = & \left(
\begin{array}{cccccccccccccccc}
 0 & 0 & 0 & 0 & 0 & 0 & 0 & 0 & 0 & 1 & 0 & 0 & 0 & 0 & 0 & 0 \\
 0 & 0 & 1 & 0 & 0 & 0 & 0 & -1 & -5 & 0 & 0 & 1 & 1 & -1 & -1 & -1 \\
 0 & 0 & 0 & 1 & 0 & 0 & 0 & -1 & -4 & 0 & 0 & 0 & 1 & -1 & -1 & -1 \\
 0 & 0 & 0 & 0 & 1 & 0 & 0 & -1 & -3 & 0 & 0 & 0 & 0 & -1 & -1 & -1 \\
 0 & 0 & 0 & 0 & 0 & 1 & 0 & -1 & -2 & 0 & 0 & 0 & 0 & 0 & -1 & -1 \\
 0 & 0 & 0 & 0 & 0 & 0 & 1 & -1 & -1 & 0 & 0 & 0 & 0 & 0 & 0 & -1 \\
 0 & 1 & 0 & 0 & 0 & 0 & 0 & -1 & -6 & -1 & 1 & 1 & 1 & -1 & -1 & -1 \\
 1 & 0 & 0 & 0 & 0 & 0 & 0 & 0 & -2 & 0 & 0 & 0 & 0 & 0 & 0 & 0 \\
 0 & 0 & 0 & 0 & 0 & 0 & 0 & 0 & 1 & 0 & 0 & 0 & 0 & 0 & 0 & 0 \\
 0 & 0 & 0 & 0 & 0 & 0 & 0 & 1 & 5 & 2 & -1 & -1 & -1 & 1 & 1 & 1 \\
 0 & 0 & 0 & 0 & 0 & 0 & 0 & 0 & 0 & 0 & 1 & 0 & 0 & 0 & 0 & 0 \\
 0 & 0 & 0 & 0 & 0 & 0 & 0 & 0 & 0 & 0 & 0 & 1 & 0 & 0 & 0 & 0 \\
 0 & 0 & 0 & 0 & 0 & 0 & 0 & 0 & 0 & 0 & 0 & 0 & 1 & 0 & 0 & 0 \\
 0 & 0 & 0 & 0 & 0 & 0 & 0 & 0 & 0 & 0 & 0 & 0 & 0 & 1 & 0 & 0 \\
 0 & 0 & 0 & 0 & 0 & 0 & 0 & 0 & 0 & 0 & 0 & 0 & 0 & 0 & 1 & 0 \\
 0 & 0 & 0 & 0 & 0 & 0 & 0 & 0 & 0 & 0 & 0 & 0 & 0 & 0 & 0 & 1 \\
\end{array}
\right), \cr
{\cal S}_{(8)/(\tilde{8})} = & \left(
\begin{array}{cccccccccccccccc}
 1 & 0 & 0 & 0 & 0 & 0 & 0 & 0 & 0 & 0 & 0 & 0 & 0 & 0 & 0 & 0 \\
 0 & 1 & 0 & 0 & 0 & 0 & 0 & 0 & 0 & 0 & 0 & 0 & 0 & 0 & 0 & 0 \\
 0 & 0 & 1 & 0 & 0 & 0 & 0 & 0 & 0 & 0 & 0 & 0 & 0 & 0 & 0 & 0 \\
 0 & 0 & 0 & 1 & 0 & 0 & 0 & 0 & 0 & 0 & 0 & 0 & 0 & 0 & 0 & 0 \\
 0 & 0 & 0 & 0 & 1 & 0 & 0 & 0 & 0 & 0 & 0 & 0 & 0 & 0 & 0 & 0 \\
 0 & 0 & 0 & 0 & 0 & 1 & 0 & 0 & 0 & 0 & 0 & 0 & 0 & 0 & 0 & 0 \\
 0 & 0 & 0 & 0 & 0 & 0 & 1 & 0 & 0 & 0 & 0 & 0 & 0 & 0 & 0 & 0 \\
 0 & 0 & 0 & 0 & 0 & 0 & 0 & 1 & 0 & 0 & 0 & 0 & 0 & 0 & 0 & 0 \\
\end{array}
\right).
\end{align}
\end{widetext}

\section{Statistics of Quasiparticles in the $\nu=2'$ Phase}
\label{nu2appendix}

In this Appendix, we consider the statistics of quasiparticles within a bulk 2D phase for the K-matrix, 
\begin{align}
K=
\left(
\begin{array}{cc}
 1 & 0 \\
 0 & 1 \\
\end{array}
\right),
\end{align}
that is constructed within the wire approach with tunneling interactions $g_\beta \cos({\cal M}^R_{\beta I} \varphi^R_I + {\cal M}^L_{\beta I} \varphi^L_I)$ defined by the tunneling vectors:
\begin{align}
\label{nu2vectors}
({\cal M}^R_{(2)})_{1 I} = & (3, -1),  \quad  ({\cal M}^L_{(2)})_{1 I} = (3,1), \cr
({\cal M}^R_{(2)})_{2 I} = & (2, 1),   \quad  ({\cal M}^L_{(2)})_{2 I} = (1,2),
\end{align}
where the $\varphi^{R/L}_I$ realize a 2-channel free Fermi liquid in the decoupled limit.
We colloquially refer to this phase as $\nu=2'$.
These tunneling vectors are {\it primitive} and null.
Primitivity ensures that the gapped -- a result of the nullity condition -- inter-wire vacuum determined by the cosine interactions is unique (up to the usual degeneracy of shifting the arguments of the cosine by $2\pi$).
If we happened to have chosen non-primitive vectors, this vacuum would be degenerate.
However, this type of degeneracy is {\it not} protected as it may be lifted by local perturbations along the wire.

Now, if we were to take the resulting $K^{\rm eff}$ determined by the tunneling vectors in Eqn. (\ref{nu2vectors}) seriously, the statistics of quasiparticles would be determined by:
\begin{align}
(K^{\rm eff})^{-1} = \left(
\begin{array}{cc}
 2 & -\frac{7}{5} \\
 -\frac{7}{5} & 1 \\
\end{array}
\right).
\end{align}
In distinction to the fermion statistics of the quasiparticles of the $\nu=2$ state, if $K^{\rm eff}$ can be taken seriously, we obtain non-trivial mutual statistics between distinct quasiparticles, one of which has the self-statistics of a boson and the other is fermionic.
We will show that no such excitations occur in this example, and we believe this conclusion to be representative of all such examples discussed in our paper.
Physically, instead of nucleating a true 2D phase defined by $K^{\rm eff}$, we believe only sequences of infinitesimal strips of $K^{\rm eff}$ obtain when the wires are sewn together using the tunneling vectors in Eqn. (\ref{nu2vectors}) and that the strips do not coalesce into a full 2D phase.

We now justify these remarks.
Within the wire construction, quasiparticles correspond to $2\pi$ kinks of the inter-wire cosine interaction.
Under the transport of {\it any} quasiparticle around a full loop that may contain some number of quasiparticles, the acquired statistical phase is:
\begin{align}
\theta = & c_1 [\langle 3 \varphi^R_1 - \varphi^R_2 + 3 \varphi^L_1 + \varphi^L_2 \rangle] \cr 
+ & c_2  \langle 2 \varphi^R_1 + \varphi^R_2 + \varphi^L_1 + 2\varphi^L_2 \rangle]\Big|^x_{x'},
\end{align}
where $c_1$ and $c_2$ are (not necessarily integral) constants and $|^x_{x'}$ instructs us to evaluate the expectation values at positions $x$ and $x'$ along the interface and search for a possible $2\pi \mathbb{Z}$ shift.
While in general, the $c_i$ need not be integral, there is an important constraint that they must satisfy: namely, if we remove the expectation value and combine arguments, the coefficient of each $\varphi^{R/L}_I$ term must be integral. 
This condition arises from the requirement that any quasiparticle transport be achieved by only using local (electron) operators.
We consider the limit where the only quasiparticle transport operators are those given by the vectors in Eqn. (\ref{nu2vectors}).
(It is an interesting question of how perturbations generated by single-particle tunneling operators might affect this result and our general study of the entanglement spectrum and entropy.)

We now determine whether fractional $c_i$ are allowed which would imply possible fractional (mutual) statistics in the $\nu=2'$ state.
Combining the arguments of the expectation values, we obtain:
\begin{align}
&(3c_1 + 2 c_2)R_1 + (c_2 - c_1) R_2 + (3c_1 + c_2) L_1 
\cr +& (c_1 + 2 c_2) L_2.
\end{align}
We use the second term to write: $c_2 = c_1 + N$ where $N \in \mathbb{Z}$.
Substituting this back into the above expression, we find:
\begin{align}
(5c_1 + N)R_1 + N R_2 + (4c_1 + N) L_1 + (3c_1 + 2N) L_2.
\end{align}
Writing $c_1 = p/q$ with $q$ not dividing $p$, integrality of the coefficients requires $q$ to divide 5, 4, and 3. 
The only possibility is to take $q=1$.
This means there are no fractional statistics in the $\nu=2'$ state.

\bibliography{stableentanglement}

\begin{thebibliography}{96}
\expandafter\ifx\csname natexlab\endcsname\relax\def\natexlab#1{#1}\fi
\expandafter\ifx\csname bibnamefont\endcsname\relax
  \def\bibnamefont#1{#1}\fi
\expandafter\ifx\csname bibfnamefont\endcsname\relax
  \def\bibfnamefont#1{#1}\fi
\expandafter\ifx\csname citenamefont\endcsname\relax
  \def\citenamefont#1{#1}\fi
\expandafter\ifx\csname url\endcsname\relax
  \def\url#1{\texttt{#1}}\fi
\expandafter\ifx\csname urlprefix\endcsname\relax\def\urlprefix{URL }\fi
\providecommand{\bibinfo}[2]{#2}
\providecommand{\eprint}[2][]{\url{#2}}

\bibitem[{\citenamefont{Witten}(1989)}]{wittenjones}
\bibinfo{author}{\bibfnamefont{E.}~\bibnamefont{Witten}},
  \bibinfo{journal}{Comm. Math. Phys.} \textbf{\bibinfo{volume}{121}},
  \bibinfo{pages}{351} (\bibinfo{year}{1989}).

\bibitem[{\citenamefont{Elitzur et~al.}(1989)\citenamefont{Elitzur, Moore,
  Schwimmer, and Seiberg}}]{Elitzur89}
\bibinfo{author}{\bibfnamefont{S.}~\bibnamefont{Elitzur}},
  \bibinfo{author}{\bibfnamefont{G.~W.} \bibnamefont{Moore}},
  \bibinfo{author}{\bibfnamefont{A.}~\bibnamefont{Schwimmer}},
  \bibnamefont{and} \bibinfo{author}{\bibfnamefont{N.}~\bibnamefont{Seiberg}},
  \bibinfo{journal}{Nucl. Phys. B} \textbf{\bibinfo{volume}{326}},
  \bibinfo{pages}{108} (\bibinfo{year}{1989}).

\bibitem[{\citenamefont{Wen}(1991)}]{Wengapless}
\bibinfo{author}{\bibfnamefont{X.~G.} \bibnamefont{Wen}},
  \bibinfo{journal}{Phys. Rev. B} \textbf{\bibinfo{volume}{43}},
  \bibinfo{pages}{11025} (\bibinfo{year}{1991}).

\bibitem[{\citenamefont{Plamadeala et~al.}(2013)\citenamefont{Plamadeala,
  Mulligan, and Nayak}}]{PlamadealaE8}
\bibinfo{author}{\bibfnamefont{E.}~\bibnamefont{Plamadeala}},
  \bibinfo{author}{\bibfnamefont{M.}~\bibnamefont{Mulligan}}, \bibnamefont{and}
  \bibinfo{author}{\bibfnamefont{C.}~\bibnamefont{Nayak}},
  \bibinfo{journal}{Phys. Rev. B} \textbf{\bibinfo{volume}{88}},
  \bibinfo{pages}{045131} (\bibinfo{year}{2013}).

\bibitem[{\citenamefont{Cano et~al.}(2014)\citenamefont{Cano, Cheng, Mulligan,
  Nayak, Plamadeala, and Yard}}]{generalstableequivalence}
\bibinfo{author}{\bibfnamefont{J.}~\bibnamefont{Cano}},
  \bibinfo{author}{\bibfnamefont{M.}~\bibnamefont{Cheng}},
  \bibinfo{author}{\bibfnamefont{M.}~\bibnamefont{Mulligan}},
  \bibinfo{author}{\bibfnamefont{C.}~\bibnamefont{Nayak}},
  \bibinfo{author}{\bibfnamefont{E.}~\bibnamefont{Plamadeala}},
  \bibnamefont{and} \bibinfo{author}{\bibfnamefont{J.}~\bibnamefont{Yard}},
  \bibinfo{journal}{Phys. Rev. B} \textbf{\bibinfo{volume}{89}},
  \bibinfo{pages}{115116} (\bibinfo{year}{2014}).

\bibitem[{\citenamefont{Frohlich and Zee}(1991)}]{frohlichzee91}
\bibinfo{author}{\bibfnamefont{J.}~\bibnamefont{Frohlich}} \bibnamefont{and}
  \bibinfo{author}{\bibfnamefont{A.}~\bibnamefont{Zee}},
  \bibinfo{journal}{Nucl. Phys. B} \textbf{\bibinfo{volume}{364}},
  \bibinfo{pages}{517} (\bibinfo{year}{1991}).

\bibitem[{\citenamefont{Frohlich and Thiran}(1994)}]{frohlichthiran94}
\bibinfo{author}{\bibfnamefont{J.}~\bibnamefont{Frohlich}} \bibnamefont{and}
  \bibinfo{author}{\bibfnamefont{E.}~\bibnamefont{Thiran}},
  \bibinfo{journal}{Journal of Statistical Physics}
  \textbf{\bibinfo{volume}{76}}, \bibinfo{pages}{209} (\bibinfo{year}{1994}).

\bibitem[{\citenamefont{Frohlich et~al.}(1995)\citenamefont{Frohlich,
  Chamseddine, Gabbiani, Kerler, King, Marchetti, Studer, and
  Thiran}}]{frohlichmany95}
\bibinfo{author}{\bibfnamefont{J.}~\bibnamefont{Frohlich}},
  \bibinfo{author}{\bibfnamefont{A.}~\bibnamefont{Chamseddine}},
  \bibinfo{author}{\bibfnamefont{F.}~\bibnamefont{Gabbiani}},
  \bibinfo{author}{\bibfnamefont{T.}~\bibnamefont{Kerler}},
  \bibinfo{author}{\bibfnamefont{C.}~\bibnamefont{King}},
  \bibinfo{author}{\bibfnamefont{P.}~\bibnamefont{Marchetti}},
  \bibinfo{author}{\bibfnamefont{U.}~\bibnamefont{Studer}}, \bibnamefont{and}
  \bibinfo{author}{\bibfnamefont{E.}~\bibnamefont{Thiran}},
  \bibinfo{journal}{Proc. of ICM.} \textbf{\bibinfo{volume}{94}}
  (\bibinfo{year}{1995}).

\bibitem[{fro()}]{frohlichstuderthiran}
\bibinfo{note}{J. Frohlich, U. M. Studer, and E. Thiran, {\it On Three Levels},
  Springer US, 225-232, 1994.}

\bibitem[{\citenamefont{Kao et~al.}(1999)\citenamefont{Kao, Chang, and
  Wen}}]{Kao99}
\bibinfo{author}{\bibfnamefont{H.-C.} \bibnamefont{Kao}},
  \bibinfo{author}{\bibfnamefont{C.-H.} \bibnamefont{Chang}}, \bibnamefont{and}
  \bibinfo{author}{\bibfnamefont{X.-G.} \bibnamefont{Wen}},
  \bibinfo{journal}{Phys. Rev. Lett.} \textbf{\bibinfo{volume}{83}},
  \bibinfo{pages}{5563} (\bibinfo{year}{1999}).

\bibitem[{\citenamefont{Kane et~al.}(1994)\citenamefont{Kane, Fisher, and
  Polchinski}}]{KFPrandom}
\bibinfo{author}{\bibfnamefont{C.~L.} \bibnamefont{Kane}},
  \bibinfo{author}{\bibfnamefont{M.~P.~A.} \bibnamefont{Fisher}},
  \bibnamefont{and}
  \bibinfo{author}{\bibfnamefont{J.}~\bibnamefont{Polchinski}},
  \bibinfo{journal}{Phys. Rev. Lett.} \textbf{\bibinfo{volume}{72}},
  \bibinfo{pages}{4129} (\bibinfo{year}{1994}).

\bibitem[{\citenamefont{Kane and Fisher}(1995)}]{KFrandom}
\bibinfo{author}{\bibfnamefont{C.~L.} \bibnamefont{Kane}} \bibnamefont{and}
  \bibinfo{author}{\bibfnamefont{M.~P.~A.} \bibnamefont{Fisher}},
  \bibinfo{journal}{Phys. Rev. B} \textbf{\bibinfo{volume}{51}},
  \bibinfo{pages}{13449} (\bibinfo{year}{1995}).

\bibitem[{\citenamefont{Kitaev}(2003)}]{kitaevtoriccode}
\bibinfo{author}{\bibfnamefont{A.~Y.} \bibnamefont{Kitaev}},
  \bibinfo{journal}{Annals Phys.} \textbf{\bibinfo{volume}{303}},
  \bibinfo{pages}{2} (\bibinfo{year}{2003}).

\bibitem[{\citenamefont{Bravyi and Kitaev}()}]{bravyi1998}
\bibinfo{author}{\bibfnamefont{S.~B.} \bibnamefont{Bravyi}} \bibnamefont{and}
  \bibinfo{author}{\bibfnamefont{A.~Y.} \bibnamefont{Kitaev}},
  \bibinfo{note}{arXiv: 9811052}.

\bibitem[{\citenamefont{Haldane}(1995)}]{Haldanestability}
\bibinfo{author}{\bibfnamefont{F.~D.~M.} \bibnamefont{Haldane}},
  \bibinfo{journal}{Phys. Rev. Lett.} \textbf{\bibinfo{volume}{74}},
  \bibinfo{pages}{2090} (\bibinfo{year}{1995}).

\bibitem[{\citenamefont{Wang and Wen}()}]{wangwenboundarydegen}
\bibinfo{author}{\bibfnamefont{J.}~\bibnamefont{Wang}} \bibnamefont{and}
  \bibinfo{author}{\bibfnamefont{X.-G.} \bibnamefont{Wen}},
  \bibinfo{note}{arXiv:1212.4863}.

\bibitem[{\citenamefont{Levin}(2013)}]{levinprotected}
\bibinfo{author}{\bibfnamefont{M.}~\bibnamefont{Levin}},
  \bibinfo{journal}{Phys. Rev. X} \textbf{\bibinfo{volume}{3}},
  \bibinfo{pages}{021009} (\bibinfo{year}{2013}).

\bibitem[{\citenamefont{Barkeshli et~al.}(2013)\citenamefont{Barkeshli, Jian,
  and Qi}}]{classtops}
\bibinfo{author}{\bibfnamefont{M.}~\bibnamefont{Barkeshli}},
  \bibinfo{author}{\bibfnamefont{C.-M.} \bibnamefont{Jian}}, \bibnamefont{and}
  \bibinfo{author}{\bibfnamefont{X.-L.} \bibnamefont{Qi}},
  \bibinfo{journal}{Phys. Rev. B} \textbf{\bibinfo{volume}{88}},
  \bibinfo{pages}{241103} (\bibinfo{year}{2013}).

\bibitem[{\citenamefont{Kitaev and Kong}(2012)}]{KitaevKonggappedboundaries}
\bibinfo{author}{\bibfnamefont{A.}~\bibnamefont{Kitaev}} \bibnamefont{and}
  \bibinfo{author}{\bibfnamefont{L.}~\bibnamefont{Kong}},
  \bibinfo{journal}{Comm. Math. Phys.} \textbf{\bibinfo{volume}{313}},
  \bibinfo{pages}{351} (\bibinfo{year}{2012}).

\bibitem[{\citenamefont{Fuchs et~al.}(2013)\citenamefont{Fuchs, Schweigert, and
  Valentino}}]{FuchsSchweigertValentino}
\bibinfo{author}{\bibfnamefont{J.}~\bibnamefont{Fuchs}},
  \bibinfo{author}{\bibfnamefont{C.}~\bibnamefont{Schweigert}},
  \bibnamefont{and}
  \bibinfo{author}{\bibfnamefont{A.}~\bibnamefont{Valentino}},
  \bibinfo{journal}{Comm. Math. Phys.} \textbf{\bibinfo{volume}{321}},
  \bibinfo{pages}{543} (\bibinfo{year}{2013}).

\bibitem[{\citenamefont{Kapustin and
  Saulina}(2011)}]{KapustinSaulinatopboundary}
\bibinfo{author}{\bibfnamefont{A.}~\bibnamefont{Kapustin}} \bibnamefont{and}
  \bibinfo{author}{\bibfnamefont{N.}~\bibnamefont{Saulina}},
  \bibinfo{journal}{Nucl. Phys. B} \textbf{\bibinfo{volume}{845}},
  \bibinfo{pages}{393} (\bibinfo{year}{2011}).

\bibitem[{\citenamefont{Kapustin}(2014)}]{Kapustingsdegen}
\bibinfo{author}{\bibfnamefont{A.}~\bibnamefont{Kapustin}},
  \bibinfo{journal}{Phys. Rev. B} \textbf{\bibinfo{volume}{89}},
  \bibinfo{pages}{125307} (\bibinfo{year}{2014}).

\bibitem[{\citenamefont{Lan et~al.}()\citenamefont{Lan, Wang, and
  Wen}}]{LanWangWen}
\bibinfo{author}{\bibfnamefont{T.}~\bibnamefont{Lan}},
  \bibinfo{author}{\bibfnamefont{J.}~\bibnamefont{Wang}}, \bibnamefont{and}
  \bibinfo{author}{\bibfnamefont{X.-G.} \bibnamefont{Wen}},
  \bibinfo{note}{arXiv:1408.6514}.

\bibitem[{\citenamefont{Barkeshli et~al.}()\citenamefont{Barkeshli, Berg, and
  Kivelson}}]{bosonedgephases}
\bibinfo{author}{\bibfnamefont{M.}~\bibnamefont{Barkeshli}},
  \bibinfo{author}{\bibfnamefont{E.}~\bibnamefont{Berg}}, \bibnamefont{and}
  \bibinfo{author}{\bibfnamefont{S.}~\bibnamefont{Kivelson}},
  \bibinfo{note}{arXiv:1402.6321}.

\bibitem[{\citenamefont{Levin and Wen}(2006)}]{LevinWenentropy}
\bibinfo{author}{\bibfnamefont{M.}~\bibnamefont{Levin}} \bibnamefont{and}
  \bibinfo{author}{\bibfnamefont{X.-G.} \bibnamefont{Wen}},
  \bibinfo{journal}{Phys.Rev.Lett.} \textbf{\bibinfo{volume}{96}},
  \bibinfo{pages}{110405} (\bibinfo{year}{2006}).

\bibitem[{\citenamefont{Kitaev and Preskill}(2006)}]{KitaevPreskillentropy}
\bibinfo{author}{\bibfnamefont{A.}~\bibnamefont{Kitaev}} \bibnamefont{and}
  \bibinfo{author}{\bibfnamefont{J.}~\bibnamefont{Preskill}},
  \bibinfo{journal}{Phys.Rev.Lett.} \textbf{\bibinfo{volume}{96}},
  \bibinfo{pages}{110404} (\bibinfo{year}{2006}).

\bibitem[{\citenamefont{Hamma et~al.}(2005)\citenamefont{Hamma, Ionicioiu, and
  Zanardi}}]{hammakitaev}
\bibinfo{author}{\bibfnamefont{A.}~\bibnamefont{Hamma}},
  \bibinfo{author}{\bibfnamefont{R.}~\bibnamefont{Ionicioiu}},
  \bibnamefont{and} \bibinfo{author}{\bibfnamefont{P.}~\bibnamefont{Zanardi}},
  \bibinfo{journal}{Physics Letters A} \textbf{\bibinfo{volume}{337}},
  \bibinfo{pages}{22} (\bibinfo{year}{2005}).

\bibitem[{\citenamefont{Li and Haldane}(2008)}]{lihaldane}
\bibinfo{author}{\bibfnamefont{H.}~\bibnamefont{Li}} \bibnamefont{and}
  \bibinfo{author}{\bibfnamefont{F.~D.~M.} \bibnamefont{Haldane}},
  \bibinfo{journal}{Phys. Rev. Lett.} \textbf{\bibinfo{volume}{101}},
  \bibinfo{pages}{010504} (\bibinfo{year}{2008}).

\bibitem[{\citenamefont{Fidkowski}(2010)}]{Fidkowski2010}
\bibinfo{author}{\bibfnamefont{L.}~\bibnamefont{Fidkowski}},
  \bibinfo{journal}{Phys. Rev. Lett.} \textbf{\bibinfo{volume}{104}},
  \bibinfo{pages}{130502} (\bibinfo{year}{2010}).

\bibitem[{\citenamefont{Qi et~al.}(2012)\citenamefont{Qi, Katsura, and
  Ludwig}}]{qikatsuraludwig}
\bibinfo{author}{\bibfnamefont{X.-L.} \bibnamefont{Qi}},
  \bibinfo{author}{\bibfnamefont{H.}~\bibnamefont{Katsura}}, \bibnamefont{and}
  \bibinfo{author}{\bibfnamefont{A.~W.~W.} \bibnamefont{Ludwig}},
  \bibinfo{journal}{Phys. Rev. Lett.} \textbf{\bibinfo{volume}{108}},
  \bibinfo{pages}{196402} (\bibinfo{year}{2012}).

\bibitem[{\citenamefont{Chandran et~al.}(2011)\citenamefont{Chandran, Hermanns,
  Regnault, and Bernevig}}]{chandran2011}
\bibinfo{author}{\bibfnamefont{A.}~\bibnamefont{Chandran}},
  \bibinfo{author}{\bibfnamefont{M.}~\bibnamefont{Hermanns}},
  \bibinfo{author}{\bibfnamefont{N.}~\bibnamefont{Regnault}}, \bibnamefont{and}
  \bibinfo{author}{\bibfnamefont{B.~A.} \bibnamefont{Bernevig}},
  \bibinfo{journal}{Physical Review B} \textbf{\bibinfo{volume}{84}},
  \bibinfo{pages}{205136} (\bibinfo{year}{2011}).

\bibitem[{\citenamefont{Swingle and Senthil}(2012)}]{swinglesenthil}
\bibinfo{author}{\bibfnamefont{B.}~\bibnamefont{Swingle}} \bibnamefont{and}
  \bibinfo{author}{\bibfnamefont{T.}~\bibnamefont{Senthil}},
  \bibinfo{journal}{Phys.Rev. B} \textbf{\bibinfo{volume}{86}},
  \bibinfo{pages}{045117} (\bibinfo{year}{2012}).

\bibitem[{\citenamefont{Chandran et~al.}(2014)\citenamefont{Chandran, Khemani,
  and Sondhi}}]{ChandranKhemaniSondhi}
\bibinfo{author}{\bibfnamefont{A.}~\bibnamefont{Chandran}},
  \bibinfo{author}{\bibfnamefont{V.}~\bibnamefont{Khemani}}, \bibnamefont{and}
  \bibinfo{author}{\bibfnamefont{S.}~\bibnamefont{Sondhi}},
  \bibinfo{journal}{Phys. Rev. Lett.} \textbf{\bibinfo{volume}{113}},
  \bibinfo{pages}{060501} (\bibinfo{year}{2014}).

\bibitem[{\citenamefont{Casini et~al.}(2014)\citenamefont{Casini, Huerta, and
  Rosabal}}]{CasiniHuertaRosabal}
\bibinfo{author}{\bibfnamefont{H.}~\bibnamefont{Casini}},
  \bibinfo{author}{\bibfnamefont{M.}~\bibnamefont{Huerta}}, \bibnamefont{and}
  \bibinfo{author}{\bibfnamefont{J.~A.} \bibnamefont{Rosabal}},
  \bibinfo{journal}{Phys. Rev. D} \textbf{\bibinfo{volume}{89}},
  \bibinfo{pages}{085012} (\bibinfo{year}{2014}).

\bibitem[{\citenamefont{Ohmori and Tachikawa}()}]{OhmoriTachikawa}
\bibinfo{author}{\bibfnamefont{K.}~\bibnamefont{Ohmori}} \bibnamefont{and}
  \bibinfo{author}{\bibfnamefont{Y.}~\bibnamefont{Tachikawa}},
  \bibinfo{note}{arXiv:1406.4167}.

\bibitem[{\citenamefont{Nayak et~al.}(2008)\citenamefont{Nayak, Simon, Stern,
  Freedman, and Sarma}}]{nonabelianreview}
\bibinfo{author}{\bibfnamefont{C.}~\bibnamefont{Nayak}},
  \bibinfo{author}{\bibfnamefont{S.~H.} \bibnamefont{Simon}},
  \bibinfo{author}{\bibfnamefont{A.}~\bibnamefont{Stern}},
  \bibinfo{author}{\bibfnamefont{M.}~\bibnamefont{Freedman}}, \bibnamefont{and}
  \bibinfo{author}{\bibfnamefont{S.~D.} \bibnamefont{Sarma}},
  \bibinfo{journal}{Rev. Mod. Phys.} \textbf{\bibinfo{volume}{80}},
  \bibinfo{pages}{1083} (\bibinfo{year}{2008}).

\bibitem[{\citenamefont{Dong et~al.}(2008)\citenamefont{Dong, Fradkin, and
  Leigh}}]{fradkintopologiesent}
\bibinfo{author}{\bibfnamefont{S.}~\bibnamefont{Dong}},
  \bibinfo{author}{\bibfnamefont{E.}~\bibnamefont{Fradkin}}, \bibnamefont{and}
  \bibinfo{author}{\bibfnamefont{R.~G.} \bibnamefont{Leigh}},
  \bibinfo{journal}{JHEP} \textbf{\bibinfo{volume}{0805}}, \bibinfo{pages}{016}
  (\bibinfo{year}{2008}).

\bibitem[{\citenamefont{Zhang et~al.}(2012)\citenamefont{Zhang, Grover, Turner,
  Oshikawa, and Vishwanath}}]{groverbraiding}
\bibinfo{author}{\bibfnamefont{Y.}~\bibnamefont{Zhang}},
  \bibinfo{author}{\bibfnamefont{T.}~\bibnamefont{Grover}},
  \bibinfo{author}{\bibfnamefont{A.}~\bibnamefont{Turner}},
  \bibinfo{author}{\bibfnamefont{M.}~\bibnamefont{Oshikawa}}, \bibnamefont{and}
  \bibinfo{author}{\bibfnamefont{A.}~\bibnamefont{Vishwanath}},
  \bibinfo{journal}{Phys. Rev. B} \textbf{\bibinfo{volume}{85}},
  \bibinfo{pages}{235151} (\bibinfo{year}{2012}).

\bibitem[{\citenamefont{Papanikolaou et~al.}(2007)\citenamefont{Papanikolaou,
  Raman, and Fradkin}}]{PapanikolaouRamanFradkin}
\bibinfo{author}{\bibfnamefont{S.}~\bibnamefont{Papanikolaou}},
  \bibinfo{author}{\bibfnamefont{K.~S.} \bibnamefont{Raman}}, \bibnamefont{and}
  \bibinfo{author}{\bibfnamefont{E.}~\bibnamefont{Fradkin}},
  \bibinfo{journal}{Phys. Rev. B} \textbf{\bibinfo{volume}{76}},
  \bibinfo{pages}{224421} (\bibinfo{year}{2007}).

\bibitem[{\citenamefont{Regnault et~al.}(2009)\citenamefont{Regnault, Bernevig,
  and Haldane}}]{Regnault2009}
\bibinfo{author}{\bibfnamefont{N.}~\bibnamefont{Regnault}},
  \bibinfo{author}{\bibfnamefont{B.~A.} \bibnamefont{Bernevig}},
  \bibnamefont{and} \bibinfo{author}{\bibfnamefont{F.~D.~M.}
  \bibnamefont{Haldane}}, \bibinfo{journal}{Phys. Rev. Lett.}
  \textbf{\bibinfo{volume}{103}}, \bibinfo{pages}{016801}
  (\bibinfo{year}{2009}).

\bibitem[{\citenamefont{Thomale et~al.}(2010)\citenamefont{Thomale, Sterdyniak,
  Regnault, and Bernevig}}]{Thomale2010}
\bibinfo{author}{\bibfnamefont{R.}~\bibnamefont{Thomale}},
  \bibinfo{author}{\bibfnamefont{A.}~\bibnamefont{Sterdyniak}},
  \bibinfo{author}{\bibfnamefont{N.}~\bibnamefont{Regnault}}, \bibnamefont{and}
  \bibinfo{author}{\bibfnamefont{B.~A.} \bibnamefont{Bernevig}},
  \bibinfo{journal}{Phys. Rev. Lett.} \textbf{\bibinfo{volume}{104}},
  \bibinfo{pages}{180502} (\bibinfo{year}{2010}).

\bibitem[{\citenamefont{L{\"a}uchli et~al.}(2010)\citenamefont{L{\"a}uchli,
  Bergholtz, Suorsa, and Haque}}]{lauchli2010}
\bibinfo{author}{\bibfnamefont{A.~M.} \bibnamefont{L{\"a}uchli}},
  \bibinfo{author}{\bibfnamefont{E.~J.} \bibnamefont{Bergholtz}},
  \bibinfo{author}{\bibfnamefont{J.}~\bibnamefont{Suorsa}}, \bibnamefont{and}
  \bibinfo{author}{\bibfnamefont{M.}~\bibnamefont{Haque}},
  \bibinfo{journal}{Physical review letters} \textbf{\bibinfo{volume}{104}},
  \bibinfo{pages}{156404} (\bibinfo{year}{2010}).

\bibitem[{\citenamefont{Regnault and Bernevig}(2011)}]{regnault2011}
\bibinfo{author}{\bibfnamefont{N.}~\bibnamefont{Regnault}} \bibnamefont{and}
  \bibinfo{author}{\bibfnamefont{B.~A.} \bibnamefont{Bernevig}},
  \bibinfo{journal}{Physical Review X} \textbf{\bibinfo{volume}{1}},
  \bibinfo{pages}{021014} (\bibinfo{year}{2011}).

\bibitem[{\citenamefont{Sterdyniak
  et~al.}(2011{\natexlab{a}})\citenamefont{Sterdyniak, Regnault, and
  Bernevig}}]{sterdyniak2011}
\bibinfo{author}{\bibfnamefont{A.}~\bibnamefont{Sterdyniak}},
  \bibinfo{author}{\bibfnamefont{N.}~\bibnamefont{Regnault}}, \bibnamefont{and}
  \bibinfo{author}{\bibfnamefont{B.}~\bibnamefont{Bernevig}},
  \bibinfo{journal}{Physical review letters} \textbf{\bibinfo{volume}{106}},
  \bibinfo{pages}{100405} (\bibinfo{year}{2011}{\natexlab{a}}).

\bibitem[{\citenamefont{Hermanns et~al.}(2011)\citenamefont{Hermanns, Chandran,
  Regnault, and Bernevig}}]{hermanns2011}
\bibinfo{author}{\bibfnamefont{M.}~\bibnamefont{Hermanns}},
  \bibinfo{author}{\bibfnamefont{A.}~\bibnamefont{Chandran}},
  \bibinfo{author}{\bibfnamefont{N.}~\bibnamefont{Regnault}}, \bibnamefont{and}
  \bibinfo{author}{\bibfnamefont{B.~A.} \bibnamefont{Bernevig}},
  \bibinfo{journal}{Physical Review B} \textbf{\bibinfo{volume}{84}},
  \bibinfo{pages}{121309} (\bibinfo{year}{2011}).

\bibitem[{\citenamefont{Sterdyniak
  et~al.}(2011{\natexlab{b}})\citenamefont{Sterdyniak, Bernevig, Regnault, and
  Haldane}}]{sterdyniak2011a}
\bibinfo{author}{\bibfnamefont{A.}~\bibnamefont{Sterdyniak}},
  \bibinfo{author}{\bibfnamefont{B.}~\bibnamefont{Bernevig}},
  \bibinfo{author}{\bibfnamefont{N.}~\bibnamefont{Regnault}}, \bibnamefont{and}
  \bibinfo{author}{\bibfnamefont{F.}~\bibnamefont{Haldane}},
  \bibinfo{journal}{New Journal of Physics} \textbf{\bibinfo{volume}{13}},
  \bibinfo{pages}{105001} (\bibinfo{year}{2011}{\natexlab{b}}).

\bibitem[{\citenamefont{Papi{\'c} et~al.}(2011)\citenamefont{Papi{\'c},
  Bernevig, and Regnault}}]{papic2011}
\bibinfo{author}{\bibfnamefont{Z.}~\bibnamefont{Papi{\'c}}},
  \bibinfo{author}{\bibfnamefont{B.}~\bibnamefont{Bernevig}}, \bibnamefont{and}
  \bibinfo{author}{\bibfnamefont{N.}~\bibnamefont{Regnault}},
  \bibinfo{journal}{Physical Review Letters} \textbf{\bibinfo{volume}{106}},
  \bibinfo{pages}{056801} (\bibinfo{year}{2011}).

\bibitem[{\citenamefont{Schliemann}(2011)}]{schliemann2011}
\bibinfo{author}{\bibfnamefont{J.}~\bibnamefont{Schliemann}},
  \bibinfo{journal}{Physical Review B} \textbf{\bibinfo{volume}{83}},
  \bibinfo{pages}{115322} (\bibinfo{year}{2011}).

\bibitem[{\citenamefont{Sterdyniak et~al.}(2012)\citenamefont{Sterdyniak,
  Chandran, Regnault, Bernevig, and Bonderson}}]{sterdyniak2012}
\bibinfo{author}{\bibfnamefont{A.}~\bibnamefont{Sterdyniak}},
  \bibinfo{author}{\bibfnamefont{A.}~\bibnamefont{Chandran}},
  \bibinfo{author}{\bibfnamefont{N.}~\bibnamefont{Regnault}},
  \bibinfo{author}{\bibfnamefont{B.~A.} \bibnamefont{Bernevig}},
  \bibnamefont{and}
  \bibinfo{author}{\bibfnamefont{P.}~\bibnamefont{Bonderson}},
  \bibinfo{journal}{Physical Review B} \textbf{\bibinfo{volume}{85}},
  \bibinfo{pages}{125308} (\bibinfo{year}{2012}).

\bibitem[{\citenamefont{Dubail et~al.}(2012{\natexlab{a}})\citenamefont{Dubail,
  Read, and Rezayi}}]{dubail2012}
\bibinfo{author}{\bibfnamefont{J.}~\bibnamefont{Dubail}},
  \bibinfo{author}{\bibfnamefont{N.}~\bibnamefont{Read}}, \bibnamefont{and}
  \bibinfo{author}{\bibfnamefont{E.}~\bibnamefont{Rezayi}},
  \bibinfo{journal}{Physical Review B} \textbf{\bibinfo{volume}{85}},
  \bibinfo{pages}{115321} (\bibinfo{year}{2012}{\natexlab{a}}).

\bibitem[{\citenamefont{Dubail et~al.}(2012{\natexlab{b}})\citenamefont{Dubail,
  Read, and Rezayi}}]{Dubail2012more}
\bibinfo{author}{\bibfnamefont{J.}~\bibnamefont{Dubail}},
  \bibinfo{author}{\bibfnamefont{N.}~\bibnamefont{Read}}, \bibnamefont{and}
  \bibinfo{author}{\bibfnamefont{E.~H.} \bibnamefont{Rezayi}},
  \bibinfo{journal}{Phys. Rev. B} \textbf{\bibinfo{volume}{86}},
  \bibinfo{pages}{245310} (\bibinfo{year}{2012}{\natexlab{b}}).

\bibitem[{\citenamefont{Rodr{\'\i}guez
  et~al.}(2012)\citenamefont{Rodr{\'\i}guez, Simon, and
  Slingerland}}]{Simon2012}
\bibinfo{author}{\bibfnamefont{I.~D.} \bibnamefont{Rodr{\'\i}guez}},
  \bibinfo{author}{\bibfnamefont{S.~H.} \bibnamefont{Simon}}, \bibnamefont{and}
  \bibinfo{author}{\bibfnamefont{J.~K.} \bibnamefont{Slingerland}},
  \bibinfo{journal}{Phys. Rev. Lett.} \textbf{\bibinfo{volume}{108}},
  \bibinfo{pages}{256806} (\bibinfo{year}{2012}).

\bibitem[{\citenamefont{Rodr{\'\i}guez
  et~al.}(2013)\citenamefont{Rodr{\'\i}guez, Davenport, Simon, and
  Slingerland}}]{Simon2013}
\bibinfo{author}{\bibfnamefont{I.~D.} \bibnamefont{Rodr{\'\i}guez}},
  \bibinfo{author}{\bibfnamefont{S.~C.} \bibnamefont{Davenport}},
  \bibinfo{author}{\bibfnamefont{S.~H.} \bibnamefont{Simon}}, \bibnamefont{and}
  \bibinfo{author}{\bibfnamefont{J.~K.} \bibnamefont{Slingerland}},
  \bibinfo{journal}{Phys. Rev. B} \textbf{\bibinfo{volume}{88}},
  \bibinfo{pages}{155307} (\bibinfo{year}{2013}).

\bibitem[{\citenamefont{Prodan et~al.}(2010)\citenamefont{Prodan, Hughes, and
  Bernevig}}]{Prodan2010}
\bibinfo{author}{\bibfnamefont{E.}~\bibnamefont{Prodan}},
  \bibinfo{author}{\bibfnamefont{T.~L.} \bibnamefont{Hughes}},
  \bibnamefont{and} \bibinfo{author}{\bibfnamefont{B.~A.}
  \bibnamefont{Bernevig}}, \bibinfo{journal}{Phys. Rev. Lett.}
  \textbf{\bibinfo{volume}{105}}, \bibinfo{pages}{115501}
  (\bibinfo{year}{2010}).

\bibitem[{\citenamefont{Pollmann et~al.}(2010)\citenamefont{Pollmann, Turner,
  Berg, and Oshikawa}}]{Pollman2010}
\bibinfo{author}{\bibfnamefont{F.}~\bibnamefont{Pollmann}},
  \bibinfo{author}{\bibfnamefont{A.~M.} \bibnamefont{Turner}},
  \bibinfo{author}{\bibfnamefont{E.}~\bibnamefont{Berg}}, \bibnamefont{and}
  \bibinfo{author}{\bibfnamefont{M.}~\bibnamefont{Oshikawa}},
  \bibinfo{journal}{Phys. Rev. B} \textbf{\bibinfo{volume}{81}},
  \bibinfo{pages}{064439} (\bibinfo{year}{2010}).

\bibitem[{\citenamefont{Turner et~al.}(2010)\citenamefont{Turner, Zhang, and
  Vishwanath}}]{Turner2010}
\bibinfo{author}{\bibfnamefont{A.~M.} \bibnamefont{Turner}},
  \bibinfo{author}{\bibfnamefont{Y.}~\bibnamefont{Zhang}}, \bibnamefont{and}
  \bibinfo{author}{\bibfnamefont{A.}~\bibnamefont{Vishwanath}},
  \bibinfo{journal}{Phys. Rev. B} \textbf{\bibinfo{volume}{82}},
  \bibinfo{pages}{241102} (\bibinfo{year}{2010}).

\bibitem[{\citenamefont{Hughes et~al.}(2011)\citenamefont{Hughes, Prodan, and
  Bernevig}}]{Hughes2011}
\bibinfo{author}{\bibfnamefont{T.~L.} \bibnamefont{Hughes}},
  \bibinfo{author}{\bibfnamefont{E.}~\bibnamefont{Prodan}}, \bibnamefont{and}
  \bibinfo{author}{\bibfnamefont{B.~A.} \bibnamefont{Bernevig}},
  \bibinfo{journal}{Phys. Rev. B} \textbf{\bibinfo{volume}{83}},
  \bibinfo{pages}{245132} (\bibinfo{year}{2011}).

\bibitem[{\citenamefont{Alexandradinata
  et~al.}(2011)\citenamefont{Alexandradinata, Hughes, and Bernevig}}]{alex2011}
\bibinfo{author}{\bibfnamefont{A.}~\bibnamefont{Alexandradinata}},
  \bibinfo{author}{\bibfnamefont{T.~L.} \bibnamefont{Hughes}},
  \bibnamefont{and} \bibinfo{author}{\bibfnamefont{B.~A.}
  \bibnamefont{Bernevig}}, \bibinfo{journal}{Physical Review B}
  \textbf{\bibinfo{volume}{84}}, \bibinfo{pages}{195103}
  (\bibinfo{year}{2011}).

\bibitem[{\citenamefont{Turner et~al.}(2011)\citenamefont{Turner, Pollmann, and
  Berg}}]{turner2011a}
\bibinfo{author}{\bibfnamefont{A.~M.} \bibnamefont{Turner}},
  \bibinfo{author}{\bibfnamefont{F.}~\bibnamefont{Pollmann}}, \bibnamefont{and}
  \bibinfo{author}{\bibfnamefont{E.}~\bibnamefont{Berg}},
  \bibinfo{journal}{Physical Review B} \textbf{\bibinfo{volume}{83}},
  \bibinfo{pages}{075102} (\bibinfo{year}{2011}).

\bibitem[{\citenamefont{Gilbert et~al.}(2012)\citenamefont{Gilbert, Bernevig,
  and Hughes}}]{gilbert2012}
\bibinfo{author}{\bibfnamefont{M.~J.} \bibnamefont{Gilbert}},
  \bibinfo{author}{\bibfnamefont{B.~A.} \bibnamefont{Bernevig}},
  \bibnamefont{and} \bibinfo{author}{\bibfnamefont{T.~L.}
  \bibnamefont{Hughes}}, \bibinfo{journal}{Physical Review B}
  \textbf{\bibinfo{volume}{86}}, \bibinfo{pages}{041401}
  (\bibinfo{year}{2012}).

\bibitem[{\citenamefont{Fang et~al.}(2013)\citenamefont{Fang, Gilbert, and
  Bernevig}}]{fang2013}
\bibinfo{author}{\bibfnamefont{C.}~\bibnamefont{Fang}},
  \bibinfo{author}{\bibfnamefont{M.~J.} \bibnamefont{Gilbert}},
  \bibnamefont{and} \bibinfo{author}{\bibfnamefont{B.~A.}
  \bibnamefont{Bernevig}}, \bibinfo{journal}{Physical Review B}
  \textbf{\bibinfo{volume}{87}}, \bibinfo{pages}{035119}
  (\bibinfo{year}{2013}).

\bibitem[{\citenamefont{Flammia et~al.}(2009)\citenamefont{Flammia, Hamma,
  Hughes, and Wen}}]{Flammia2009}
\bibinfo{author}{\bibfnamefont{S.~T.} \bibnamefont{Flammia}},
  \bibinfo{author}{\bibfnamefont{A.}~\bibnamefont{Hamma}},
  \bibinfo{author}{\bibfnamefont{T.~L.} \bibnamefont{Hughes}},
  \bibnamefont{and} \bibinfo{author}{\bibfnamefont{X.-G.} \bibnamefont{Wen}},
  \bibinfo{journal}{Phys. Rev. Lett.} \textbf{\bibinfo{volume}{103}},
  \bibinfo{pages}{261601} (\bibinfo{year}{2009}).

\bibitem[{\citenamefont{Yao and Qi}(2010)}]{Yao2010}
\bibinfo{author}{\bibfnamefont{H.}~\bibnamefont{Yao}} \bibnamefont{and}
  \bibinfo{author}{\bibfnamefont{X.-L.} \bibnamefont{Qi}},
  \bibinfo{journal}{Phys. Rev. Lett.} \textbf{\bibinfo{volume}{105}},
  \bibinfo{pages}{080501} (\bibinfo{year}{2010}).

\bibitem[{\citenamefont{Dubail and Read}(2011)}]{dubail2011}
\bibinfo{author}{\bibfnamefont{J.}~\bibnamefont{Dubail}} \bibnamefont{and}
  \bibinfo{author}{\bibfnamefont{N.}~\bibnamefont{Read}},
  \bibinfo{journal}{Physical review letters} \textbf{\bibinfo{volume}{107}},
  \bibinfo{pages}{157001} (\bibinfo{year}{2011}).

\bibitem[{\citenamefont{Mondragon-Shem and
  Hughes}(2014{\natexlab{a}})}]{mondragon2014}
\bibinfo{author}{\bibfnamefont{I.}~\bibnamefont{Mondragon-Shem}}
  \bibnamefont{and} \bibinfo{author}{\bibfnamefont{T.~L.}
  \bibnamefont{Hughes}}, \bibinfo{journal}{Journal of Statistical Mechanics:
  Theory and Experiment} \textbf{\bibinfo{volume}{2014}},
  \bibinfo{pages}{P10022} (\bibinfo{year}{2014}{\natexlab{a}}).

\bibitem[{\citenamefont{Refael and Moore}(2004)}]{refael2004}
\bibinfo{author}{\bibfnamefont{G.}~\bibnamefont{Refael}} \bibnamefont{and}
  \bibinfo{author}{\bibfnamefont{J.~E.} \bibnamefont{Moore}},
  \bibinfo{journal}{Physical review letters} \textbf{\bibinfo{volume}{93}},
  \bibinfo{pages}{260602} (\bibinfo{year}{2004}).

\bibitem[{\citenamefont{Jia et~al.}(2008)\citenamefont{Jia, Subramaniam,
  Gruzberg, and Chakravarty}}]{jia2008}
\bibinfo{author}{\bibfnamefont{X.}~\bibnamefont{Jia}},
  \bibinfo{author}{\bibfnamefont{A.~R.} \bibnamefont{Subramaniam}},
  \bibinfo{author}{\bibfnamefont{I.~A.} \bibnamefont{Gruzberg}},
  \bibnamefont{and}
  \bibinfo{author}{\bibfnamefont{S.}~\bibnamefont{Chakravarty}},
  \bibinfo{journal}{Physical Review B} \textbf{\bibinfo{volume}{77}},
  \bibinfo{pages}{014208} (\bibinfo{year}{2008}).

\bibitem[{\citenamefont{Chen et~al.}(2012)\citenamefont{Chen, Hsu, Hughes, and
  Fradkin}}]{chen2012}
\bibinfo{author}{\bibfnamefont{X.}~\bibnamefont{Chen}},
  \bibinfo{author}{\bibfnamefont{B.}~\bibnamefont{Hsu}},
  \bibinfo{author}{\bibfnamefont{T.~L.} \bibnamefont{Hughes}},
  \bibnamefont{and} \bibinfo{author}{\bibfnamefont{E.}~\bibnamefont{Fradkin}},
  \bibinfo{journal}{Physical Review B} \textbf{\bibinfo{volume}{86}},
  \bibinfo{pages}{134201} (\bibinfo{year}{2012}).

\bibitem[{\citenamefont{Mondragon-Shem
  et~al.}(2013)\citenamefont{Mondragon-Shem, Khan, and Hughes}}]{mondragon2013}
\bibinfo{author}{\bibfnamefont{I.}~\bibnamefont{Mondragon-Shem}},
  \bibinfo{author}{\bibfnamefont{M.}~\bibnamefont{Khan}}, \bibnamefont{and}
  \bibinfo{author}{\bibfnamefont{T.~L.} \bibnamefont{Hughes}},
  \bibinfo{journal}{Physical Review Letters} \textbf{\bibinfo{volume}{110}},
  \bibinfo{pages}{046806} (\bibinfo{year}{2013}).

\bibitem[{\citenamefont{Pouranvari and Yang}(2013)}]{pouranvari2013}
\bibinfo{author}{\bibfnamefont{M.}~\bibnamefont{Pouranvari}} \bibnamefont{and}
  \bibinfo{author}{\bibfnamefont{K.}~\bibnamefont{Yang}},
  \bibinfo{journal}{Physical Review B} \textbf{\bibinfo{volume}{88}},
  \bibinfo{pages}{075123} (\bibinfo{year}{2013}).

\bibitem[{\citenamefont{Mondragon-Shem and
  Hughes}(2014{\natexlab{b}})}]{mondragon2014a}
\bibinfo{author}{\bibfnamefont{I.}~\bibnamefont{Mondragon-Shem}}
  \bibnamefont{and} \bibinfo{author}{\bibfnamefont{T.~L.}
  \bibnamefont{Hughes}}, \bibinfo{journal}{Phys. Rev. B}
  \textbf{\bibinfo{volume}{90}}, \bibinfo{pages}{104204}
  (\bibinfo{year}{2014}{\natexlab{b}}).

\bibitem[{\citenamefont{Pouranvari and Yang}(2014)}]{pouranvari2014}
\bibinfo{author}{\bibfnamefont{M.}~\bibnamefont{Pouranvari}} \bibnamefont{and}
  \bibinfo{author}{\bibfnamefont{K.}~\bibnamefont{Yang}},
  \bibinfo{journal}{Physical Review B} \textbf{\bibinfo{volume}{89}},
  \bibinfo{pages}{115104} (\bibinfo{year}{2014}).

\bibitem[{\citenamefont{Kane et~al.}(2002)\citenamefont{Kane, Mukhopadhyay, and
  Lubensky}}]{KaneMukhopadhyayLubensky}
\bibinfo{author}{\bibfnamefont{C.~L.} \bibnamefont{Kane}},
  \bibinfo{author}{\bibfnamefont{R.}~\bibnamefont{Mukhopadhyay}},
  \bibnamefont{and} \bibinfo{author}{\bibfnamefont{T.~C.}
  \bibnamefont{Lubensky}}, \bibinfo{journal}{Phys. Rev. Lett.}
  \textbf{\bibinfo{volume}{88}}, \bibinfo{pages}{036401}
  (\bibinfo{year}{2002}).

\bibitem[{\citenamefont{Teo and Kane}(2014)}]{TeoKanewires}
\bibinfo{author}{\bibfnamefont{J.~C.~Y.} \bibnamefont{Teo}} \bibnamefont{and}
  \bibinfo{author}{\bibfnamefont{C.~L.} \bibnamefont{Kane}},
  \bibinfo{journal}{Phys. Rev. B} \textbf{\bibinfo{volume}{89}},
  \bibinfo{pages}{085101} (\bibinfo{year}{2014}).

\bibitem[{\citenamefont{Neupert et~al.}()\citenamefont{Neupert, Chamon, Mudry,
  and Thomale}}]{thomalewires}
\bibinfo{author}{\bibfnamefont{T.}~\bibnamefont{Neupert}},
  \bibinfo{author}{\bibfnamefont{C.}~\bibnamefont{Chamon}},
  \bibinfo{author}{\bibfnamefont{C.}~\bibnamefont{Mudry}}, \bibnamefont{and}
  \bibinfo{author}{\bibfnamefont{R.}~\bibnamefont{Thomale}},
  \bibinfo{note}{arXiv:1403.0953}.

\bibitem[{\citenamefont{Lundgren et~al.}(2013)\citenamefont{Lundgren, Fuji,
  Furukawa, and Oshikawa}}]{lundgrenentanglement}
\bibinfo{author}{\bibfnamefont{R.}~\bibnamefont{Lundgren}},
  \bibinfo{author}{\bibfnamefont{Y.}~\bibnamefont{Fuji}},
  \bibinfo{author}{\bibfnamefont{S.}~\bibnamefont{Furukawa}}, \bibnamefont{and}
  \bibinfo{author}{\bibfnamefont{M.}~\bibnamefont{Oshikawa}},
  \bibinfo{journal}{Phys. Rev. B} \textbf{\bibinfo{volume}{88}},
  \bibinfo{pages}{245137} (\bibinfo{year}{2013}).

\bibitem[{\citenamefont{Chen and Fradkin}(2013)}]{chenentanglement}
\bibinfo{author}{\bibfnamefont{X.}~\bibnamefont{Chen}} \bibnamefont{and}
  \bibinfo{author}{\bibfnamefont{E.}~\bibnamefont{Fradkin}},
  \bibinfo{journal}{Journal of Statistical Mechanics: Theory and Experiment}
  \textbf{\bibinfo{volume}{2013}}, \bibinfo{pages}{P08013}
  (\bibinfo{year}{2013}).

\bibitem[{\citenamefont{Furukawa and Kim}(2011)}]{FurukawaKim}
\bibinfo{author}{\bibfnamefont{S.}~\bibnamefont{Furukawa}} \bibnamefont{and}
  \bibinfo{author}{\bibfnamefont{Y.~B.} \bibnamefont{Kim}},
  \bibinfo{journal}{Phys. Rev. B} \textbf{\bibinfo{volume}{83}},
  \bibinfo{pages}{085112} (\bibinfo{year}{2011}).

\bibitem[{\citenamefont{Calabrese and Cardy}(2004)}]{CardyCalabreseentropy}
\bibinfo{author}{\bibfnamefont{P.}~\bibnamefont{Calabrese}} \bibnamefont{and}
  \bibinfo{author}{\bibfnamefont{J.}~\bibnamefont{Cardy}}, \bibinfo{journal}{J.
  Stat. Mech.} \textbf{\bibinfo{volume}{2004}}, \bibinfo{pages}{06002}
  (\bibinfo{year}{2004}).

\bibitem[{\citenamefont{Zhou et~al.}(2006)\citenamefont{Zhou, Barthel,
  Fj{\ae}restad, and Schollw{\"o}ck}}]{schollwockbcentangle}
\bibinfo{author}{\bibfnamefont{H.-Q.} \bibnamefont{Zhou}},
  \bibinfo{author}{\bibfnamefont{T.}~\bibnamefont{Barthel}},
  \bibinfo{author}{\bibfnamefont{J.~O.} \bibnamefont{Fj{\ae}restad}},
  \bibnamefont{and}
  \bibinfo{author}{\bibfnamefont{U.}~\bibnamefont{Schollw{\"o}ck}},
  \bibinfo{journal}{Phys. Rev. A} \textbf{\bibinfo{volume}{74}},
  \bibinfo{pages}{050305} (\bibinfo{year}{2006}).

\bibitem[{\citenamefont{Laflorencie et~al.}(2006)\citenamefont{Laflorencie,
  S{\o}rensen, Chang, and Affleck}}]{affleckbcentangle}
\bibinfo{author}{\bibfnamefont{N.}~\bibnamefont{Laflorencie}},
  \bibinfo{author}{\bibfnamefont{E.~S.} \bibnamefont{S{\o}rensen}},
  \bibinfo{author}{\bibfnamefont{M.-S.} \bibnamefont{Chang}}, \bibnamefont{and}
  \bibinfo{author}{\bibfnamefont{I.}~\bibnamefont{Affleck}},
  \bibinfo{journal}{Phys. Rev. Lett.} \textbf{\bibinfo{volume}{96}},
  \bibinfo{pages}{100603} (\bibinfo{year}{2006}).

\bibitem[{\citenamefont{Lauchli}()}]{lauchlibcentropy}
\bibinfo{author}{\bibfnamefont{A.~M.} \bibnamefont{Lauchli}},
  \bibinfo{note}{arXiv:1303.0741}.

\bibitem[{\citenamefont{Vidal}(2000)}]{vidalmonotones}
\bibinfo{author}{\bibfnamefont{G.}~\bibnamefont{Vidal}},
  \bibinfo{journal}{Journal of Modern Optics} \textbf{\bibinfo{volume}{47}},
  \bibinfo{pages}{355} (\bibinfo{year}{2000}).

\bibitem[{\citenamefont{Xie~Chen and Wen}(2010)}]{chenguwenunitaries}
\bibinfo{author}{\bibfnamefont{Z.-C.~G.} \bibnamefont{Xie~Chen}}
  \bibnamefont{and} \bibinfo{author}{\bibfnamefont{X.-G.} \bibnamefont{Wen}},
  \bibinfo{journal}{Phys. Rev. B} \textbf{\bibinfo{volume}{82}},
  \bibinfo{pages}{155138} (\bibinfo{year}{2010}).

\bibitem[{\citenamefont{Hastings}()}]{Hastingslocality}
\bibinfo{author}{\bibfnamefont{M.~B.} \bibnamefont{Hastings}},
  \bibinfo{note}{arXiv:1008.5137}.

\bibitem[{\citenamefont{Tu et~al.}(2013)\citenamefont{Tu, Zhang, and
  Qi}}]{tuzhangqimomentumone}
\bibinfo{author}{\bibfnamefont{H.-H.} \bibnamefont{Tu}},
  \bibinfo{author}{\bibfnamefont{Y.}~\bibnamefont{Zhang}}, \bibnamefont{and}
  \bibinfo{author}{\bibfnamefont{X.-L.} \bibnamefont{Qi}},
  \bibinfo{journal}{Phys. Rev. B} \textbf{\bibinfo{volume}{88}},
  \bibinfo{pages}{195412} (\bibinfo{year}{2013}).

\bibitem[{\citenamefont{Zaletel et~al.}(2014)\citenamefont{Zaletel, Mong, and
  Pollmann}}]{zaletelflux}
\bibinfo{author}{\bibfnamefont{M.~P.} \bibnamefont{Zaletel}},
  \bibinfo{author}{\bibfnamefont{R.~S.~K.} \bibnamefont{Mong}},
  \bibnamefont{and} \bibinfo{author}{\bibfnamefont{F.}~\bibnamefont{Pollmann}},
  \bibinfo{journal}{J. Stat. Mech.} \textbf{\bibinfo{volume}{2014}},
  \bibinfo{pages}{P10007} (\bibinfo{year}{2014}).

\bibitem[{\citenamefont{Giamarchi}(2004)}]{Giamarchibook}
\bibinfo{author}{\bibfnamefont{T.}~\bibnamefont{Giamarchi}},
  \emph{\bibinfo{title}{Quantum Physics in One Dimension}}, International
  Series of Monographs on Physics (Book 121) (\bibinfo{publisher}{Oxford
  University Press}, \bibinfo{year}{2004}).

\bibitem[{\citenamefont{Martin et~al.}(1997)\citenamefont{Martin, Montambaux,
  and Van}}]{SchulzcoupledLLs}
\bibinfo{editor}{\bibfnamefont{T.}~\bibnamefont{Martin}},
  \bibinfo{editor}{\bibfnamefont{G.}~\bibnamefont{Montambaux}},
  \bibnamefont{and} \bibinfo{editor}{\bibfnamefont{J.~T.~T.}
  \bibnamefont{Van}}, eds., \emph{\bibinfo{title}{Coupled Luttinger Liquids,
  Strongly Correlated Magnetic and Superconducting Systems}}, vol.
  \bibinfo{volume}{478} (\bibinfo{publisher}{Springer-Verlag},
  \bibinfo{year}{1997}).

\bibitem[{\citenamefont{Levin and Stern}(2012)}]{LevinSternclassoffracs}
\bibinfo{author}{\bibfnamefont{M.}~\bibnamefont{Levin}} \bibnamefont{and}
  \bibinfo{author}{\bibfnamefont{A.}~\bibnamefont{Stern}},
  \bibinfo{journal}{Phys. Rev. B} \textbf{\bibinfo{volume}{86}},
  \bibinfo{pages}{115131} (\bibinfo{year}{2012}).

\bibitem[{\citenamefont{Wang and Levin}(2013)}]{WangLevinweaksymmetry}
\bibinfo{author}{\bibfnamefont{C.}~\bibnamefont{Wang}} \bibnamefont{and}
  \bibinfo{author}{\bibfnamefont{M.}~\bibnamefont{Levin}},
  \bibinfo{journal}{Phys. Rev. B} \textbf{\bibinfo{volume}{88}},
  \bibinfo{pages}{245136} (\bibinfo{year}{2013}).

\bibitem[{\citenamefont{Read}(1990)}]{Read90}
\bibinfo{author}{\bibfnamefont{N.}~\bibnamefont{Read}}, \bibinfo{journal}{Phys.
  Rev. Lett.} \textbf{\bibinfo{volume}{65}}, \bibinfo{pages}{1502}
  (\bibinfo{year}{1990}).

\bibitem[{\citenamefont{Wen and Zee}(1992)}]{WenZee92}
\bibinfo{author}{\bibfnamefont{X.~G.} \bibnamefont{Wen}} \bibnamefont{and}
  \bibinfo{author}{\bibfnamefont{A.}~\bibnamefont{Zee}},
  \bibinfo{journal}{Phys. Rev. B} \textbf{\bibinfo{volume}{46}},
  \bibinfo{pages}{2290} (\bibinfo{year}{1992}).

\bibitem[{\citenamefont{Peschel}(2003)}]{peschelreduced}
\bibinfo{author}{\bibfnamefont{I.}~\bibnamefont{Peschel}}, \bibinfo{journal}{J.
  Phys. A} \textbf{\bibinfo{volume}{36}}, \bibinfo{pages}{L205}
  (\bibinfo{year}{2003}).

\bibitem[{\citenamefont{Lu and Lee}(2014)}]{LuLeesymmetryedges}
\bibinfo{author}{\bibfnamefont{Y.-M.} \bibnamefont{Lu}} \bibnamefont{and}
  \bibinfo{author}{\bibfnamefont{D.-H.} \bibnamefont{Lee}},
  \bibinfo{journal}{Phys. Rev. B} \textbf{\bibinfo{volume}{89}},
  \bibinfo{pages}{205117} (\bibinfo{year}{2014}).

\bibitem[{\citenamefont{Chamon and Wen}(1994)}]{chamonwenreconstruction}
\bibinfo{author}{\bibfnamefont{C.~d.~C.} \bibnamefont{Chamon}}
  \bibnamefont{and} \bibinfo{author}{\bibfnamefont{X.~G.} \bibnamefont{Wen}},
  \bibinfo{journal}{Phys. Rev. B} \textbf{\bibinfo{volume}{49}},
  \bibinfo{pages}{8227} (\bibinfo{year}{1994}).

\end{thebibliography}

\end{document}